\documentclass[preprint,12pt]{elsarticle}
\usepackage[utf8]{inputenc}
\usepackage{indentfirst} 
\usepackage{amsmath,amssymb}   
\usepackage{latexsym}
\usepackage{graphicx} 
\usepackage{subfigure}   
\usepackage{color}
\usepackage{CJK}
\usepackage{leftidx}
\usepackage{bm}
\usepackage{ulem}

\journal{Journal of Theoretical Biology}

\begin{document}

\begin{frontmatter}

\title{A Bubble Model for the Gating of K$_\mathrm{v}$ Channels}

\author[label1]{Zilong Song}
\author[label2,label3]{Robert  Eisenberg}
\author[label4]{Shixin Xu}
\author[label5,label6,label7]{Huaxiong Huang \corref{cor1}}
\cortext[cor1]{hhuang@uic.edu.cn,hhuang@yorku.ca}

\address[label1]{Department of Mathematics and Statistics, Utah State University,  3900 Old Main Hill, Logan, UT 84322, USA.}

\address[label2]{Department of Applied Mathematics, Illinois Institute of Technology, Chicago, IL, 60616, USA}
\address[label3]{Department of Physiology and Biophysics, Rush University, Chicago, IL, 60612, USA.}
\address[label4]{Duke Kunshan University, 8 Duke Ave, Kunshan, Jiangsu, China.}
\address[label5]{Research Centre for Mathematics, Advanced Institute of Natural Sciences, Beijing Normal University (Zhuhai), China}

\address[label6]{BNU-HKBU United International College, Zhuhai, China}
\address[label7]{Department of Mathematics and Statistics, York University, Toronto, ON, M3J 1P3, Canada.}

\begin{abstract}
    Voltage-gated K$_\mathrm{v}$ channels play fundamental roles in many  biological processes, such as the generation of the action potential. The gating mechanism of K$_\mathrm{v}$ channels is characterized experimentally by single-channel recordings and ensemble properties of the channel currents. In this work, we propose a bubble model coupled with a Poisson-Nernst-Planck (PNP) system to capture the key characteristics, particularly the delay in the opening of channels. The coupled PNP system is solved numerically by a finite-difference method and the solution is compared with an analytical approximation. We hypothesize that the stochastic behaviour of the gating phenomenon is due to randomness of the bubble and channel sizes. The predicted ensemble average of the currents under various applied voltage across the channels is consistent with experimental observations, and the Cole-Moore delay is captured by varying the holding potential. 
\end{abstract}


\end{frontmatter}

\section{Introduction}

Voltage-gated ion channels play fundamental roles in many biological activities, such as signal generation and propagation in the nervous system, pacemaker activity in the heart, and coordination of contraction in skeletal muscle \cite{biel2009,jensen2012,jacobson2010}. For example, the voltage-gated Na (Na$_\mathrm{v}$) and K (K$_\mathrm{v}$) channels are key players in the generation of action potential (AP) signals in the nervous system \cite{Hille2001}, cardiac and skeletal muscle. This rapid and transient change of membrane potential propagates long distances (meters) in the nervous system and muscle fibers as well. The opening and closing of ion channels as the voltage changes across the membrane determine the  depolarization (positive change in potential) and repolarization (negative change of membrane potential) that form the propagating AP \cite{RN29636,RN26041,RN12551,RN267}. 

The opening of ion channels follows the change in voltage with a delay and that delay is an important determinant of the conduction velocity. The conduction velocity helps determine how fast the nervous system can function. One of the objectives of the present work is to model the delay in the opening of single K$_\mathrm{v}$ channels as well as their ensemble properties. The delay in the opening of Na$_\mathrm{v}$ channels is particularly important in determining the conduction velocity of the action potential.  Therefore, understanding the mechanism of delay is of great biological importance. It is not unreasonable to expect that the delay is set by a process that is optimized as much as possible within the constraints of physics, protein structure, and evolutionary history \cite{RN229}.

Hodgkin and Huxley (HH) provided an empirical model of the generation of AP in 1952 \cite{hodgkin1952}. The conductances they used are ensemble averages of those from many channels. Understanding the molecular mechanisms that produce these conductances and the AP is one of the main goals of biophysics for the past seventy years. Recent advances in structural biology \cite{RN5856} and single-channel recording \cite{RN22885} have catalyzed our understanding of the physical mechanisms that produce these conductances. The ionic basis of selective conduction is now understood reasonably well for sodium channels \cite{RN26287,RN45982,RN45981,RN45983}.

The opening and closing of voltage-dependent channels involves many steps \cite{RN11549,RN11546,RN29127}. Some of the steps in the voltage-dependent gating of K$_\mathrm{v}$ are now known in molecular and physical detail \cite{lacroix2014,catacuzzeno2019,catacuzzeno2020,catacuzzeno2020b,bassetto2021,RN30709,catacuzzeno2021,horng2016,horng2019,Hille2001,llano1986,llano1988,kim2014,bezanilla2002}.
The first step is the response of the voltage sensor to the voltage change, and significant progress has been made in understanding the physics of that response. It is plausible \cite{hodgkin1952} that the permeability changes depend on the presence of voltage sensors in the form of charged or dipole particles, as suggested earlier in a different form \cite{RN12551,RN291}. The second step is the communication of the voltage sensor with the conduction pore of the channel. This was revealed experimentally in the single-channel ON-OFF  currents (that occur at random intervals) measured by bilayer or patch-clamp experiments from one channel protein at a time \cite{RN135,RN25640}. The development of patch-clamp experiments \cite{hamill1981} was a breakthrough in the understanding of the gating mechanisms and provided experimental verification at high resolution of many studies and models.  

In the patch-clamp experiments, the recordings of single K$_\mathrm{v}$ channels showed a delay of currents in response to a step voltage change. The ionic current was generated rapidly after the delay, and vanished when the channel closed suddenly \cite{llano1986,llano1988, Hille2001,werry2013}. The recordings also showed that the delays varied in each ON-OFF experiment: the gating transitions are stochastic. The ensemble average has a smoother transient time course for the currents (or opening and closing of channels), which resembles the classical macroscopic currents (or voltage-dependent conductances) in the HH model. 

The delay in opening was first studied  in the inaugural issue of the Biophysical Journal \cite{cole1960} in the ensemble of channels.  Cole and Moore were able to control the resting potential (i.e., their holding potential) present before the AP mechanism was turned on. The earlier work of Hodgkin and Huxley had not addressed this issue in detail because the actual resting potential of their squid nerve was substantially different from that used as a holding potential \cite{moore1960}. Hodgkin and Huxley chose to use nerve fibers with more positive resting potentials so their voltage clamp system could control the voltage throughout the nerve fiber, something not easy to do \cite{taylor1960}. Cole and Moore found the delay in the response of the nerve fiber to a change in voltage was much larger when the initial potential (also called the holding potential) was more negative. 

Given the importance of this delay (we call the Cole-Moore delay), it is striking that a molecular scale biophysical explanation has not been developed, as far as we know, until very recently \cite{RN45984}. Given the obvious evolutionary disadvantage of additional delay, it seems likely that whatever is responsible for the delay is an essential component of the ionic channels that create the AP. We expect the cause of the Cole-Moore delay to be found in many channel types where it has not been investigated in detail.

The amount of work on channel proteins that produce the AP has increased spectacularly in the last decades. The most important single advance (from a biophysical point of view) is the ON-OFF properties of the single channels, that in ensemble produce the Cole-Moore delay. Many researchers have proposed that the ON-OFF property arises from the collapse of a bubble. When the single-channel current is zero, a region of the voltage sensor with low effective dielectric constant \cite{catacuzzeno2021} acts as a hydrophobic gasket that excludes water and ions from that region of the protein, forming a dewetted region, which is known as a bubble. Direct evidence for the existence of bubbles is emerging as structural biologists exploit the magnificent capabilities of modern techniques of x-ray crystallography and cryoelectron microscopy \cite{lacroix2014,langan2020}.

Various modeling efforts have been devoted to understanding the gating mechanisms of K$_\mathrm{v}$ channels \cite{catacuzzeno2020}. In the early years, kinetic models (or called Markov models) were used for channel gating, by assuming the voltage sensor has multiple subunits which make transitions between different states \cite{schoppa1998, bezanilla1994, tytgat1992}. Formally, such kinetic models have some similarity to Hodgkin and Huxley's, as the four $n$-gates in the HH model can be interpreted as four independent subunits that control the gating \cite{Hille2001}.  Such models have been able to predict some important features of the gating mechanism of K$_\mathrm{v}$ channels (e.g., Shaker channel)  and to fit experimental data, but could not reveal much about the physics of the gating process. With the availability of more structural information about channels and advances in computing power, quantitative models using molecular dynamics (MD) have been developed in recent decades \cite{delemotte2011,delemotte2017,jensen2012}. MD simulations incorporating physical laws and interactions of atoms provide insights into the movement of the voltage sensors, intermediate states, and closure of the pore (forming a dewetted region). However, the MD approach is limited by the timescale of the simulations, resolving events in the timescale of $10^{-15}$s, and the total simulation length is orders of magnitude lower than the timespan (e.g., $10^{-3}$ s) of experimentally or biologically relevant processes. This makes it difficult to directly validate the MD results by using the macroscopic currents in experiments. To overcome these limitations, alternative multiscale or macroscopic models \cite{peyser2012,dryga2012,kim2014,horng2019} have been developed with reasonable approximations. Some models are based on the formulation of Brownian dynamics, where the voltage sensor is treated as a Brownian particle \cite{catacuzzeno2019}. Brownian models are able to predict macroscopic gating currents, where the free parameters involved have been estimated based on multiscale modeling approaches \cite{catacuzzeno2021}.

Here we take a different approach. Following the previous hypothesis of the hydrophobic region, we construct a specific macroscopic model of a bubble within the framework of Poisson-Nernst-Planck (PNP) systems and show how it produces the time course of single K$_\mathrm{v}$ channels and the ensemble properties, including the Cole-Moore delay. The PNP system and its variants have been found successful in modeling and simulation of many biological processes \cite{song2020,song2019,horng2019,cao2020,song2018,eisenberg2007,zhang2020}, such as current-voltage curves through ion channels, the selectivity of ion channels, and ion transport processes in the cell and tissue scales. In this work, a bubble is assumed to be present in the pore (or filter) region of the K$_\mathrm{v}$ channel, due to the properties of the gating sensor and channel walls. In the bubble, ions are not present and so cannot carry charge through it, whereas outside the bubble, ion transport is governed by the PNP system. The model is constructed so it can easily accommodate more specific structural information such as the shape, permanent charge (e.g., the spatial distribution of acid and base residue side chains), and dielectric properties of the voltage sensor and conduction pore of channels. We calculate the properties of a single channel containing a bubble and an ensemble average based on a simple statistical distribution of such channels to represent the macroscopic currents usually recorded in studies of the opening and closing of channels. This average does not depend on models  \cite{colquhoun1981,colquhoun1995} of single-channel kinetics. It only assumes that the opening of each channel (or voltage sensor) is independent of the others (because channels are many Debye lengths apart, shielded by the ions, water dipoles (and quadrupoles), and the ionic atmosphere of proteins and lipid bilayer). 

This manuscript is arranged as follows. Section 2 sets up the bubble model within the framework of PNP systems, followed by a nondimensionalization. In section 3, the results for a single channel are presented. The bubble model is solved by a finite-difference method and also solved with analytical approximations. The results for the profiles of quantities in the model and the macroscopic currents through the channel are cross-validated by both methods. Section 4 shows the results for ensemble properties of the K$_\mathrm{v}$ channels and the Cole-Moore delay, with certain assumptions on the statistical distributions of the bubble locations and cross-sectional area of the channel. Finally, some concluding remarks are provided in Section 5. 

\section{A bubble model for a voltage-gated Potassium channel}

\subsection{The model setup}

\begin{figure}[h]
\begin{center}
\includegraphics[width=14cm]{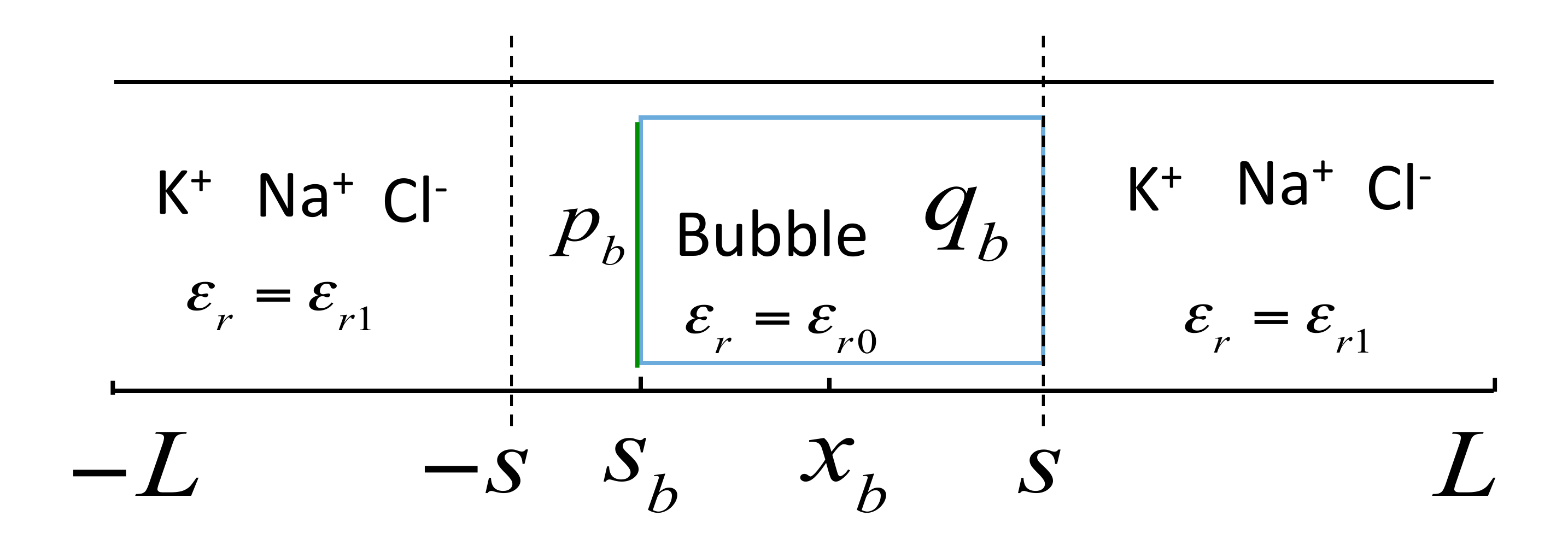}
\caption{Sketch of the K$_\mathrm{v}$ channel with the bubble in the middle region.}
\label{fig1}
\end{center}
\end{figure}

We consider a voltage-gated Potassium (K$_\mathrm{v}$) channel in one spatial dimension, as shown in Figure \ref{fig1}. The total length of channel is set as $2 L $, and the length of the middle  (filter and pore) region is $2s$. The positions $x=\pm s$ are the locations of the two edges of the middle region. The bubble, which carried negative charges with magnitude  $q_b>0$,  can occupy all or part of it, and is centered at $x=x_b$. We assume that the charge is uniformly distributed inside the bubble. The left chamber is connected to a bath environment similar to the exterior of a cell, while the right chamber is connected to one similar to interior of a cell. We further assume that the right interface of the bubble is fixed at $x=s$ and the left interface $x= s_b =  -s+2 x_b$ is mobile. We anticipate that when the voltage at the right end of the channel is elevated, the bubble shrinks and moves to the right. When the left and right interfaces coincide, the bubble vanishes. 

We consider the case with three ions species K$^+$, Na$^+$ and Cl$^-$ (sometimes called the major bio-ions) outside of the bubble, and the ions can not penetrate into the bubble. Outside of the bubble, the PNP system is used to model ion transport
\begin{equation}
\label{eq1}
\begin{aligned}
&  -\epsilon_0 \partial_x ( \epsilon_r \partial_x\phi) = e_0 (c_1+c_2- c_3), \quad -L<x<s_b, ~ s<x<L\\
& \frac{\partial c_i}{\partial t}  = - \partial_x J_i = D_i \partial_x \left( \partial_x c_i + \frac{e_0 z_i}{k_B T}  c_i \partial_x \phi \right),\quad i =1,2,3,
\end{aligned}
\end{equation}
where $c_1$, $c_2$ and $c_3$ are the concentrations of K$^+$, Na$^+$ and Cl$^-$ with valences $z_1=1$, $z_2=1$, $z_3= -1$, $\phi$ is the electric potential, $\epsilon_r$ is the dielectric constant, $D_i$ ($i=1,2,3$) are diffusion coefficients, and $\epsilon_0, e_0, k_B, T$ are constants given in Appendix A. 

The boundary conditions are given by 
\begin{equation}
\label{eq8}
\begin{aligned}
& \phi (-L, t)=0 , \quad \phi(L, t) =V_0+ V_1 H(t-t_1),\\
& c_i (-L, t) = c_i^L, \quad c_i (L, t) = c_i^R, \quad i=1,2,3,
\end{aligned}
\end{equation}
where $V_0$ is the initial (holding) membrane potential when the bubble is in equilibrium (or resting state), $V_1$ is the voltage jump at $t=t_1$, $H(t)$ is a Heaviside function, and $c_i^L$ and $c_i^R$ ($i=1,2,3$) are given bath concentrations at the left and right ends \cite{llano1988}, which are electro-neutral. In the experiment, the holding potential $V_0$ is not the same as the Nernst potential of K$^+$. The leak current is allowed to flow through a different pathway while maintaining $V_0$. 

Inside the bubble, we have
\begin{equation}
\label{eq2}
\begin{aligned}
&  -\epsilon_0 \partial_x ( \epsilon_r \partial_x\phi) =- \frac{q_b}{V_b}, \quad s_b(t)<x<s,
\end{aligned}
\end{equation}
where $V_b = (s- s_b)A$ is the volume of the bubble and $A$ is the cross sectional area for the bubble region. The dielectric constant is defined as
\begin{equation}
\label{eq3}
\begin{aligned}
&  \epsilon_r= \begin{cases}
\epsilon_{r0}, \quad [s_b,s] \\ 
\epsilon_{r1}, \quad \mathrm{others}.
\end{cases}
\end{aligned}
\end{equation}

In addition, we assume that there exists a dipole on the left interface of the bubble $x=s_b$, responsible for maintaining a voltage difference on the two sides of the bubble. Since the membrane potential is not 0 at equilibrium, the presence of the dipole with a suitable dipole strength $p_b$ guarantees that the bubble is in equilibrium initially. We can rewrite the equation of $\phi$ in a compact form in the entire domain
\begin{equation}
\label{eq4}
\begin{aligned}
&  -\epsilon_0 \partial_x ( \epsilon_r \partial_x\phi) = e_0 (c_1+c_2-c_3) - \frac{q_b}{V_b} + p_b \partial_x (\delta (x-s_b)), ~ -L<x<L,
\end{aligned}
\end{equation}
with the interpretation that $c_i = 0$ ($i=1,2,3$) in the bubble and $q_b=0$ outside of the bubble.

The total electric force on the bubble is 
\begin{equation}
\label{eq5}
\begin{aligned}
& \int_A \int_{s_b}^s -\frac{q_b}{V_b} \left(-\partial_x \phi\right) dx dA= q_b \frac{\phi (s) - \phi_(s_b)}{s-s_b},
\end{aligned}
\end{equation}
then the motion of the bubble is modeled by
\begin{equation}
\label{eq6}
\begin{aligned}
& \frac{d x_b}{d t} =\frac{q_bD_b}{k_B T}  \frac{\phi (s) - \phi_(s_b)}{s-s_b},
\end{aligned}
\end{equation}
where $D_b$ ($\ll D_i$) is the diffusion coefficient of the bubble. Using the relationship $s_b = -s + 2 x_b$, we can rewrite equation (\ref{eq6}) as
\begin{equation}
\label{eq7}
\begin{aligned}
& \frac{d s_b}{d t} = \frac{2q_bD_b}{k_B T}   \frac{\phi (s) - \phi_(s_b)}{s-s_b} .
\end{aligned}
\end{equation}

At the two interfaces $x=s_b$ and $s$, the electric potential and electric displacement are continuous, and there is no ionic flux across the bubble interfaces. Mathematically, we set
\begin{equation}
\label{eq9}
\begin{aligned}
&  [\phi]=0, \quad [\epsilon_r \partial_x \phi ]= 0, \quad J_i  = 0, \; (i=1,2,3),
\end{aligned}
\end{equation}
where square brackets mean the jump across the interface, e.g., $[\phi(s)] = \phi(s+) - \phi(s-)$. If we include the effect of dipole ($p_b$ in equation (\ref{eq4})) on the interface, we obtain a nonzero jump $[\phi]$ at $x= s_b$. When the two interfaces coincide (i.e., $s_b=s$), the bubble collapses. And we assume that the dipole disappears (i.e., it is treated as an intrinsic property of the bubble) and the interface conditions are replaced by continuity conditions
\begin{equation}
\label{eq10}
\begin{aligned}
&  [\phi]=0, \quad [\epsilon_r \partial_x \phi ]= 0,\quad [c_i] =0, \quad  [J_i]  = 0, \; (i=1,2,3).
\end{aligned}
\end{equation}

In summary, we have a system of equations for ion transport coupled with the motion of the bubble, given by (\ref{eq4}), $(\ref{eq1})_2$ and (\ref{eq7}), together with boundary and interface conditions (\ref{eq8},\ref{eq9},\ref{eq10}). The total current is conserved in this model, by including three different types of current, given in Appendix B. This is a special case of the continuity of total current for Maxwell equations \cite{eisenberg2017,eisenberg2018}, and is also similar to the case of a PNP system for electric eels \cite{song2020}.\\

{\noindent \bf Remark 1.}  If the dipole does not vanish (i.e., it is treated as property of the channel or channel wall) after the bubble collapses, we will have nonzero jump $[\phi]$ related to the dipole, and $[c_i] = 0$ ($i=1,2,3$) are replaced by continuity of electro-chemical potentials.

\subsection{Nondimensionalization}

In this subsection, we nondimensionalize our model, which will be used in the calculations in the subsequent sections. We adopt the following scales
\begin{equation}
\label{eq11}
\begin{aligned}
& \tilde{x} = \frac{x}{L}, \quad \tilde{s} = \frac{s}{L},\quad  \tilde{x}_b = \frac{x_b}{L}, \quad \tilde{V}_b = \frac{V_b}{ L A},\\
& \tilde{\phi} = \frac{\phi}{k_B T/e_0}, \quad \tilde{V}_0 = \frac{V_0}{k_B T/e_0}, \quad \tilde{V}_1 = \frac{V_1}{k_B T/e_0},  \\
& \tilde{c}_i = \frac{c_i}{c_0}, \quad \tilde{c}_i^L = \frac{c_i^L}{c_0},\quad \tilde{c}_i^R = \frac{c_i^R}{c_0}, \quad \tilde{D}_i = \frac{D_i}{D_0},(i=1,2,3), \\
&  \tilde{D}_b = \frac{D_b}{D_0}, \quad \tilde{p}_b= \frac{p_b}{e_0 c_0 L^2}, \quad \tilde{q}_b = \frac{q_b}{e_0},\\
& \tilde{t} =\frac{t}{t_0}, \quad t_0 = \frac{L^2}{D_0}, \quad \tilde{J} = \frac{J}{J_0 },\quad J_0= \frac{D_0 c_0}{L}.
\end{aligned}
\end{equation}
Some typical values in the above scales and the following boundary conditions are based on \cite{llano1988} and given in Appendix A.

Substituting (\ref{eq11}) into the system in the previous subsection, we obtain a dimensionless system for variables with tilde (like $\tilde{\phi}$). In order to simply the notations, we drop the tilde and use the quantities (like $\phi$) in the dimensionless system. We have the following set of equations in nondimensional form
\begin{equation}
\label{eq12}
\begin{aligned}
&  -\epsilon \partial_x ( \epsilon_r \partial_x\phi) = c_1+c_2 -c_3- \frac{1}{\beta} \frac{{q}_b}{(s-s_b)}  + p_b \partial_x (\delta(x-s_b)), \quad -1<x<1\\
& \frac{\partial c_i}{\partial t}  = - \partial_x J_i = D_i \partial_x \left( \partial_x c_i + z_i c_i  \partial_x \phi \right),\quad i =1,2, 3,\quad -1<x<s_b, \, s<x<1,
\end{aligned}
\end{equation}
with the interpretation that $c_i=0$ ($i=1,2,3$) in the bubble $x\in [s_b,s]$ and $q_b =0$ outside of the bubble. Here the two dimensionless parameters are defined by
\begin{equation}
\label{eq13}
\begin{aligned}
& \epsilon = \frac{\epsilon_0 k_B T}{e_0^2  c_0 L^2}, \quad \beta = LA c_0.
\end{aligned}
\end{equation}
The dielectric constant remains the same
\begin{equation}
\label{eq14}
\begin{aligned}
&  \epsilon_r= \begin{cases}
\epsilon_{r0}, \quad [s_b,s] \\ 
\epsilon_{r1}, \quad \mathrm{others}
\end{cases}
\end{aligned}
\end{equation}
The motion of the bubble is given by 
\begin{equation}
\label{eq15}
\begin{aligned}
& \frac{d s_b}{d t} =2 D_b q_b  \frac{\phi (s) - \phi_(s_b)}{s-s_b}. 
\end{aligned}
\end{equation}

Boundary conditions are given by
\begin{equation}
\label{eq16}
\begin{aligned}
& \phi (-1,t)= 0, \quad \phi(1,t) =V_0 + V_1 * H(t-t_1),\\
& c_i (-1,t) = c_i^L, \quad c_i (1,t) = c_i^R, \; (i=1,2,3).
\end{aligned}
\end{equation}
Interface conditions are
\begin{equation}
\label{eq17}
\begin{aligned}
& [\phi]=0 , \quad [\epsilon_r \partial_x \phi ]= 0, \quad J_i  = 0, (i=1,2,3), \quad \textrm{at} ~~ x= s_b,s.
\end{aligned}
\end{equation}
After the bubble collapses (for the case that the dipole disappears), we have
\begin{equation}
\label{eq18}
\begin{aligned}
& [\phi]=0 , \quad [\epsilon_r \partial_x \phi ]= 0,   \quad [c_i] =0, \quad [J_i]  = 0, (i=1,2,3), \quad \textrm{at} ~~ x=s_b= s .
\end{aligned}
\end{equation}


\section{Results for a single channel}

We first compute the initial state when the bubble is in equilibrium by solving the system of equations with a numerical method, followed by the results of the non-equilibrium state including the motion of the bubble and time evolution of the concentrations and electric potential. After the bubble collapses, the ionic fluxes reaches a steady state. In addition, we also present the results obtained with an approximate solution (and numerical evidence) for the intermediate quasi-static states and the final steady state.

\subsection{Initial state and strength of dipole}

We examine the case that the bubble initially occupies the entire middle region and stays at equilibrium, i.e., $s_b = -s$. When $V_0=0$, the bubble is in equilibrium due to symmetry. If $V_0\ne 0$, one the other hand, equilibrium is achieved for an appropriate dipole strength $p_b$.  

Near the interface $x=s_b$, the effect of the other terms is small compared the dipole, and  equation (\ref{eq12}) becomes
\begin{equation}
\label{eq19}
\begin{aligned}
&  -\epsilon \partial_x ( \epsilon_r \partial_x\phi) =  p_b \partial_x (\delta(x-s_b)),
\end{aligned}
\end{equation}
and integrating once gives
\begin{equation}
\label{eq20}
\begin{aligned}
&  -\epsilon \epsilon_r \partial_x\phi (x)  =  p_b\delta(x-s_b) +C.
\end{aligned}
\end{equation}
By integrating again and taking the limit of $x\rightarrow s_b$, we obtain
\begin{equation}
\label{eq21}
\begin{aligned}
& [\phi(s_b )] = -\frac{p_b}{\epsilon} \left(\frac{1}{2 \epsilon_{r0}} + \frac{1}{2 \epsilon_{r1}}\right).
\end{aligned}
\end{equation}
Therefore, for a given $V_0$, we find the following relationship 
\begin{equation}
\label{eq22}
\begin{aligned}
& -\frac{p_b}{\epsilon} \left(\frac{1}{2 \epsilon_{r0}} + \frac{1}{2 \epsilon_{r1}}\right)=V_0,
\end{aligned}
\end{equation}
and then the bubble will be in equilibrium as in the symmetric case with $V_0=0$.

For the equilibrium profile, the fluxes are 0 and one can not distinguish the effects of the two positive ions Na$^+$ and K$^+$. We can group the two positive ion species and treat them as a single species.  The boundary conditions for $c_1 + c_2$ and $c_3$ will be the same, and hence we will have exact symmetry for this equilibrium case. The equilibrium profiles can be determined analytically, and we take $V_0=0$ and $p_b=0$ in the derivation. Since the bubble is in equilibrium, inside the bubble we have (note $s_b = -s$)
\begin{equation}
\label{eq23}
\begin{aligned}
& \phi (x) = B_1 x^2 + \phi(0), \quad B_1= \frac{{q}_b}{4s \epsilon \epsilon_{r0} \beta}.
\end{aligned}
\end{equation}
Taking the derivative and together with interface conditions at $x=s$, we have 
\begin{equation}
\label{eq24}
\begin{aligned}
&\epsilon_{r1} \phi'(s+) =  \epsilon_{r0} \phi'(s-)= \epsilon_{r0} 2 B_1  s = \frac{q_b}{2 \epsilon \beta}.
\end{aligned}
\end{equation}
Due to symmetry, we only consider the right chamber $s<x<1$. It is easy to verify that the PNP system (\ref{eq12}) in equilibrium reduces to
\begin{equation}
\label{eq25}
\begin{aligned}
& \epsilon \epsilon_{r1} \phi'' = c_3-(c_1+c_2) =  e^{\phi} - e^{-\phi},
\end{aligned}
\end{equation}
where $c_3^R=1$ has been used. Integrating once gives
\begin{equation}
\label{eq26}
\begin{aligned}
& \frac{1}{2}\epsilon \epsilon_{r1} [(\phi'(x))^2 - (\phi'(s+))^2 ] =e^{\phi}+ e^{-\phi} - (e^{\phi_s}+ e^{-\phi_s}),
\end{aligned}
\end{equation}
where $\phi_s = \phi(s)$. Then, by combining with (\ref{eq23}), we get 
\begin{equation}
\label{eq27}
\begin{aligned}
 (\phi'(x))^2 = G(\phi)& = \left(\frac{q_b}{2 \epsilon \beta \epsilon_{r1}}\right)^2 + \frac{2}{\epsilon \epsilon_{r1}}\left(e^{\phi}+ e^{-\phi} - (e^{\phi_s}+ e^{-\phi_s})\right),
\end{aligned}
\end{equation}
which leads to the solution
\begin{equation}
\label{eq28}
\begin{aligned}
& x=  \int_{\phi_s}^\phi \frac{1}{\sqrt{G(\phi)}} d\phi + s.
\end{aligned}
\end{equation}
The unknown constant $\phi_s$ in the solution can be determined by the condition
\begin{equation}
\label{eq29}
\begin{aligned}
1=  \int_{\phi_s}^{0} \frac{1}{\sqrt{G(\phi)}} d\phi + s .
\end{aligned}
\end{equation}

{\noindent \bf Remark 2.} Because of symmetry, we can estimate $\phi_s$ from the above derivation as
\begin{equation}
\label{eq30}
\begin{aligned}
\phi_s \approx -\ln \left(\frac{q_b^2}{8 \epsilon \epsilon_{r1} \beta^2}\right)
\end{aligned}
\end{equation}
for $q_b$ in a certain range. 

\begin{figure}[h]
\begin{center}
\includegraphics[width=6cm]{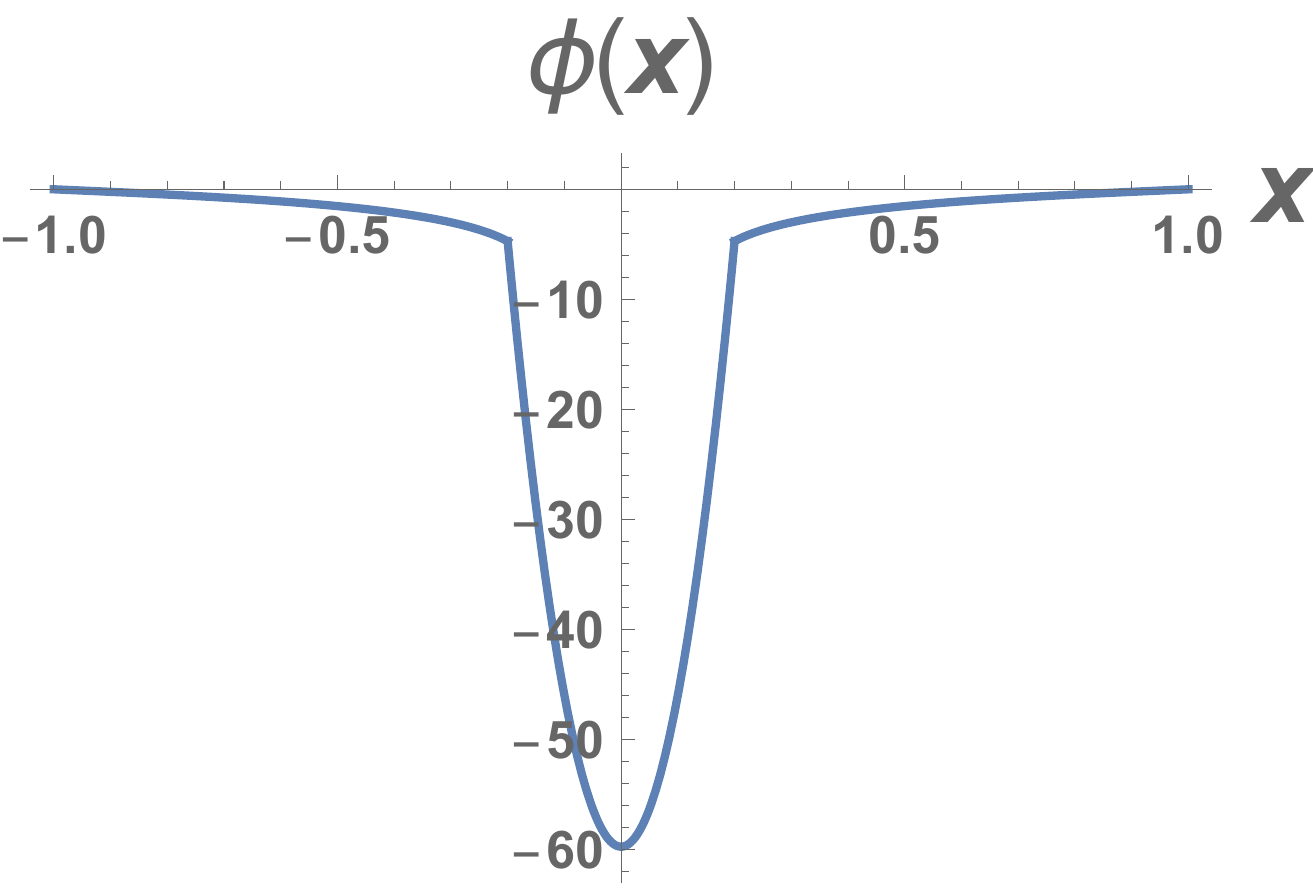} \includegraphics[width=6cm]{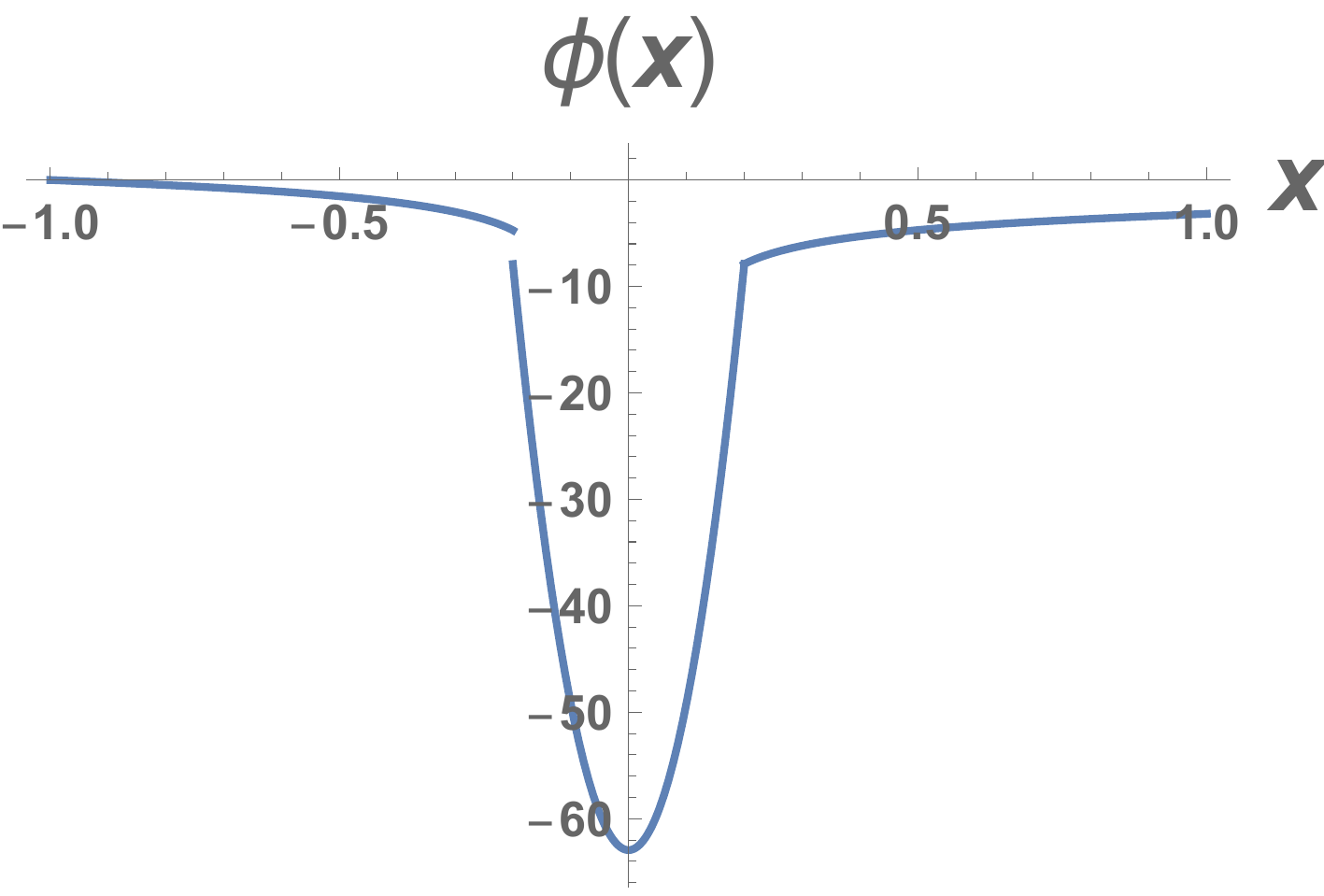} \\
\includegraphics[width=4.4cm]{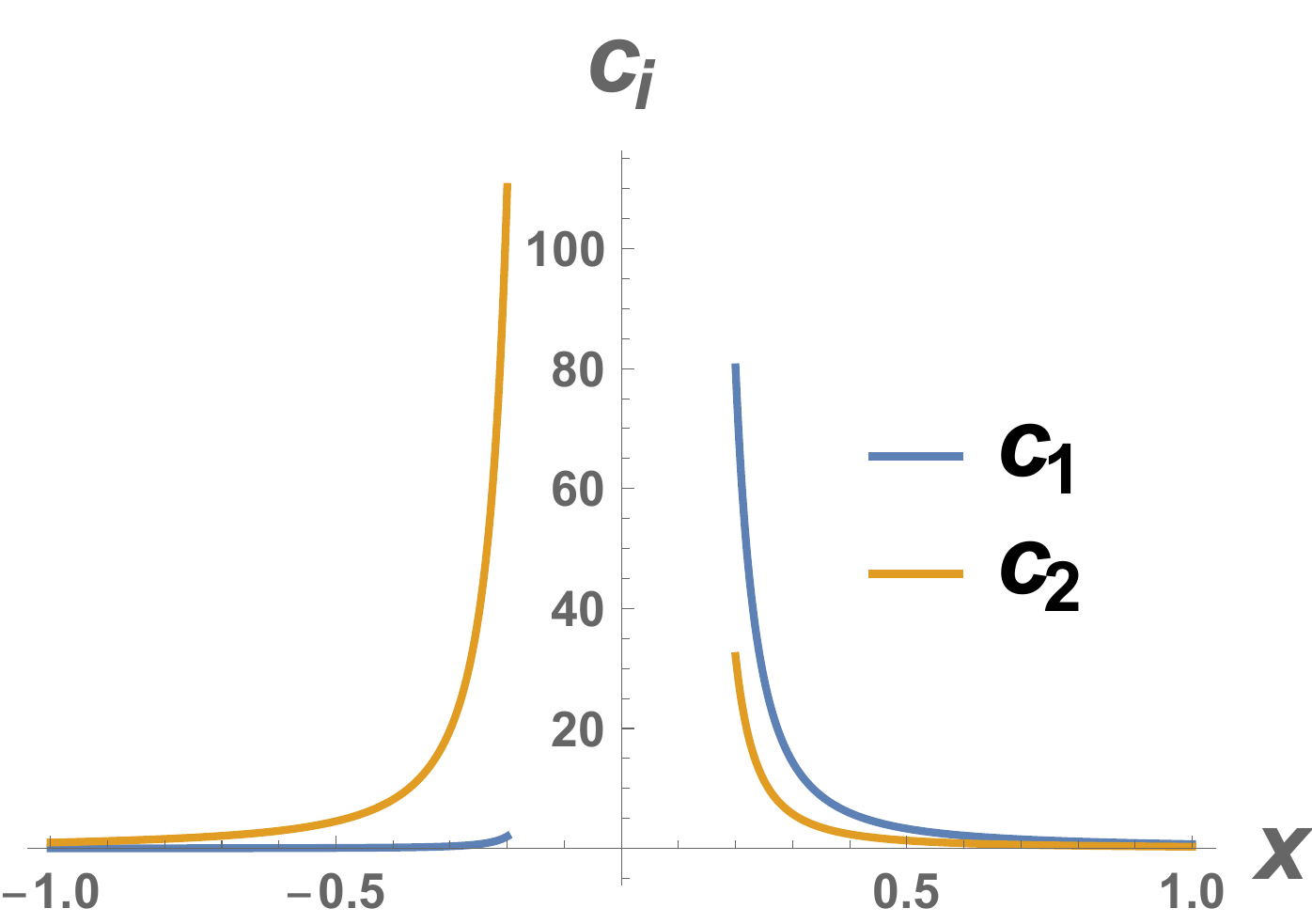} \includegraphics[width=4.4cm]{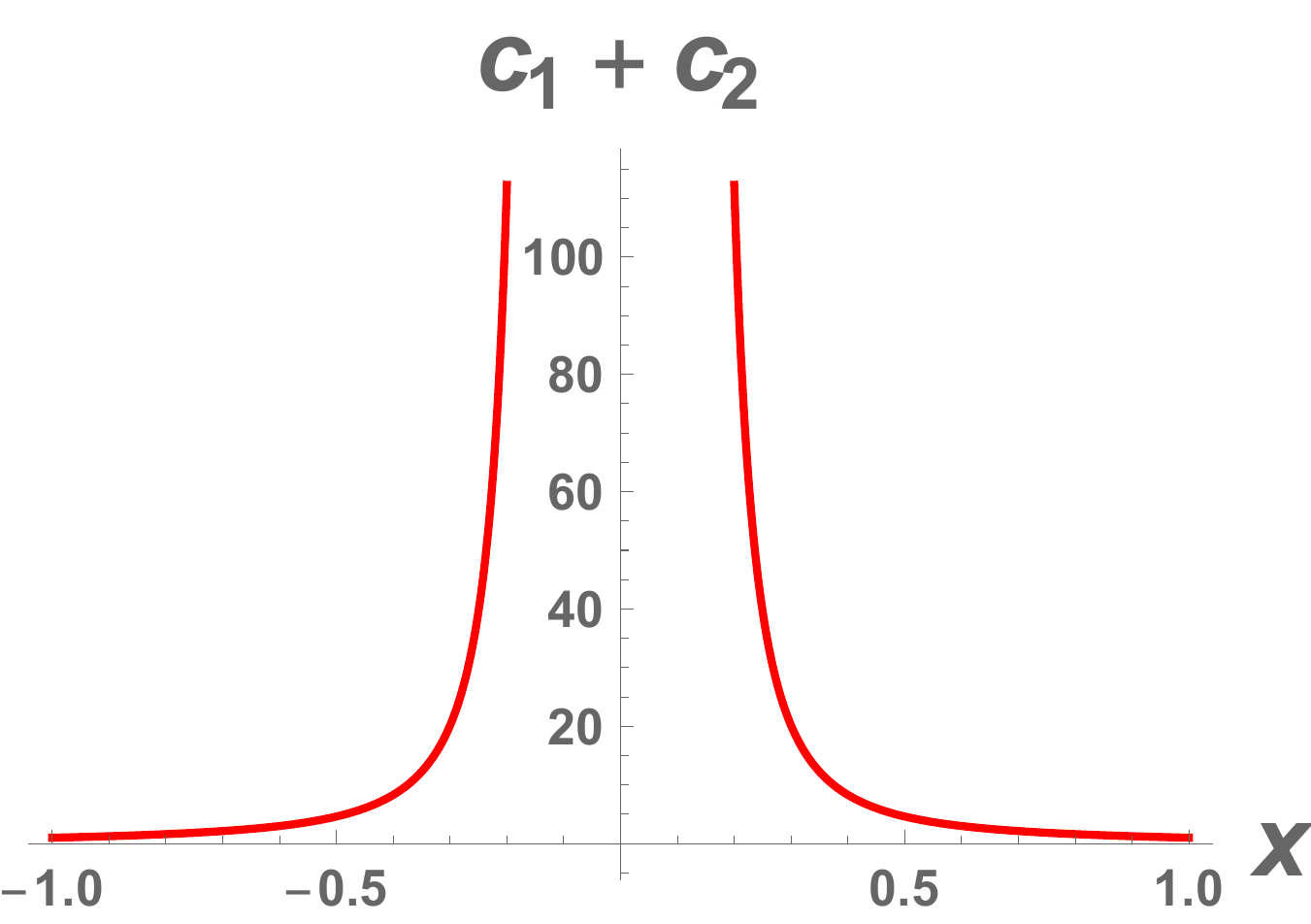} \includegraphics[width=4.4cm]{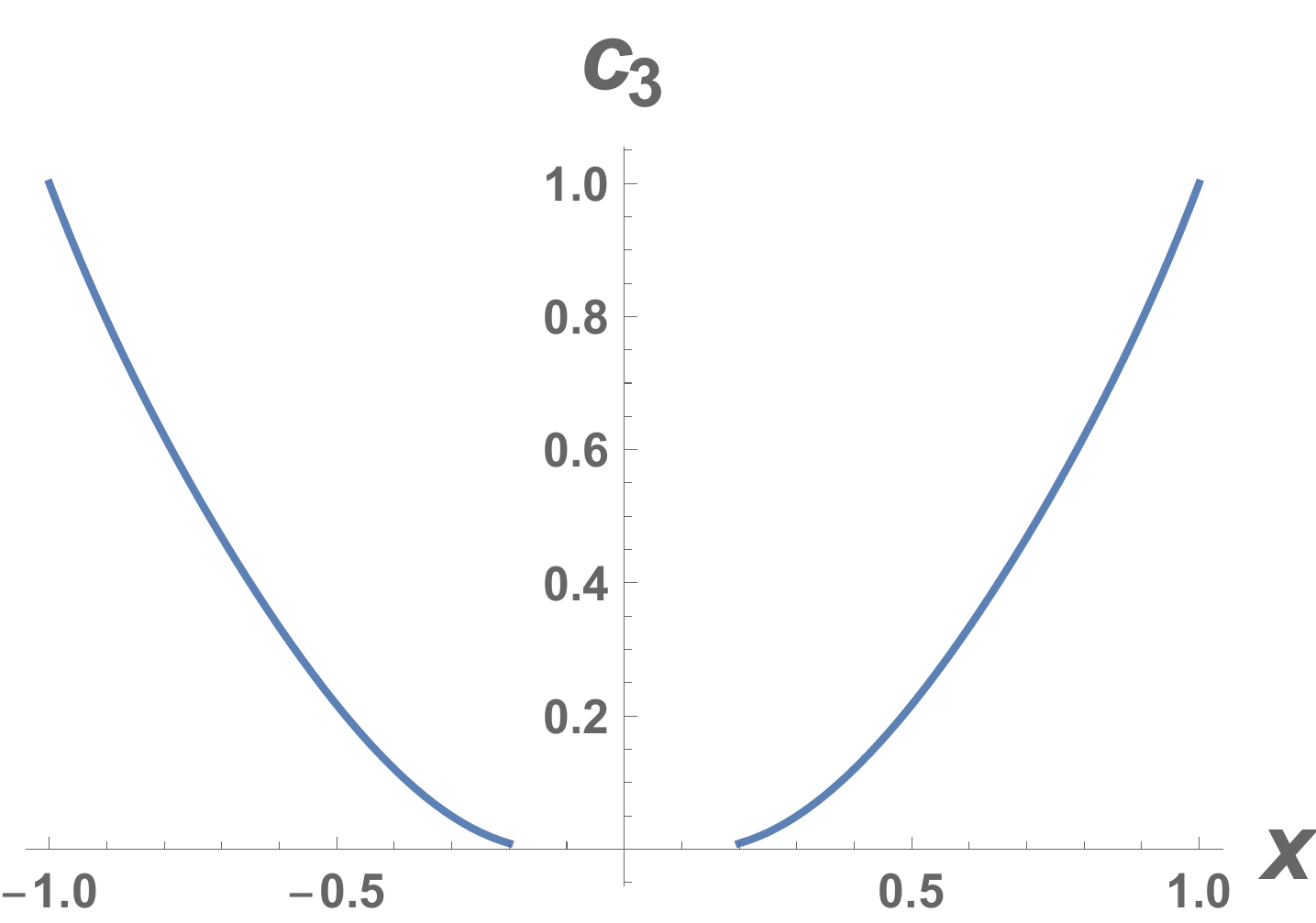} 
\caption{Equilibrium electric potential $\phi$ and concentrations $c_1,c_2,c_3$ when $q_b=2$.}
\label{fig2}
\end{center}
\end{figure}

For the numerical results obtained in this paper, we vary the potentials at the two end of the domain while fixing the magnitude of permanent charge as $q_b=2$.  Most of the other parameter values used for the computation are also fixed and given in Appendix A.  

In Figure \ref{fig2}(a), the electric potential $\phi$ is plotted for the case of $V_0=0$. When $V_0=-3.18$ (i.e., -80 mV), we obtain $p_b \approx 0.044$ using (\ref{eq22}) and the electric potential is plotted in Figure \ref{fig2}(b), where the jump at the interface $x=s_b$ is due to the presence of the dipole. The concentrations $c_1$, $c_2$ and $c_3$ (which can be computed from solution of $\phi$) are shown in Figures \ref{fig2}(c-e). It can be seen that $c_1+c_2$ is symmetric (Figure \ref{fig2}(d)) as expected. The initial membrane potential $V_0$ is balanced by the jump of $\phi$ due to the presence of the dipole. In the non-equilibrium case (before the bubble collapses), we ignore both the initial membrane potential $V_0$ and the dipole, so that the value of $\phi$ is continuous at the interfaces. The solutions in Figure \ref{fig2} will be verified by numerical simulations in the subsequent subsections.

\subsection{The dynamics of the bubble motion and channel currents}

In this part, we present numerical solutions of the PNP system and bubble motion.  Inside the bubble, there exist no ions and their concentrations $c_i$ ($i=1,2,3$) are zero. For convenience, the PNP system is solved inside the bubble by assigning small diffusion coefficients ($D_1 = D_2 = 10^{-15}$). 

\begin{figure}[h]
\begin{center}
\includegraphics[width=5cm]{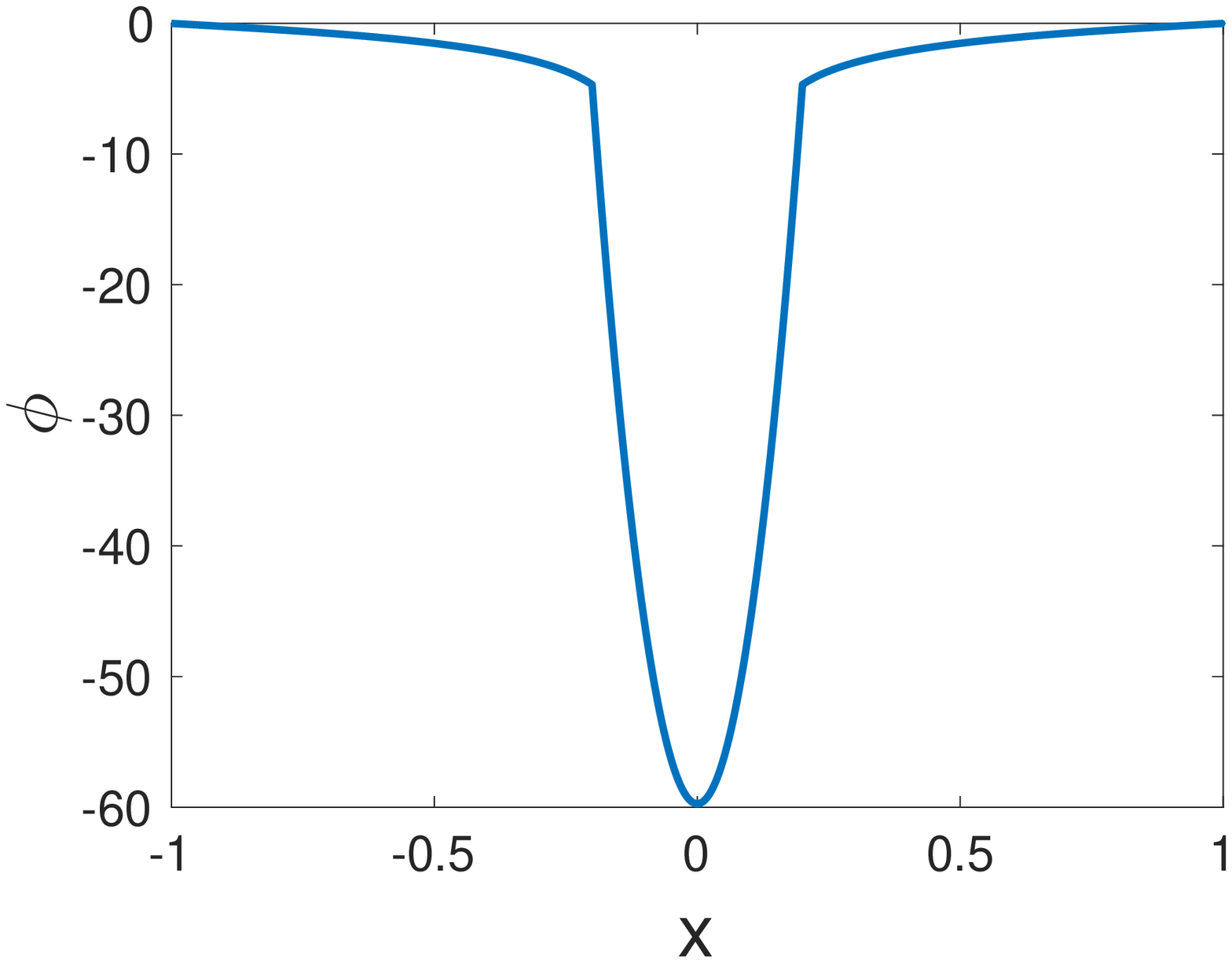} \includegraphics[width=5cm]{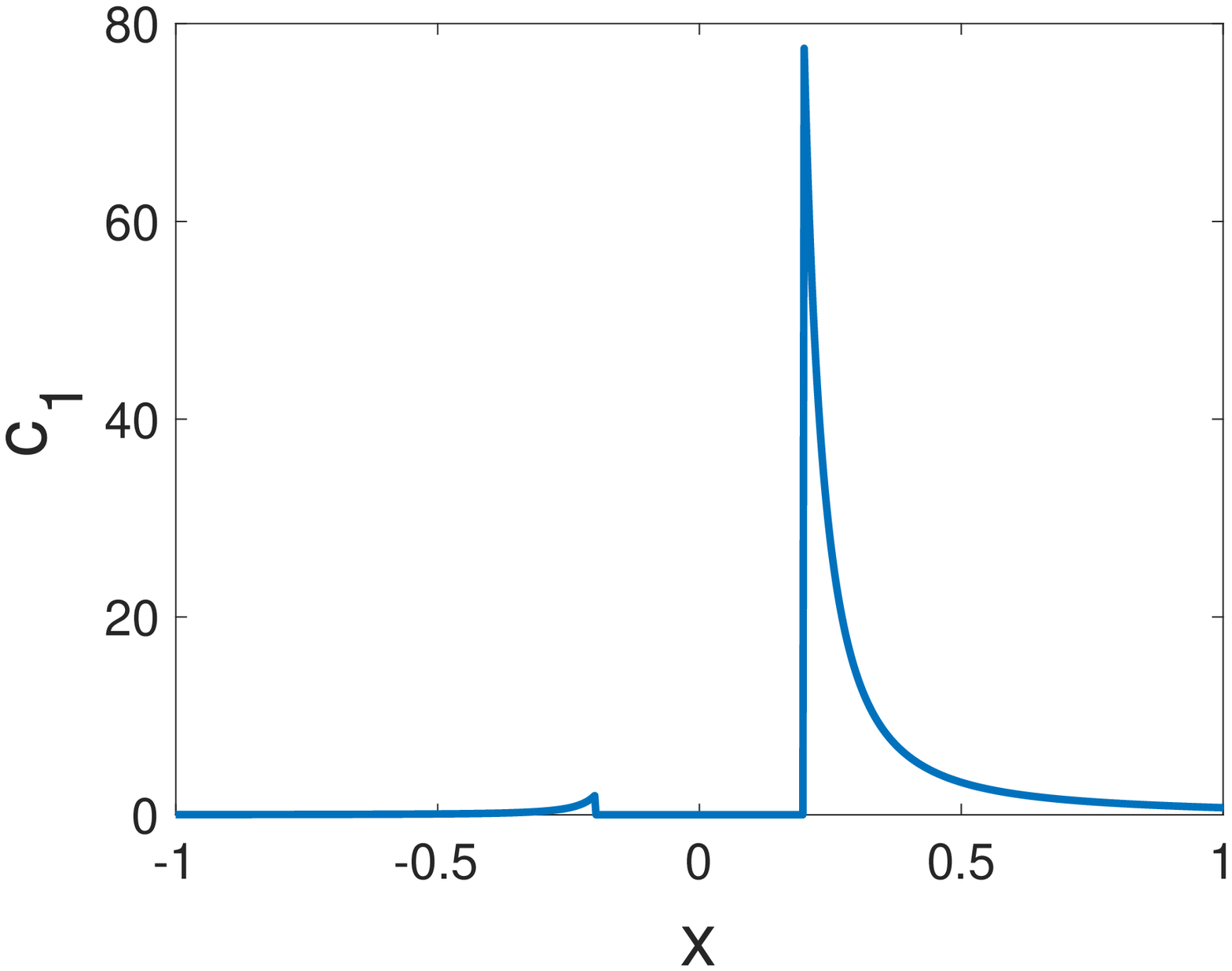}
\includegraphics[width=5cm]{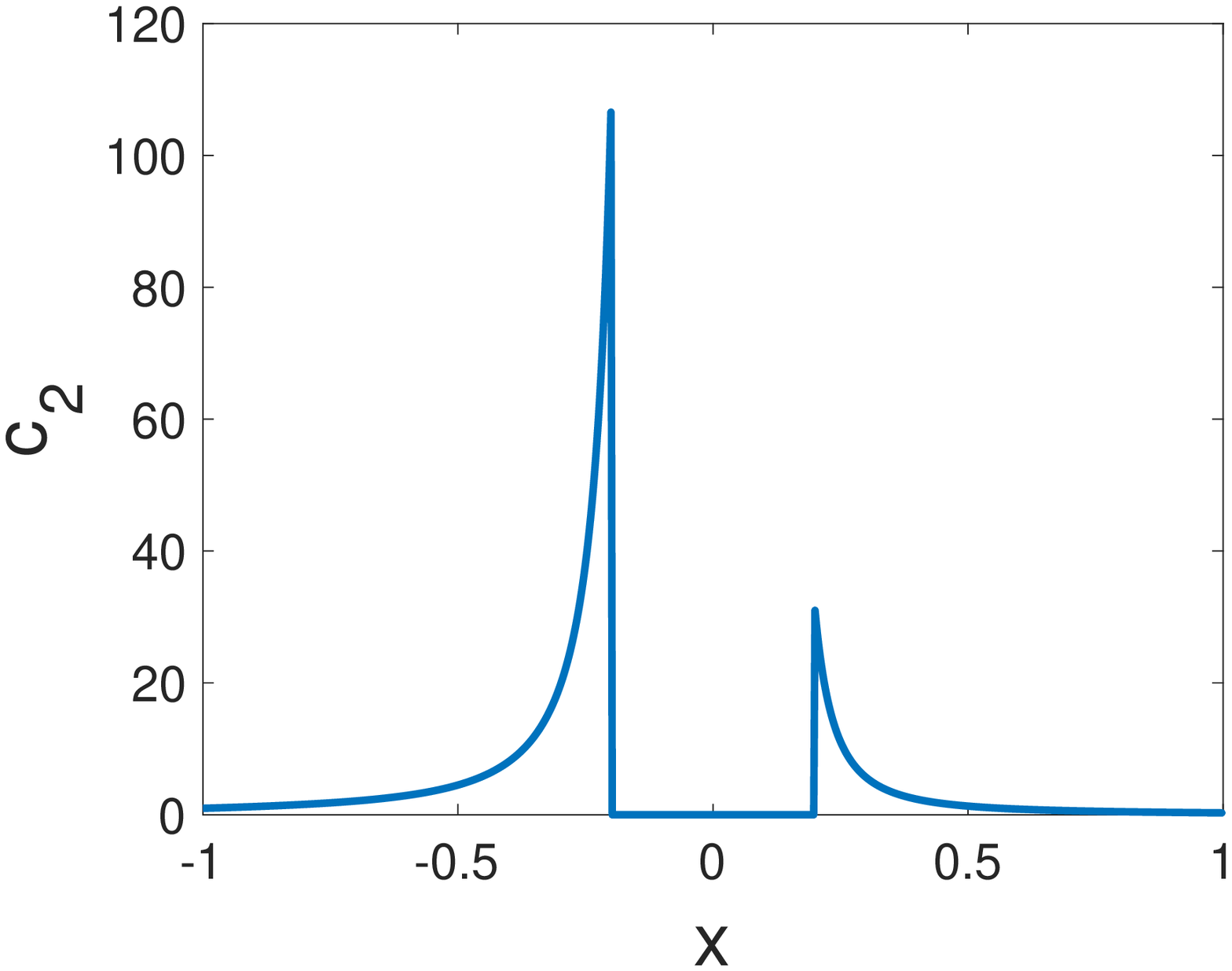} \includegraphics[width=5cm]{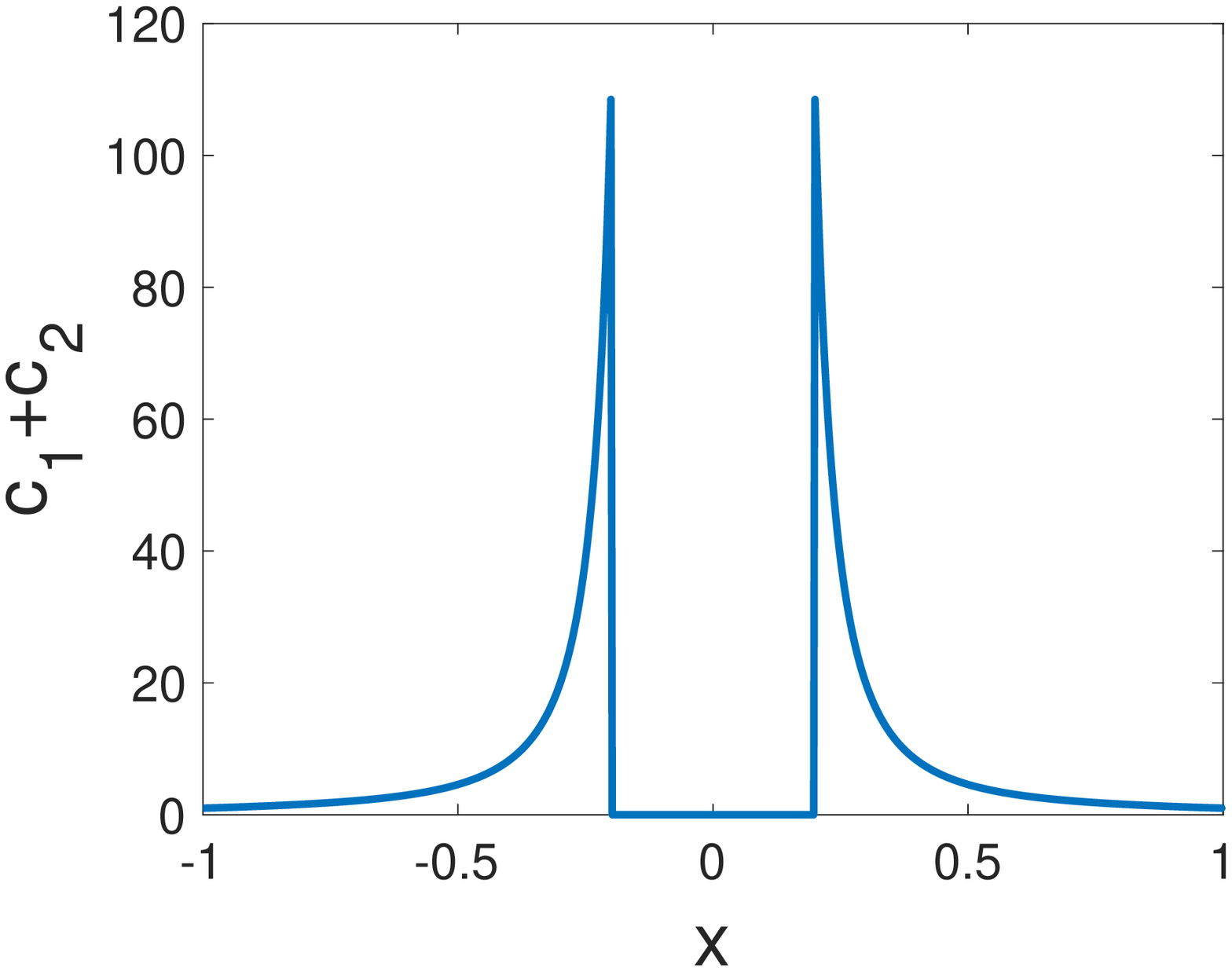} 
\includegraphics[width=5cm]{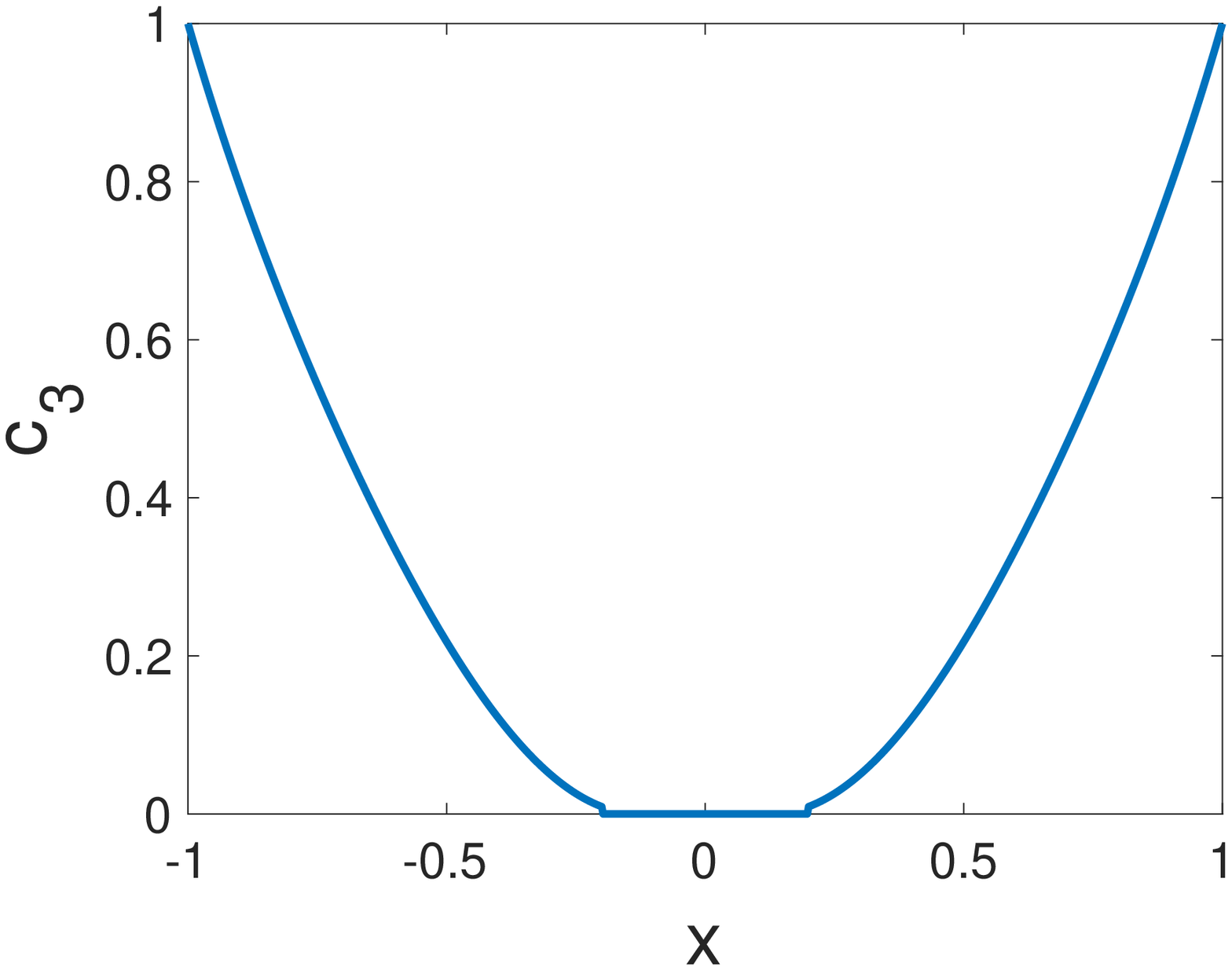} \includegraphics[width=5cm]{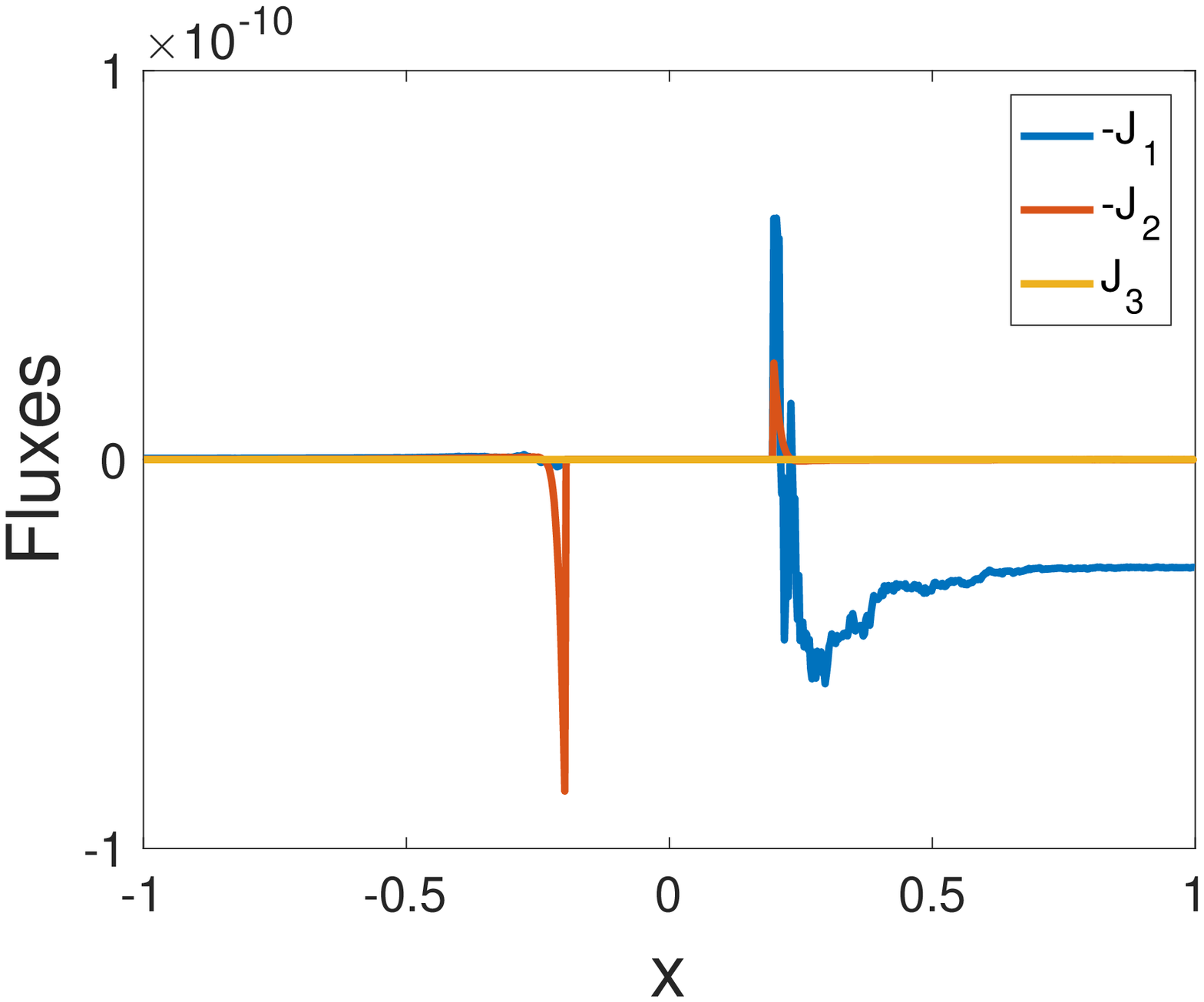} 
\caption{Equilibrium electric potential $\phi$, concentrations $c_1$, $c_2$, $c_1+c_2$, $c_3$, and ionic fluxes $J_i$ ($i=1,2,3$) with $h=0.0025$.}
\label{fig3}
\end{center}
\end{figure}

The finite difference method is used to solve the system, with a uniform mesh $h= x_k- x_{k-1}$. A temporal semi-implicit discrete scheme is used with $t_n =n \Delta t$ and $x_k=x_0+k h$, given by
\begin{equation}
\label{eq31}
\begin{aligned}
&  -\epsilon \frac{\epsilon_{r,k-1/2}}{h^2} \phi_{k-1}^{n+1} +  \epsilon \frac{\epsilon_{r,k-1/2}+ \epsilon_{r,k+1/2}}{h^2} \phi_{k}^{n+1}  - \epsilon \frac{\epsilon_{r,k+1/2}}{h^2} \phi_{k+1}^{n+1}  \\
& \hspace{5cm} - c_{1,k}^{n+1}  - c_{2,k}^{n+1}  + c_{3,k}^{n+1}  = -q_k^{n+1}, \\
& \frac{c_{i,k}^{n+1} - c_{i,k}^n}{\Delta t} = -\frac{J_{i, k+1/2}^{n+1} -J_{i,k-1/2}^{n+1}}{h}, \quad i=1,2,3,\\
&J_{i, k+1/2}^{n+1}= -D_{i,k+1/2} \frac{c_{i,k+1}^{n+1} - c_{i,k}^{n+1}}{h}  - D_{i,k+1/2} z_i c_{k+1/2}^{n} \frac{\phi_{k+1}^{n+1} -\phi_{k}^{n+1} }{h},
\end{aligned}
\end{equation}
where harmonic average is used for the diffusion coefficient  
\begin{equation}
\label{eq32}
\begin{aligned}
&  D_{i,k+1/2} = \frac{2}{\frac{1}{D_{i,k}} + \frac{1}{D_{i,k+1}}},\quad i=1,2,3.
\end{aligned}
\end{equation}
In this way, we ensure that the ionic fluxes are small near the interface as approximations of $J_i=0$ ($i=1,2,3$). When the bubble collapses, the diffusion coefficient is guaranteed to be the same as that outside of the bubble, and the continuity conditions are recovered. The quantities $\epsilon_{r,k+1/2}$ and $q_k^n$ in (\ref{eq31}) are defined in Appendix B. The discrete scheme also preserves the continuity of the total current, as in the original continuous model.

\begin{figure}[h]
\begin{center}
\includegraphics[width=6cm]{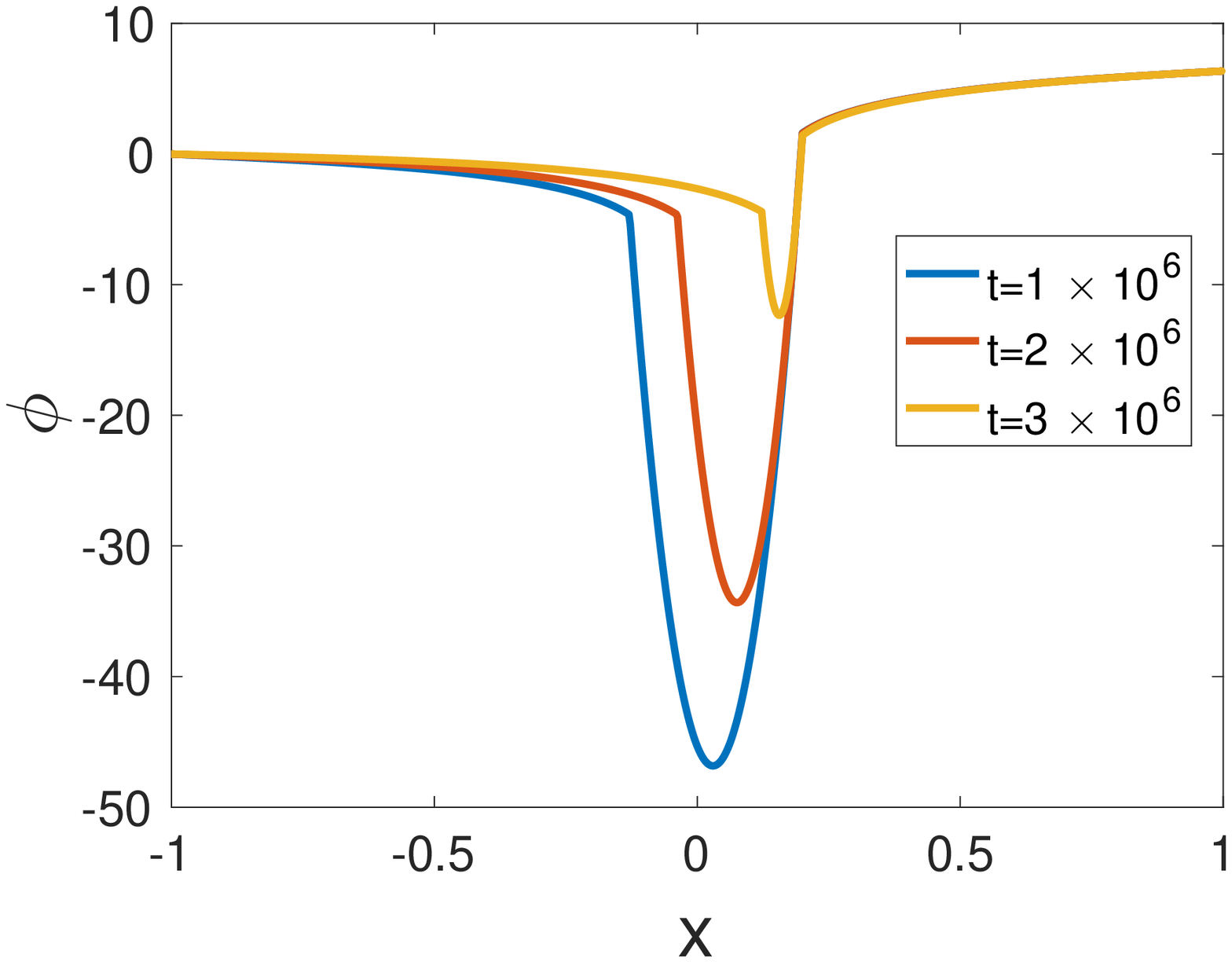} \includegraphics[width=6cm]{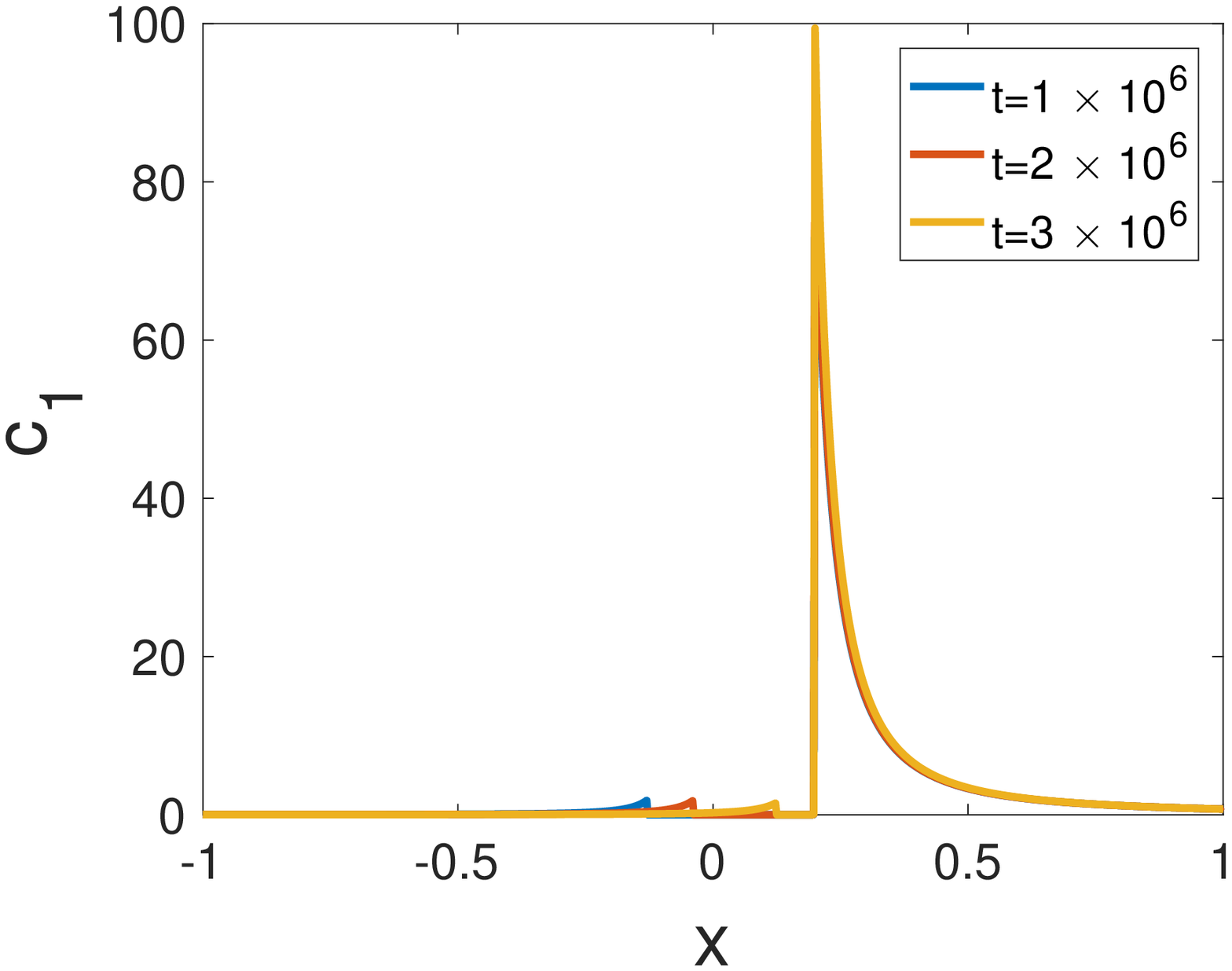}
\includegraphics[width=6cm]{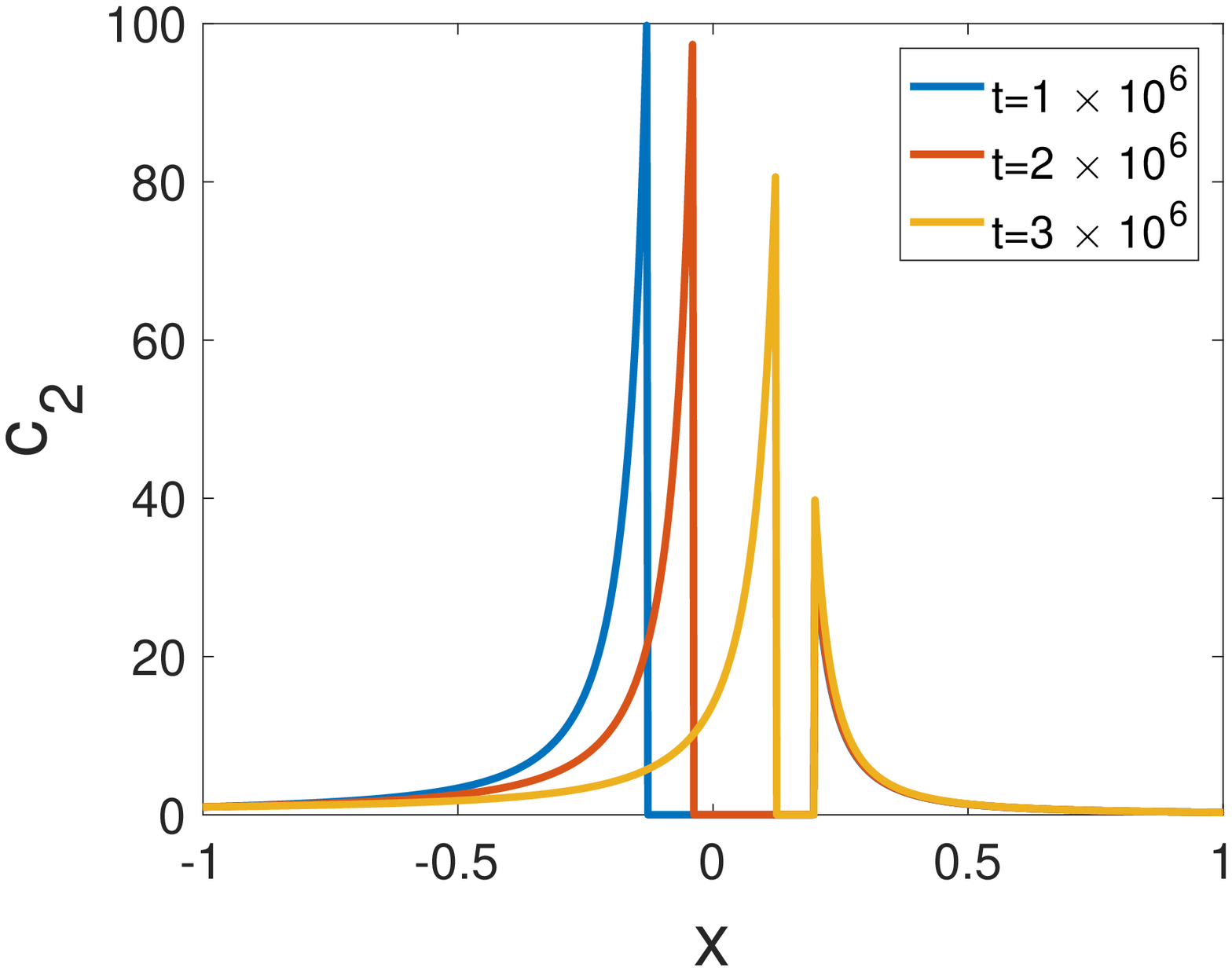}\includegraphics[width=6cm]{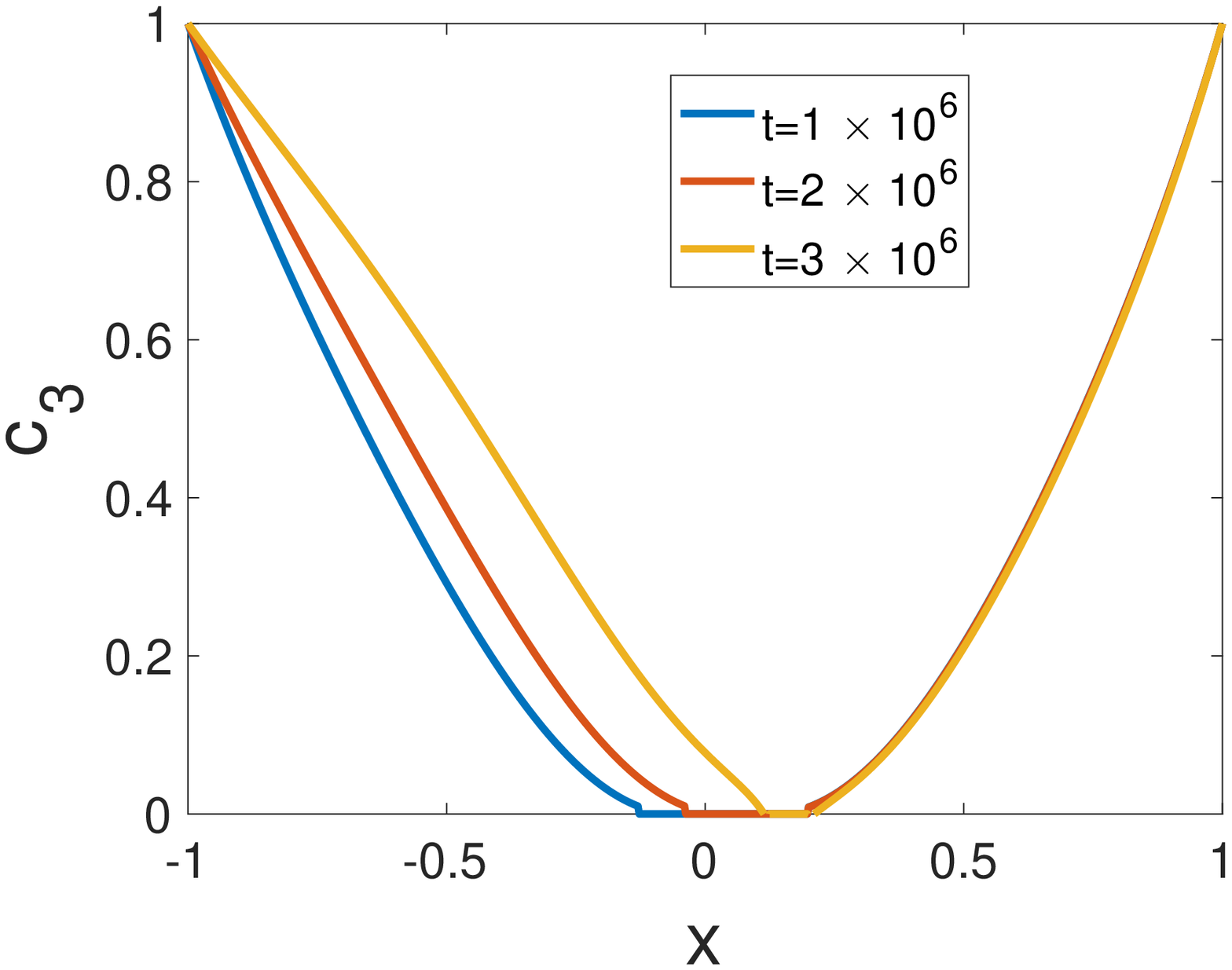}
\caption{Electric potential $\phi$ and ionic concentrations $c_1,c_2,c_3$ at three different times with $V_1=6.36$.}
\label{fig4}
\end{center}
\end{figure}

For $q_b = 2$, we first compute the initial equilibrium when the bubble occupies the entire middle region.  The initial condition at $t=0$ is set as
\begin{equation}
\label{eq33}
\begin{aligned}
& \phi(x,0) = 0, \quad -1< x < 1, \\
& c_i(x,0) =\begin{cases}
& c_i^L  \quad -1 <x< s_b=-s, \\
& 0, \quad  s_b<x< s,\\
& c_i^R, \quad s<x<1,
\end{cases}
\end{aligned}
\end{equation}
where $i=1,2,3$. We also set $V_0=0$ and $p_b =0$ in the computation so that $\phi$ is continuous.  The computation is carried out until the system reaches a steady state. For a given mesh size $h=0.0025$, Figure \ref{fig3} shows the numerical solution of electric potential $\phi$, concentrations $c_i$ and ionic fluxes $J_i$ ($i=1,2,3$), which are in good agreement with the analytical results in the previous subsection.

\begin{figure}[h]
\begin{center}
\includegraphics[width=4.4cm]{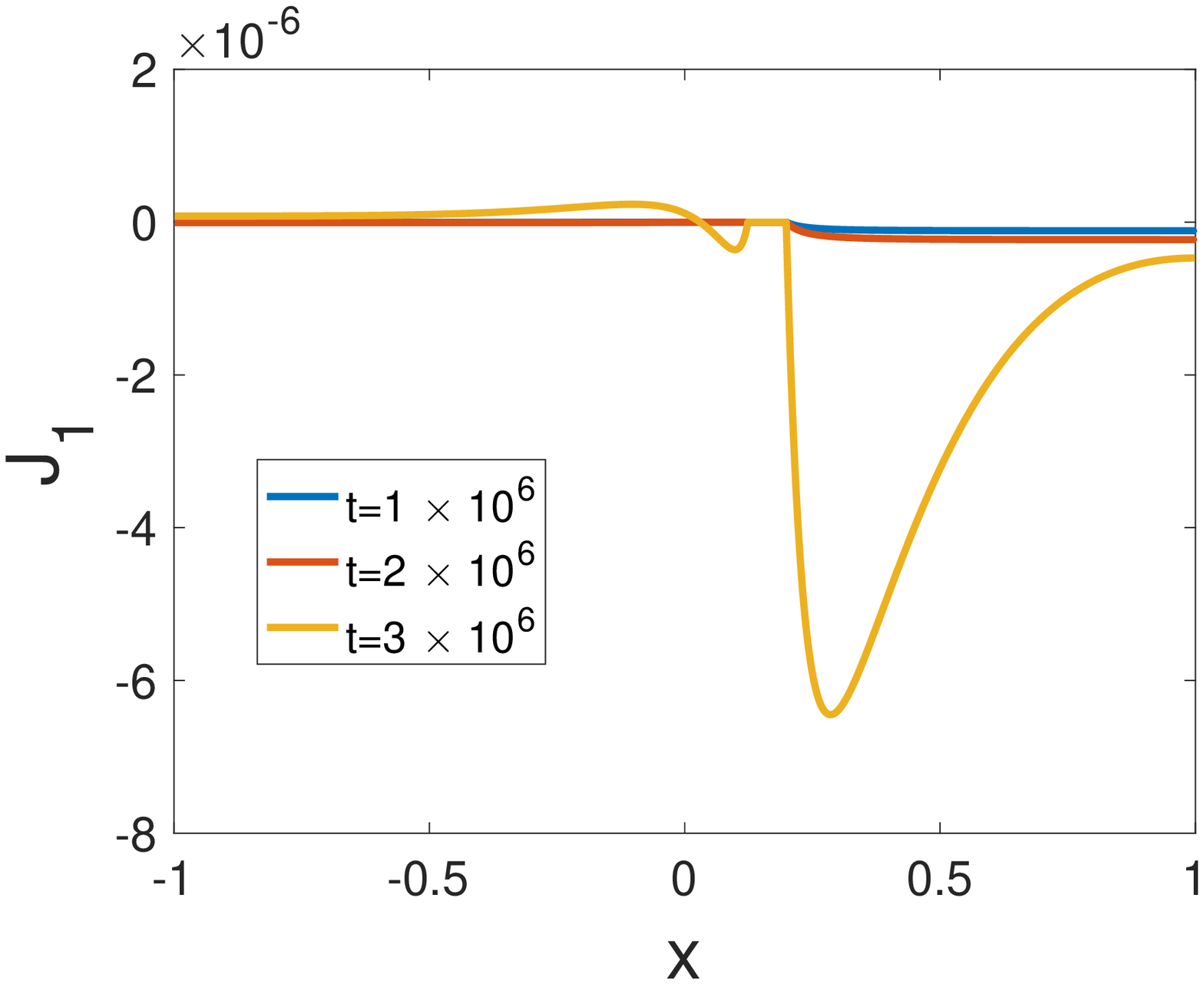} \includegraphics[width=4.4cm]{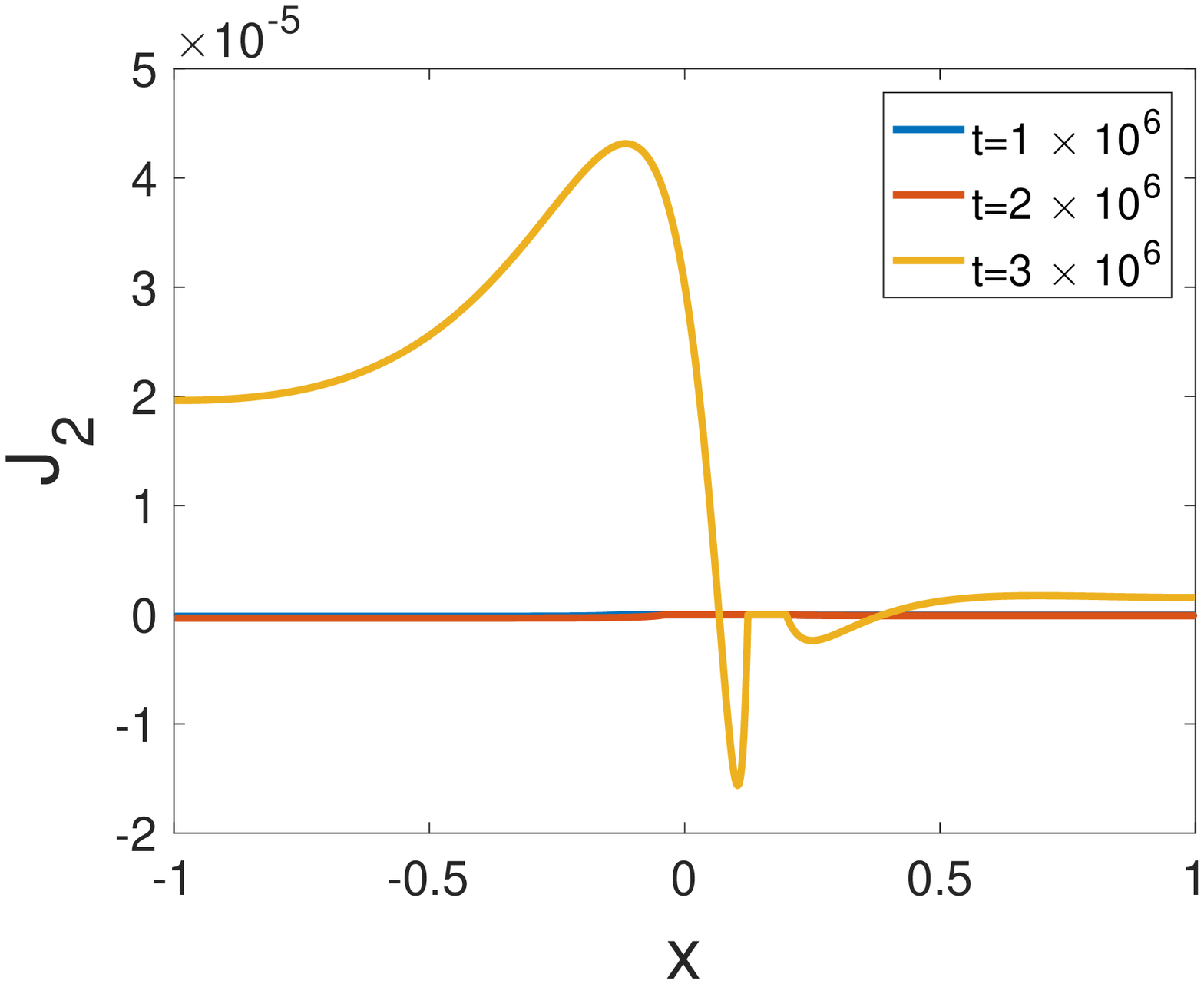}
\includegraphics[width=4.4cm]{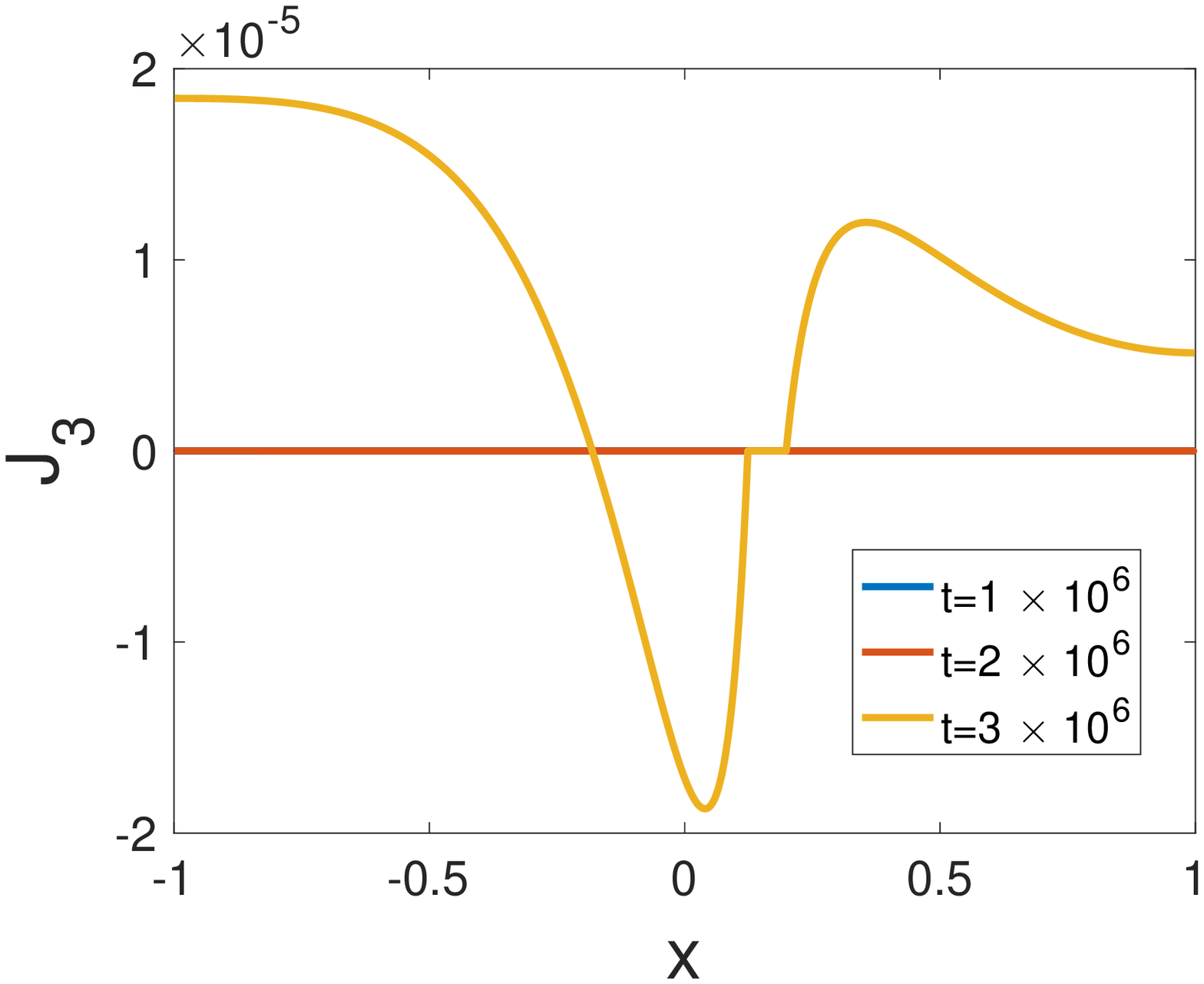}
\caption{The fluxes $J_i$ ($i=1,2,3$) at three different times with $V_1=6.36$.}
\label{fig5}
\end{center}
\end{figure}

\begin{figure}[h]
\begin{center}
\includegraphics[width=6cm]{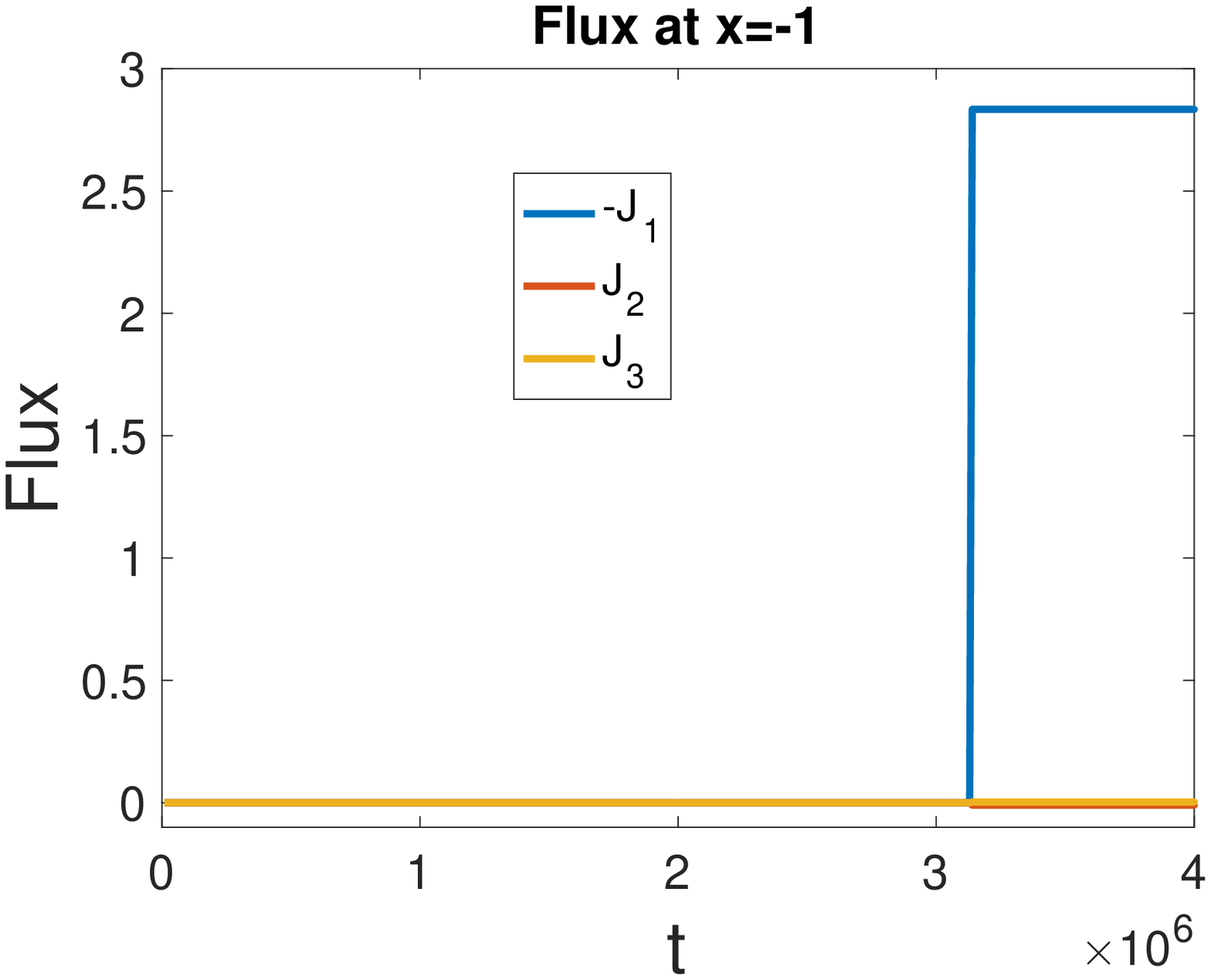} \includegraphics[width=6cm]{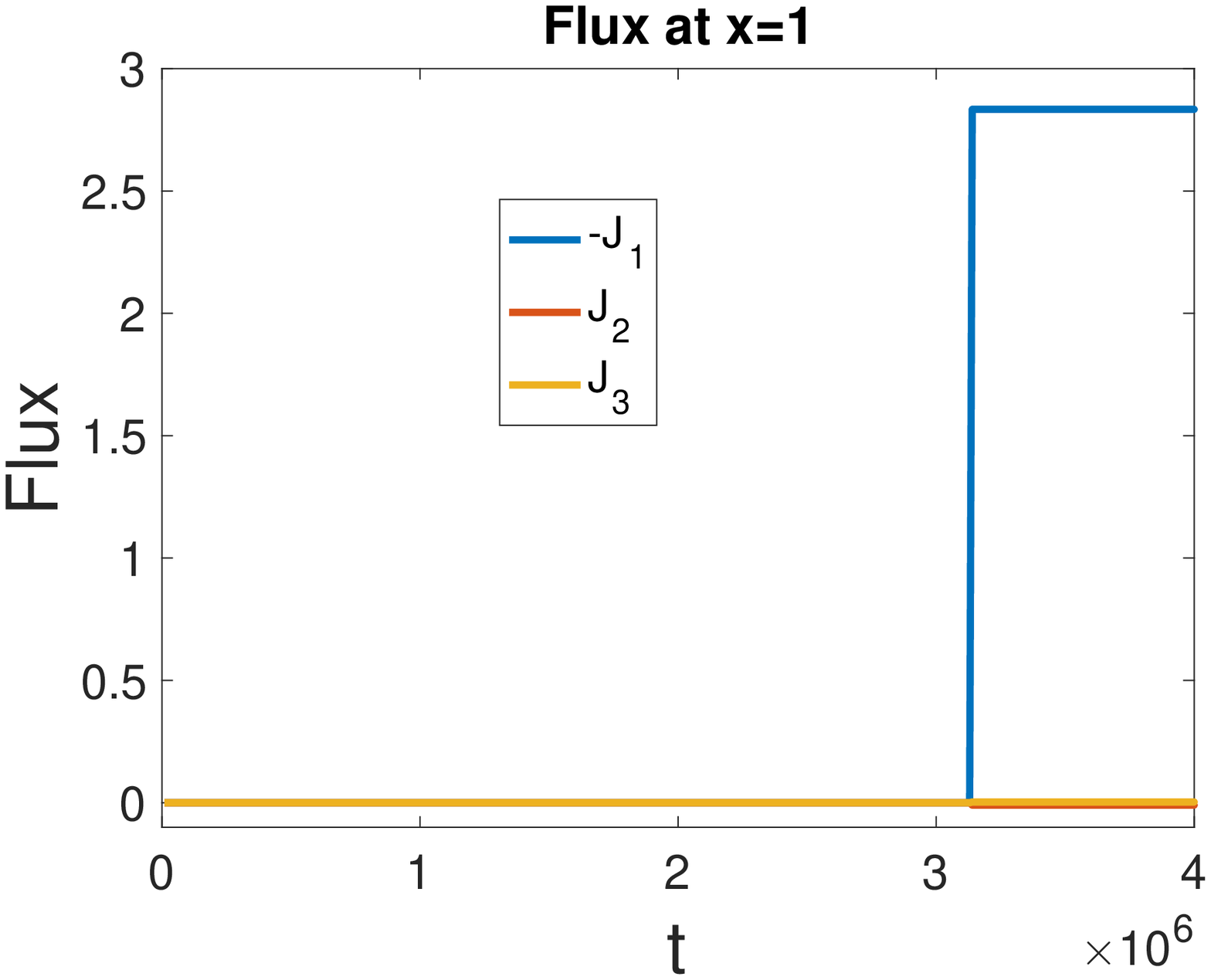}
\caption{The dynamics of ionic fluxes $J_i$ ($i=1,2,3$) with $V_1=6.36$.}
\label{fig6}
\end{center}
\end{figure}

\begin{figure}[h]
\begin{center}
\includegraphics[width=5cm]{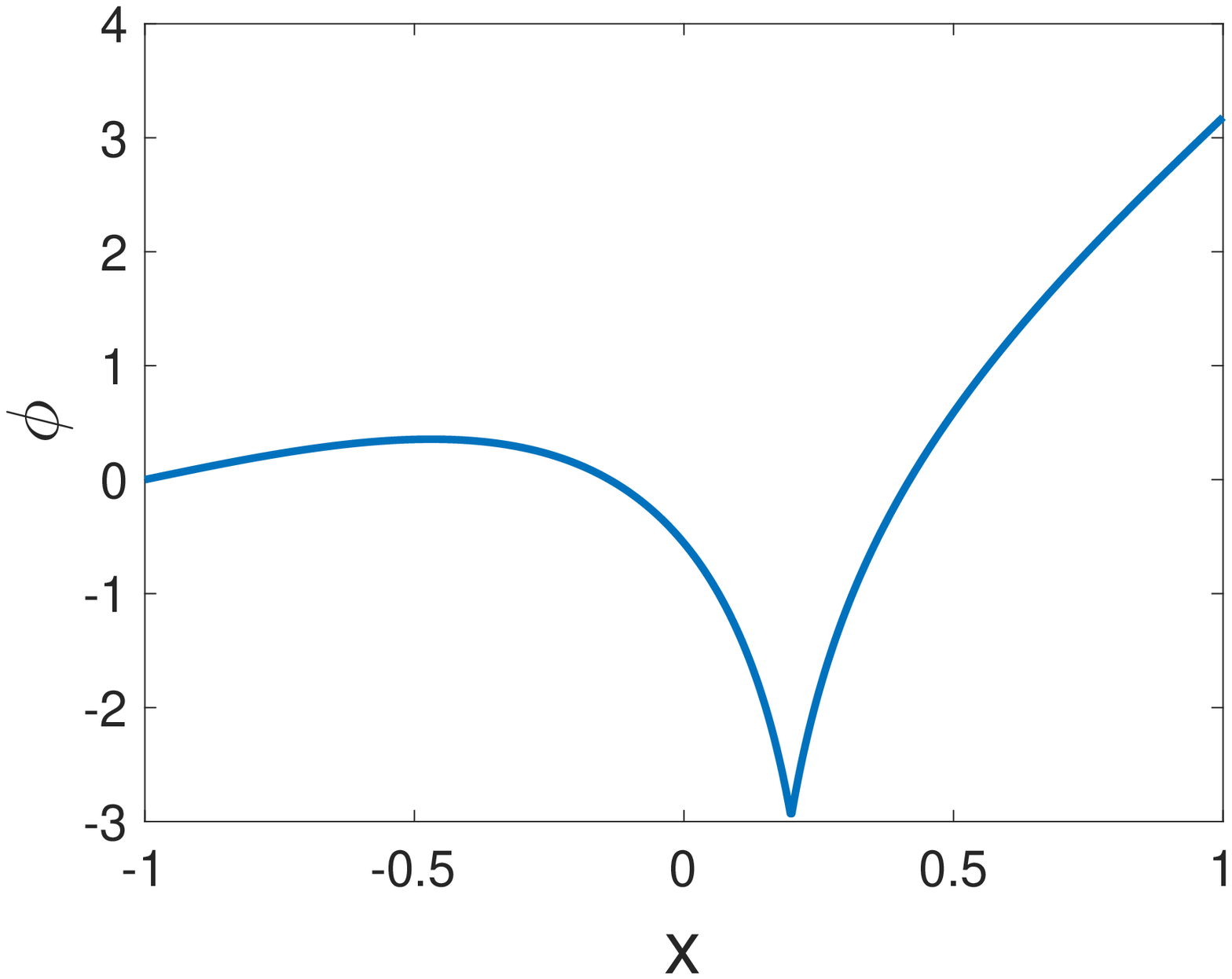} \includegraphics[width=5cm]{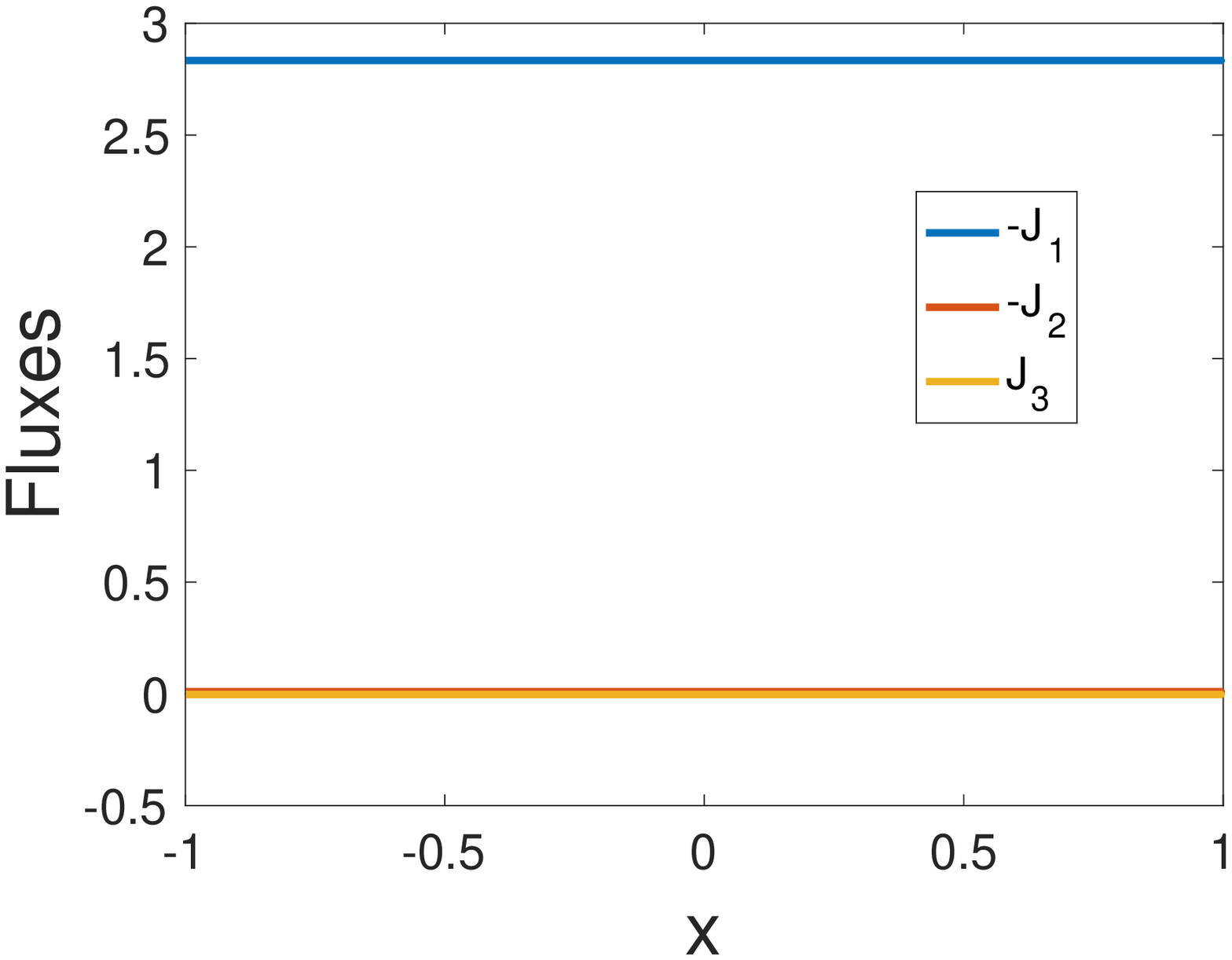}  \\\includegraphics[width=4.4cm]{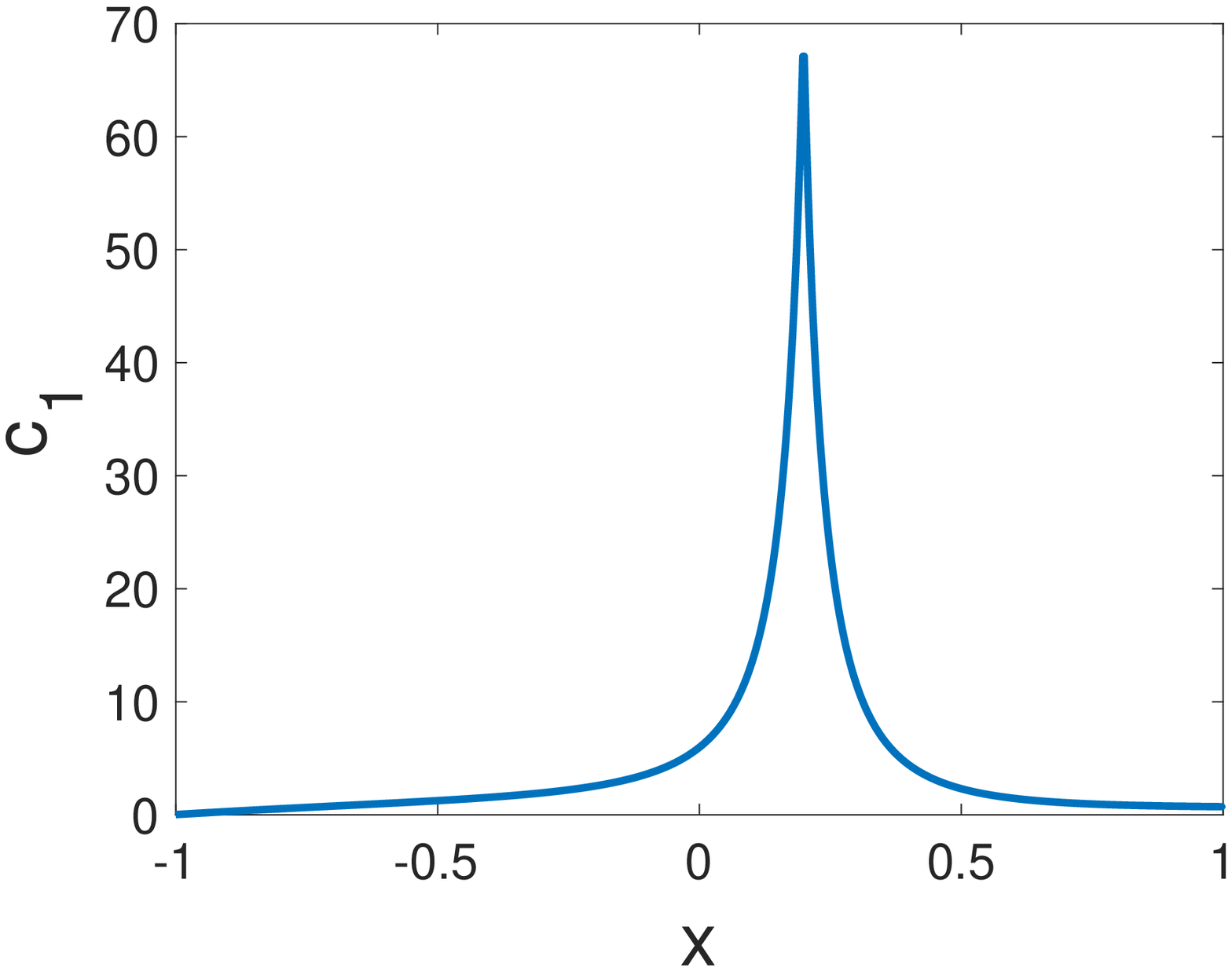}
\includegraphics[width=4.4cm]{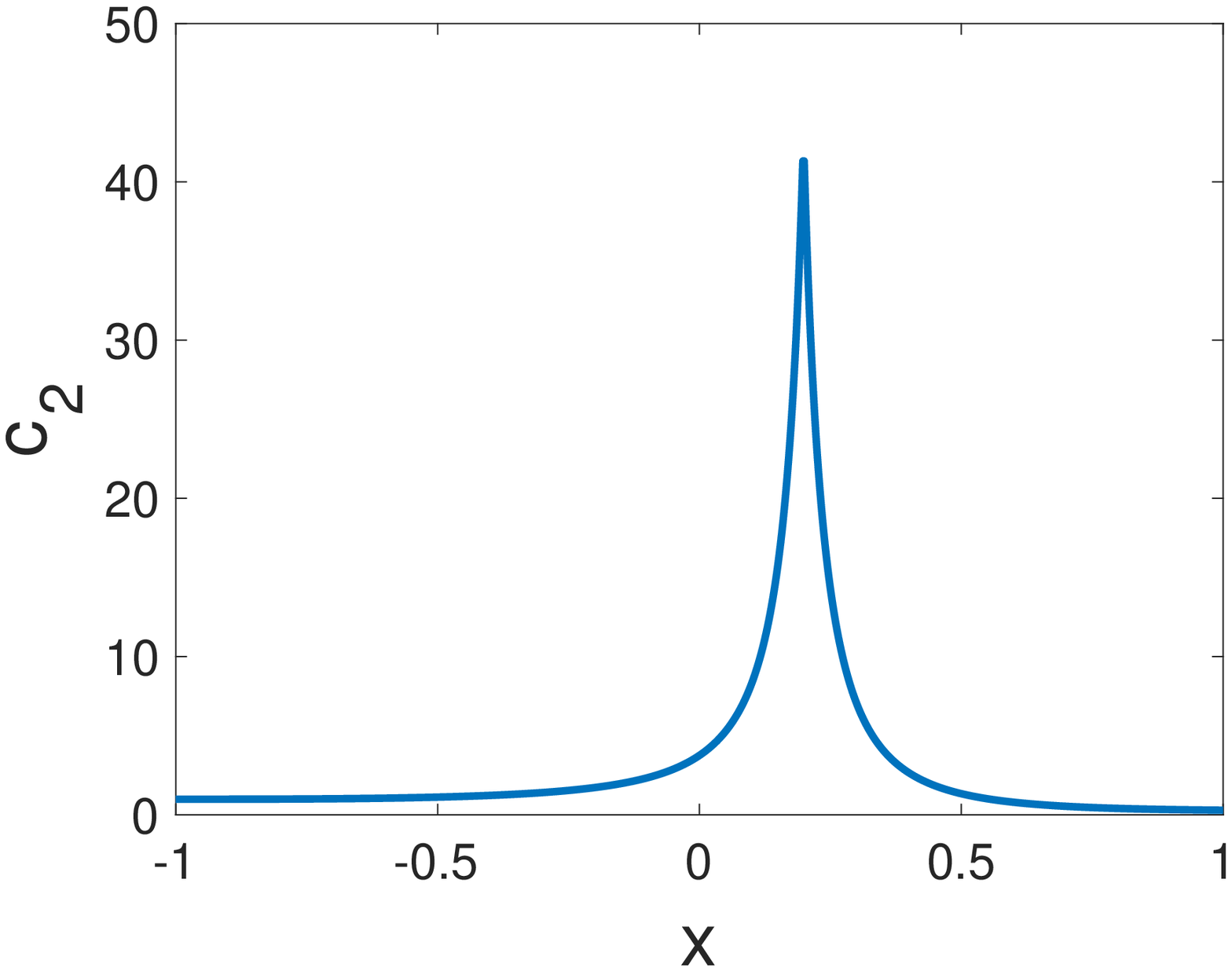} \includegraphics[width=4.4cm]{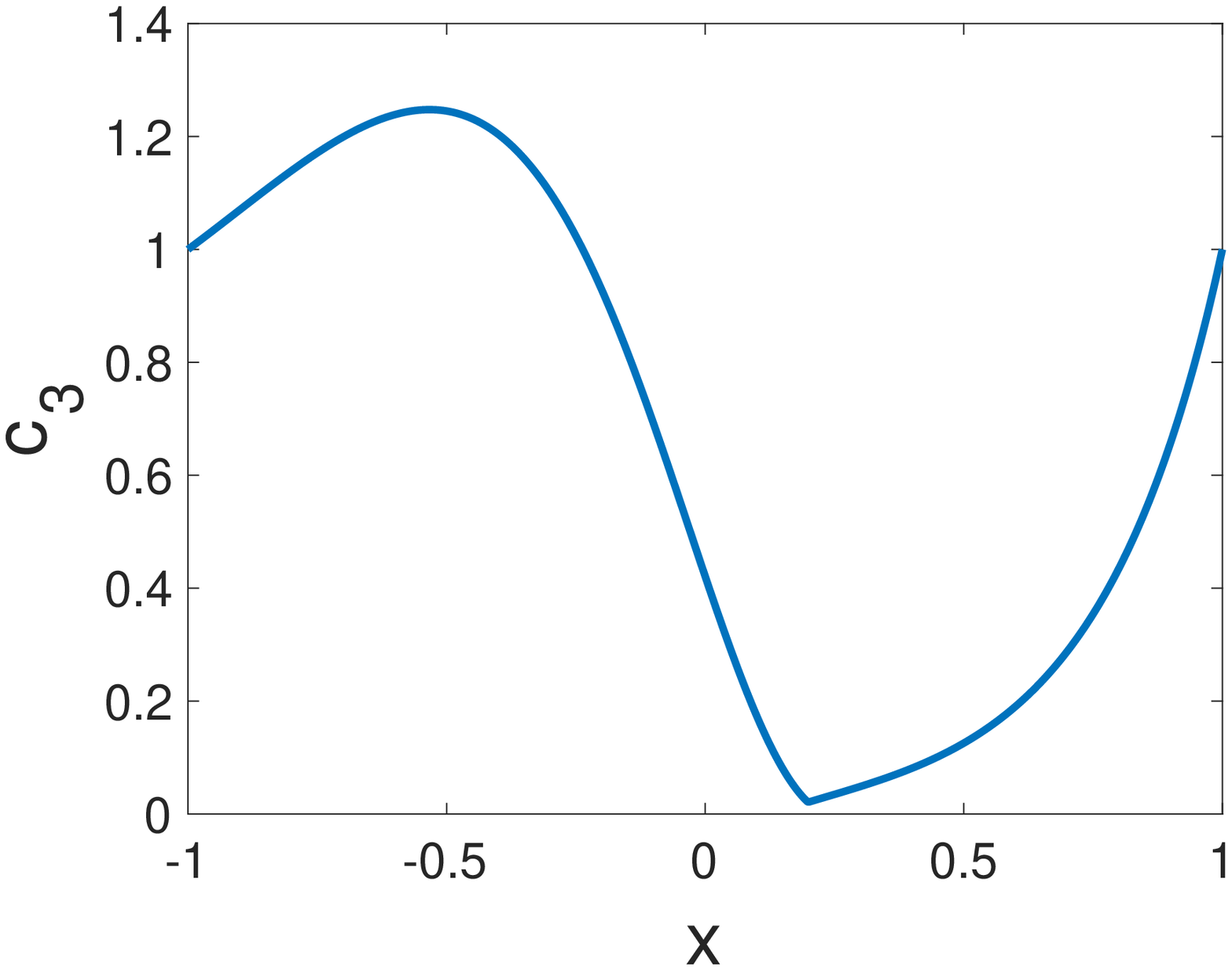}
\caption{The steady state with $V_1=6.36$.}
\label{fig7}
\end{center}
\end{figure}

Next, we present the results on the bubble motion and dynamic behaviour of the PNP system. We start from the equilibrium state at $t=0$ and increase the electric potential from zero to $\phi(1,t) = V_1 = 6.36$ (i.e., 160 mV) for $0<t<t^\ast$ (we already set $V_0 =0$ and $p_b=0$), where $t^\ast$ is the unknown time when the bubble collapses. Figure \ref{fig4} shows $\phi,c_i$ ($i=1,2,3$) at three different times. The minimum value for $\phi$ inside the bubble gradually increases in Figure \ref{fig4}(a), and the interface $s_b$ moves to the right as indicated by Figure \ref{fig4}.  Figure \ref{fig5} shows the ionic fluxes at three different times, which are small. After the bubble collapses and the dipole disappears, we reset $\phi(1,t) = V_0 + V_1 = 3.18$.  Figure \ref{fig6} shows the three ionic fluxes at $x=\pm 1$. It can be seen that they remain small until the bubble collapses (i.e., $s_b =s$) at $t^\ast= 3.13 \times 10^6$, which is 17.6 ms in dimensional unit. The ionic fluxes and $\phi,c_i$ ($i=1,2,3$) reach a steady state soon after the bubble collapses, as shown in Figure \ref{fig7}. At steady state, the dimensionless ionic flux $J_1$ and the dimensional current $I$ are found to be
\begin{equation}
\label{eq34}
\begin{aligned}
J_1 \approx - 2.834, \quad I = |J_1| e_0 A J_0 \approx  10 \,\mathrm{pA}.
\end{aligned}
\end{equation}

{\noindent \bf Remark 3.} The value of the steady state current $I$ obtained above is close to that given in Figure 2(a) of \cite{llano1988}. When the voltage jump $V_1$ is reset to zero after system reaches a steady state, the ionic fluxes reduce to zero immediately, indicating the closure of the ion channel. In this sense, our proposed model provides a plausible gating mechanism once the bubble is generated. However, the mechanism of the bubble generation is not considered here and will be the subject of a future study. 



\subsection{Quasi-static equilibrium}

Since the motion of the bubble is extremely slow compared with the diffusive timescale of the ions,  ionic fluxes are essentially zero (Figure \ref{fig5}) before the bubble collapses. Therefore, we can use quasi-static solution with zero ionic fluxes as an approximation of the intermediate states. A hybrid method can be used to determine the solution of intermediate states by first obtaining an analytical solution (in terms of integrals), where integration constants involved can be determined easily using a numerical method afterwards.

\begin{figure}[h]
\begin{center}
\includegraphics[width=5cm]{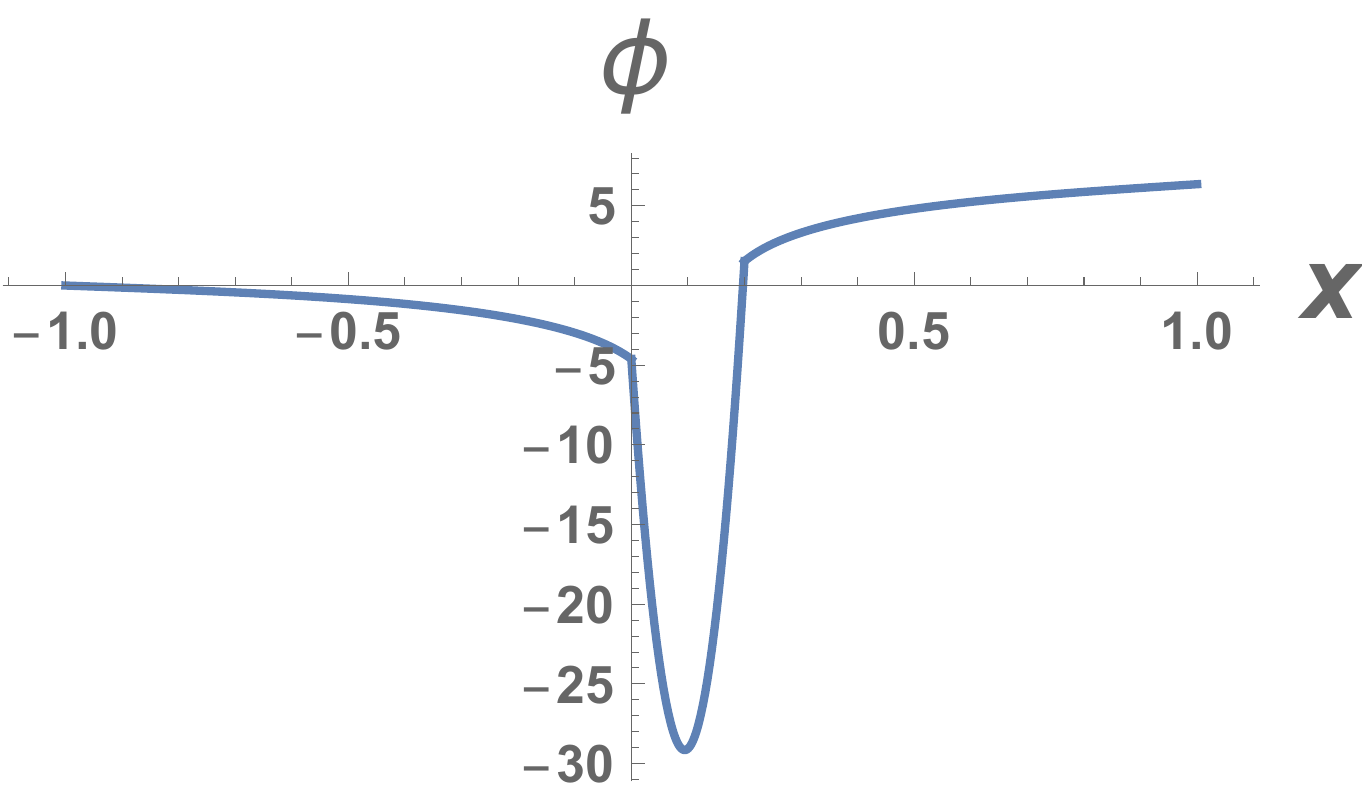} \includegraphics[width=5cm]{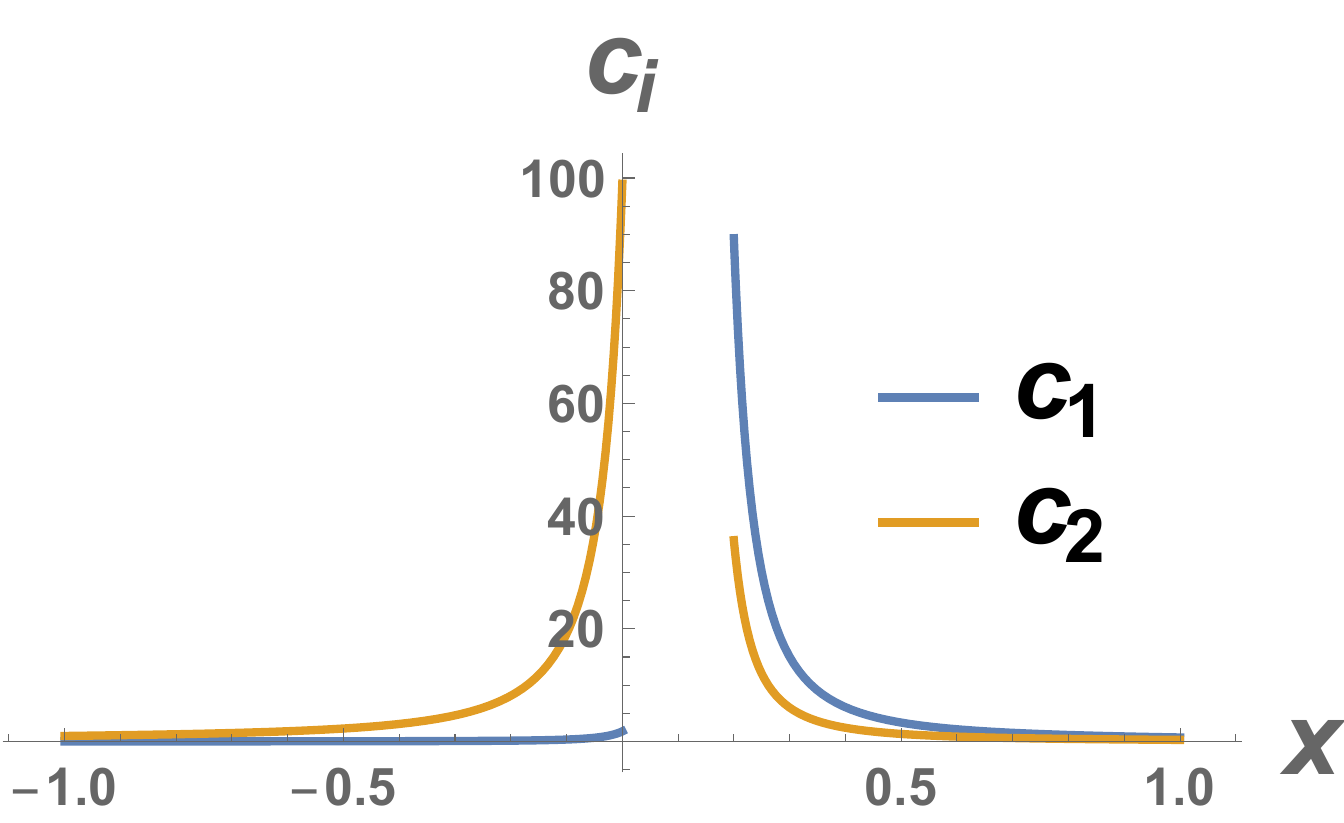}\\
\includegraphics[width=5cm]{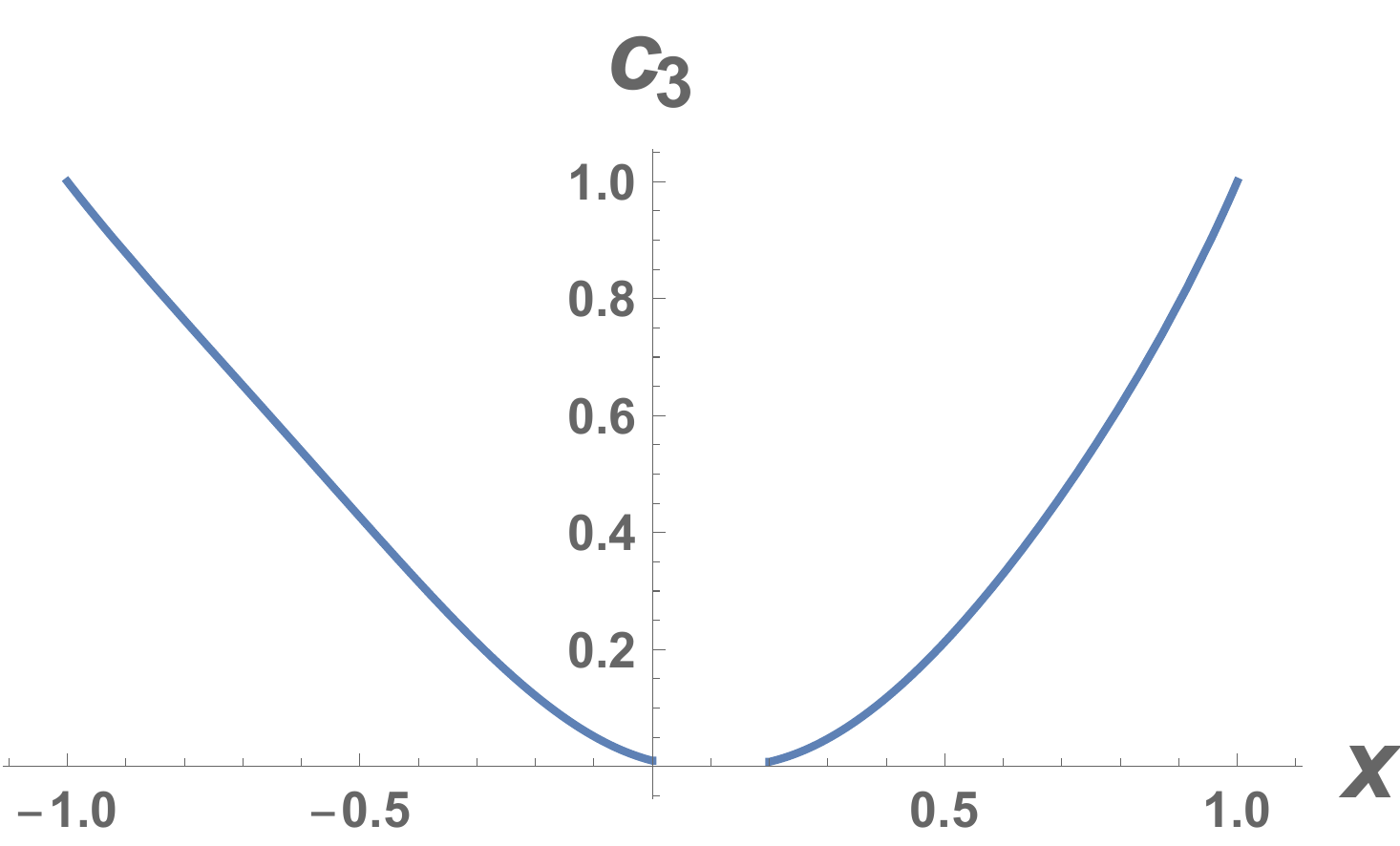} \includegraphics[width=5cm]{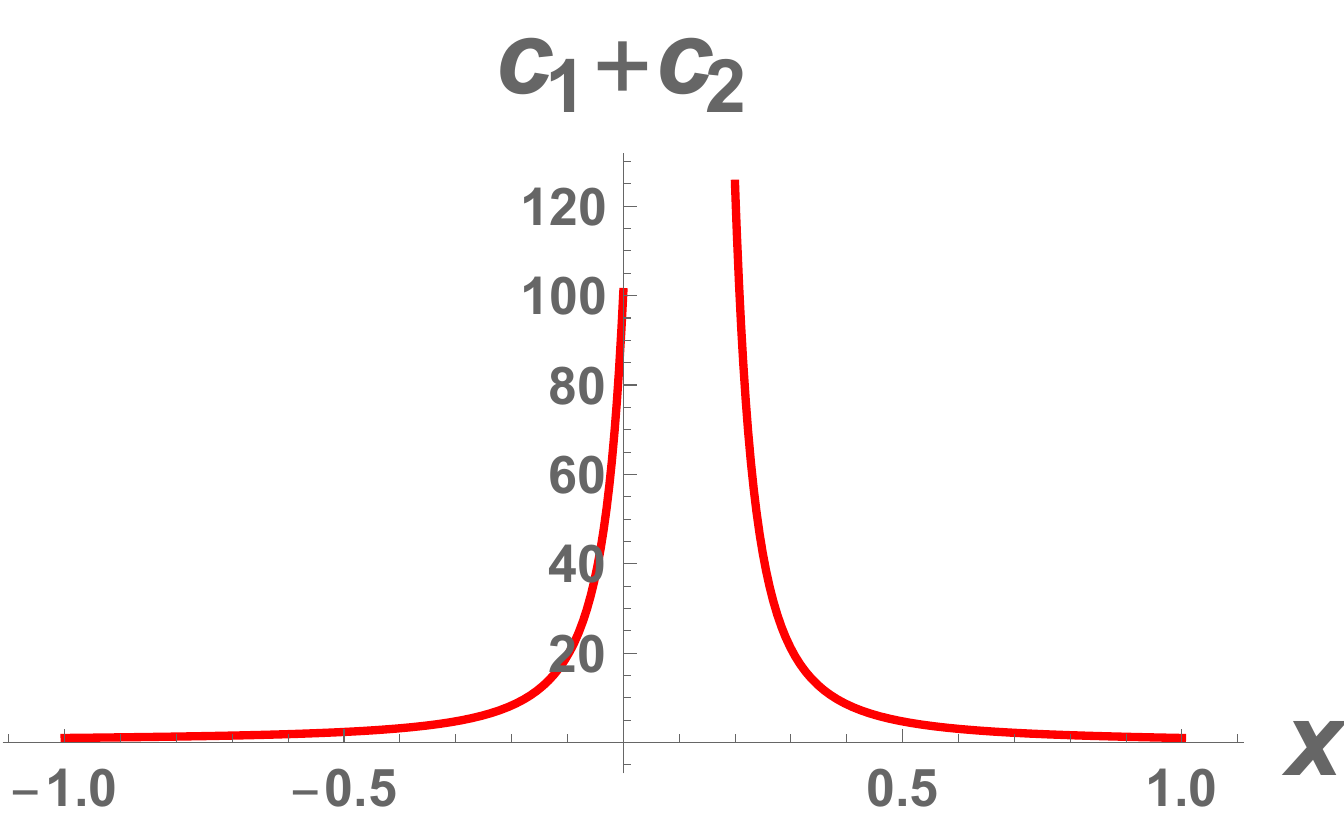}
\caption{The quasi-static solution of $\phi$, $c_1$, $c_2$, $c_3$ and $c_1+c_2$ for $s_b =0$ and $V_1 = 6.36$.}
\label{fig8}
\end{center}
\end{figure}

Given boundary condition $V_1$ and interface position $s_b$, solving the quasi-static equilibrium is similar to that for solving the initial state. We set $V_0=0$ and $p_b = 0$ so that the continuity condition of $\phi$ can be used at interface $s_b$. Inside the bubble, we have 
\begin{equation}
\label{eq35}
\begin{aligned}
& \phi (x) = B_1 (x-s)^2 + \tilde{\phi}_s (x-s)+\phi_s, \quad B_1= \frac{{q}_b}{2(s-s_b) \epsilon \epsilon_{r0} \beta},
\end{aligned}
\end{equation}
where $\phi_s,\tilde{\phi}_s$ are to be determined. The solutions of $\phi$ outside of the bubble can be written as
\begin{equation}
\label{eq36}
\begin{aligned}
& x=  \int_{\phi_s}^\phi \frac{1}{\sqrt{G_1(\phi; \phi_s,\tilde{\phi}_s)}} d\phi + s, \quad s<x<1,\\
& x=  -\int_{\phi_{s_b}}^\phi \frac{1}{\sqrt{G_2(\phi; \phi_s,\tilde{\phi}_s)}} d\phi + s_b, \quad -1<x<s_b,
\end{aligned}
\end{equation}
where $\phi_{s_b} = \phi(s_b)$ can be expressed by $\phi_s$ and $\tilde{\phi}_s$. The derivation for $G_1(\phi), G_2(\phi)$ are given in Appendix C. For given parameter values including $V_1$ and $s_b$, the two unknowns $\phi_s$ and $\tilde{\phi}_s$ can be determined by the two boundary conditions $\phi(1)=V_1$ and $\phi(-1)=0$, i.e., 
\begin{equation}
\label{eq37}
\begin{aligned}
1=  \int_{\phi_s}^{V_1} \frac{1}{\sqrt{G_1(\phi; \phi_s,\tilde{\phi}_s)}} d\phi + s, \quad  -1=  -\int_{\phi_{s_b}}^0 \frac{1}{\sqrt{G_2(\phi; \phi_s,\tilde{\phi}_s)}} d\phi + s_b,
\end{aligned}
\end{equation}
and $\phi(x),c_i(x)$ ($i=1,2,3$) can be obtained afterwards. As an example, for $q_b=2$, $V_1=6.36$ and $s_b=0$, solutions of $\phi$ and $c_i$ ($i=1,2,3$) can be computed using the procedure outlined above and plotted in Figure \ref{fig8}.

For fixed $V_1$, we can treat $\phi (s), \phi(s_b)$ as functions of the parameter $s_b$, which can be determined by 
\begin{equation}
\label{eq38}
\begin{aligned}
& \frac{d s_b}{d t} =2 D_b q_b  \frac{\phi (s) - \phi_(s_b)}{s-s_b} =2 D_b q_b  f(s_b) .
\end{aligned}
\end{equation}
Integrating in time, we obtain $t^\ast$, the time delay after the voltage jump and before the bubble collapses, 
\begin{equation}
\label{eq39}
\begin{aligned}
&t^\ast= \int_{-s}^{s} \frac{1}{2 D_b q_b  f(x)} dx.
\end{aligned}
\end{equation}

\begin{figure}[h]
\begin{center}
\includegraphics[width=6cm]{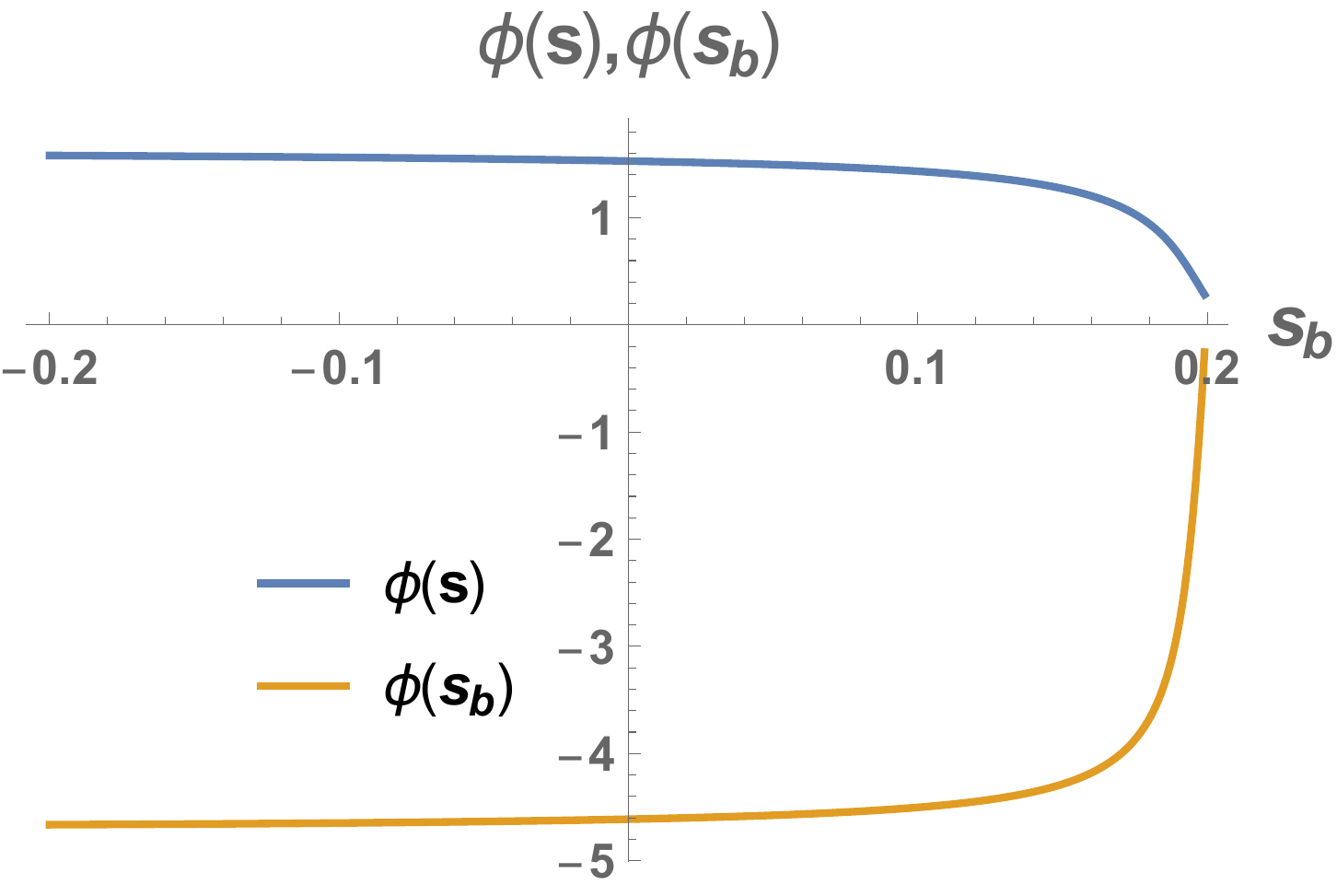} \includegraphics[width=6cm]{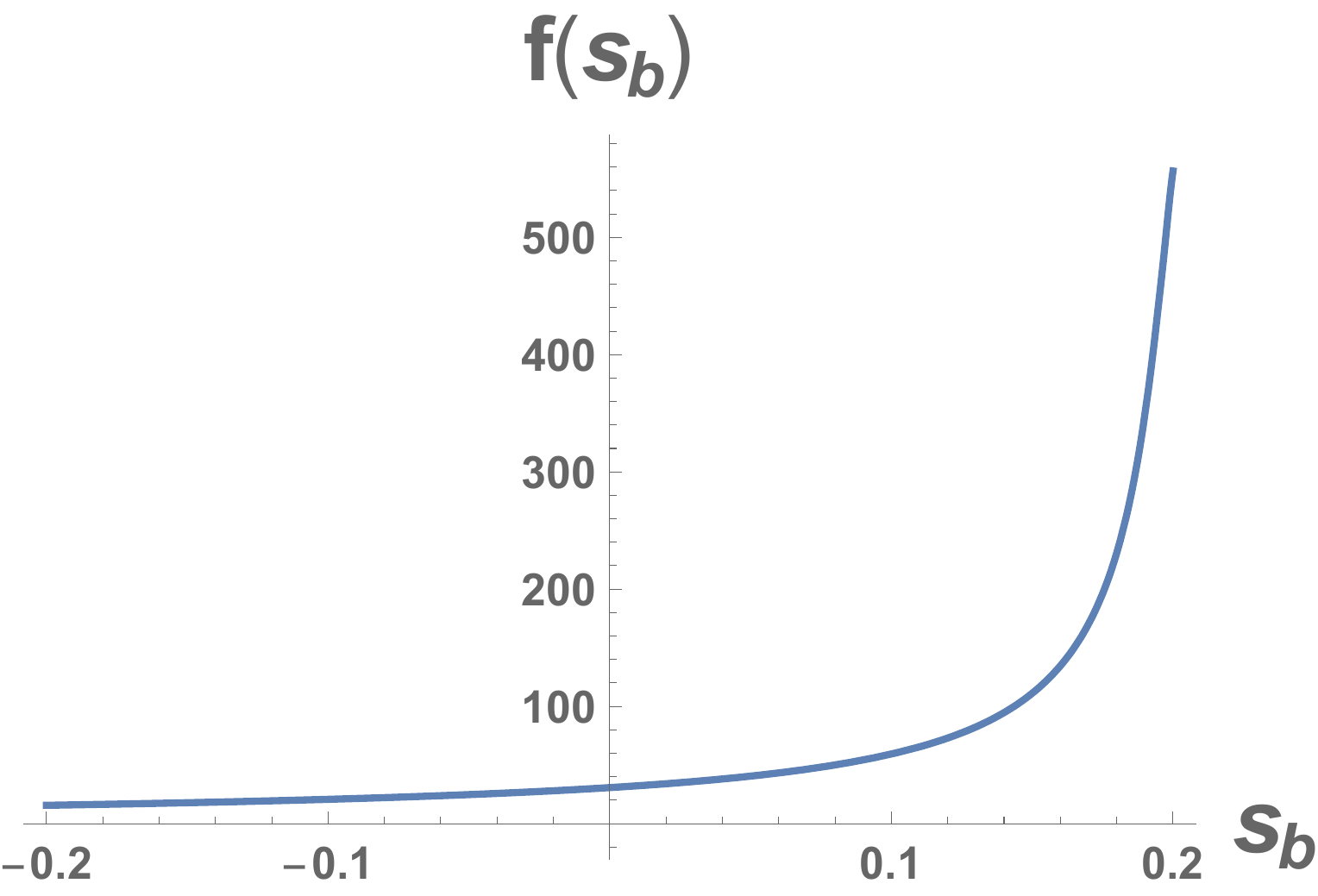}
\caption{The dependence of $\phi(s)$ and $\phi(s_b)$ on $s_b$ and the function $f(s_b)$ for $V_1 = 6.36$.}
\label{fig9}
\end{center}
\end{figure}

Figure \ref{fig9}(a) shows the dependence of quantities $\phi(s)$ and $\phi(s_b)$ on $s_b$ and Figure \ref{fig9}(b) shows the function $f(s_b)$. From (\ref{eq39}), we find that $t^\ast \approx 3.26 \times 10^{6}$, which is $18.3 $ ms in dimensional unit, which is slightly longer than that obtained using the finite difference method (17.6 ms) previously. 

\subsection{The steady state after the collapse of the bubble}

After the bubble collapses, interface  conditions $J_i  = 0$ ($i=1,2,3$) are replaced by continuity conditions $[J_i]=0$ and $q_b$ becomes a point charge (a delta function). Due to the presence of $q_b$, the concentration of $c_3$ is approximately zero near $x=s$, and we assume $J_3/D_3 \approx 0 $. The system at the steady state can be approximated by
\begin{equation}
\label{eq40}
\begin{aligned}
 -\epsilon \epsilon_{r1} \phi''(x) & = c_1+c_2 -c_3 - \frac{q_b}{\beta} \delta (x-s),\\
 -J_1 &=  c_1'+ c_1   \phi',\\
 -\frac{J_2}{D_2} & =  c_2'+ c_2   \phi',\\
  0& = c_3'-c_3 \phi',
\end{aligned}
\end{equation}
where $D_1 = 1$ has been used.  If we combine the effects of $c_1,c_2$ and define $J_p=J_1 + \frac{J_2}{D_2}$, then the system can be reduced to a single equation of $\phi$ (see the derivation in Appendix C)
\begin{equation}
\label{eq41}
\begin{aligned}
 & -\epsilon \epsilon_{r1} \phi''(x) =\frac{1}{2} \epsilon \epsilon_{r1} \left[ (\phi'(x))^2  -  (\phi'(1))^2 \right] - J_p (x-1) - 2 (e^{\phi-V}-1), \\
 & \hspace{8cm} \mathrm{for} \quad s<x<1,\\
  & -\epsilon \epsilon_{r1} \phi''(x) = \frac{1}{2} \epsilon \epsilon_{r1} \left[ (\phi'(x))^2  -  (\phi'(-1))^2 \right] - J_p (x+1) - 2 (e^{\phi} -1), \\
 & \hspace{8cm} \mathrm{for} \quad -1<x<s,
\end{aligned}
\end{equation}
where $V= V_0 + V_1$ and $c_3^L= c_3^R=1$ have been used. The point charge (delta function) at $x=s$ gives the jump condition
\begin{equation}
\label{eq42}
\begin{aligned}
\, [\phi'] = \frac{q_b}{\beta \epsilon \epsilon_{r1}}.
\end{aligned}
\end{equation}

\begin{figure}[h]
\begin{center}
\includegraphics[width=6cm]{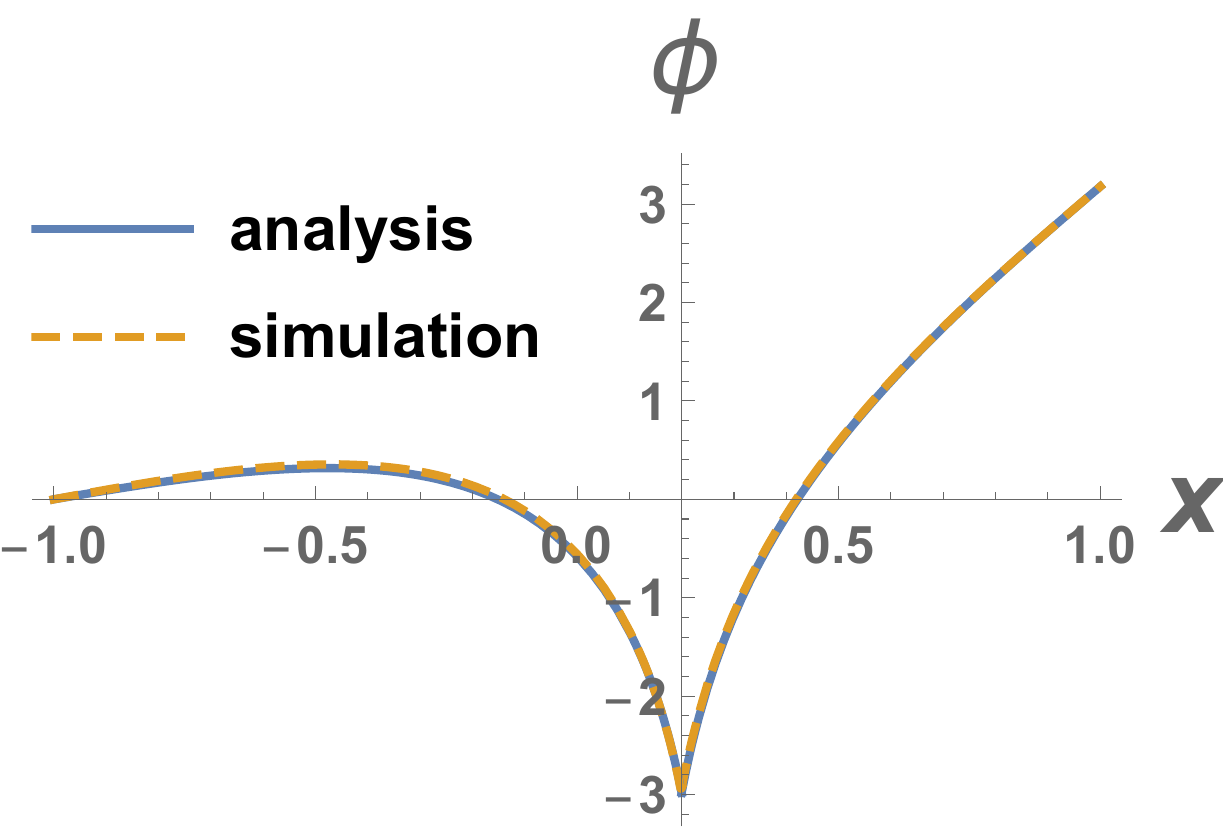} \includegraphics[width=6cm]{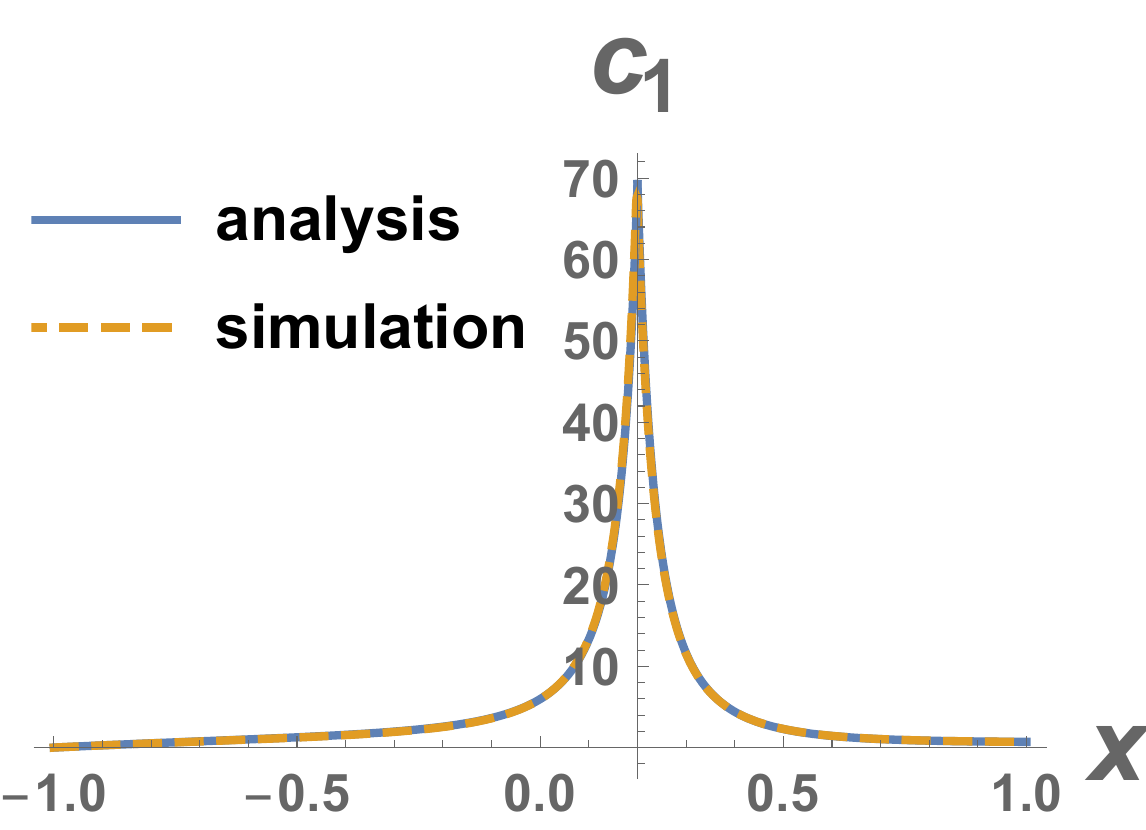} \\\includegraphics[width=6cm]{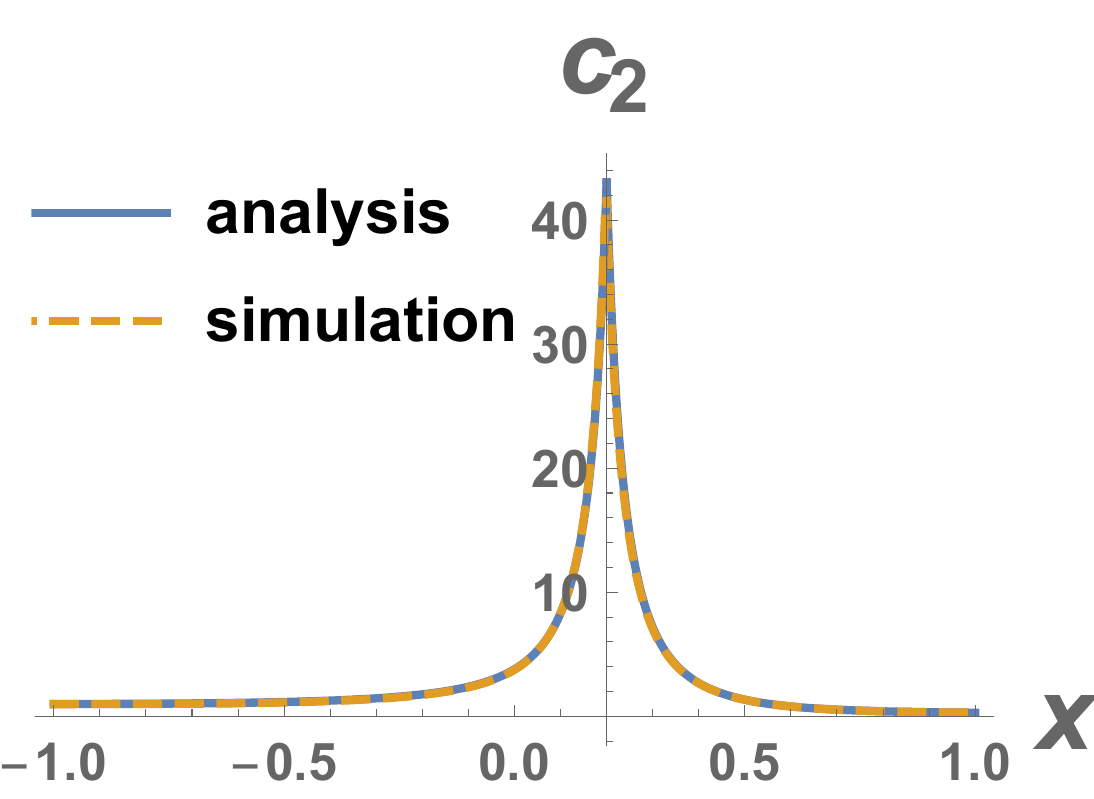}\includegraphics[width=6cm]{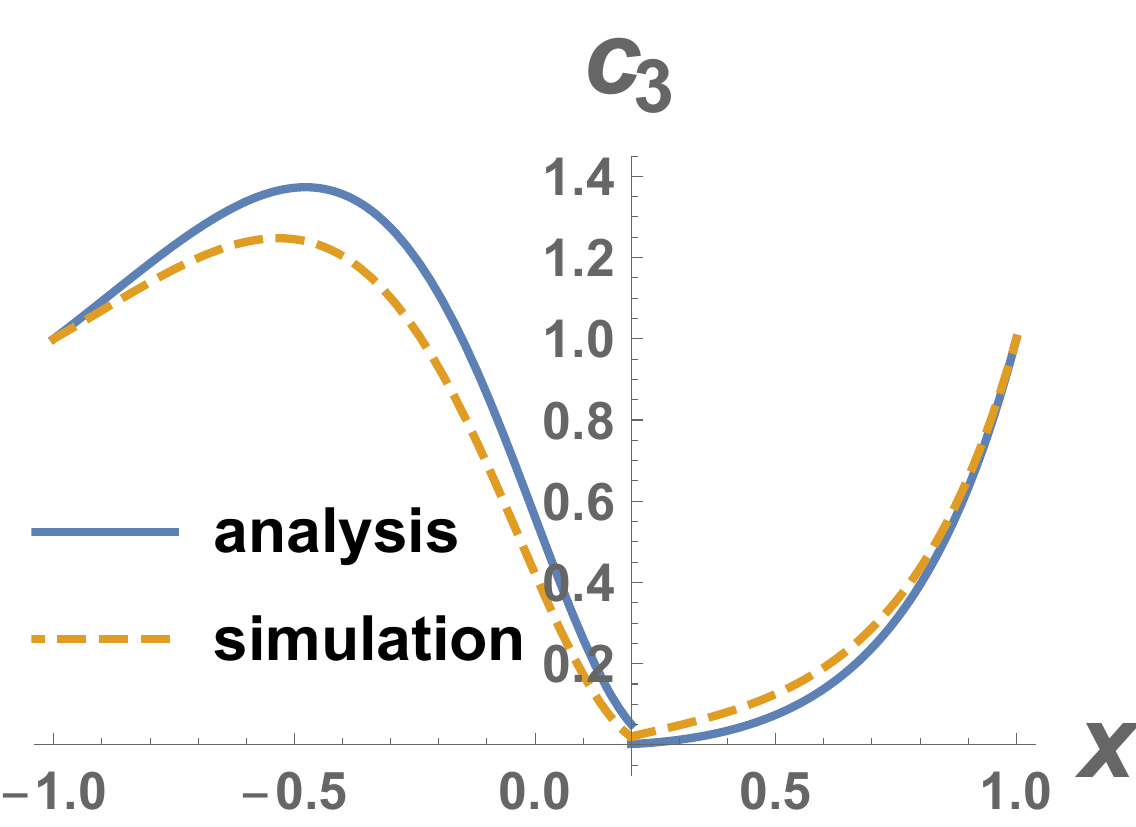}
\caption{Comparison of $\phi$ and $c_i$ ($i=1,2,3$) at steady state with $V_1 = 6.36$ between semi-analytical approximation (solid lines) and the finite-difference solution (dashed lines). }
\label{fig10}
\end{center}
\end{figure}

Given $\phi'(1)$, $\phi'(-1)$, and $J_p$, the solutions can be easily  determined numerically, in the two regions $s<x<1$ and $-1<x<s$. The three constants $\phi'(1),\phi'(-1),J_p$ can be determined by condition (\ref{eq42}), $[\phi]=0$ and $[c_1+c_2] = 0$ at $x=s$ (in practice the numerical procedure is more stable if the ratio $(c_1(s+)+c_2(s+))/(c_1(s-)+c_2(s-)) = 1$ is used instead of $[c_1+c_2] = 0$). Once $\phi(x)$ is obtained, $c_1$ and flux $J_1$ can be computed by equation $(\ref{eq40})_{2}$ and the continuity condition $[c_1] =0$ at $x=s$. Similarly, $c_2$ and flux $J_2$ can be computed by equation $(\ref{eq40})_{3}$ and the continuity condition $[c_2] =0$ at $x=s$.

\begin{figure}[h]
\begin{center}
\includegraphics[width=5cm]{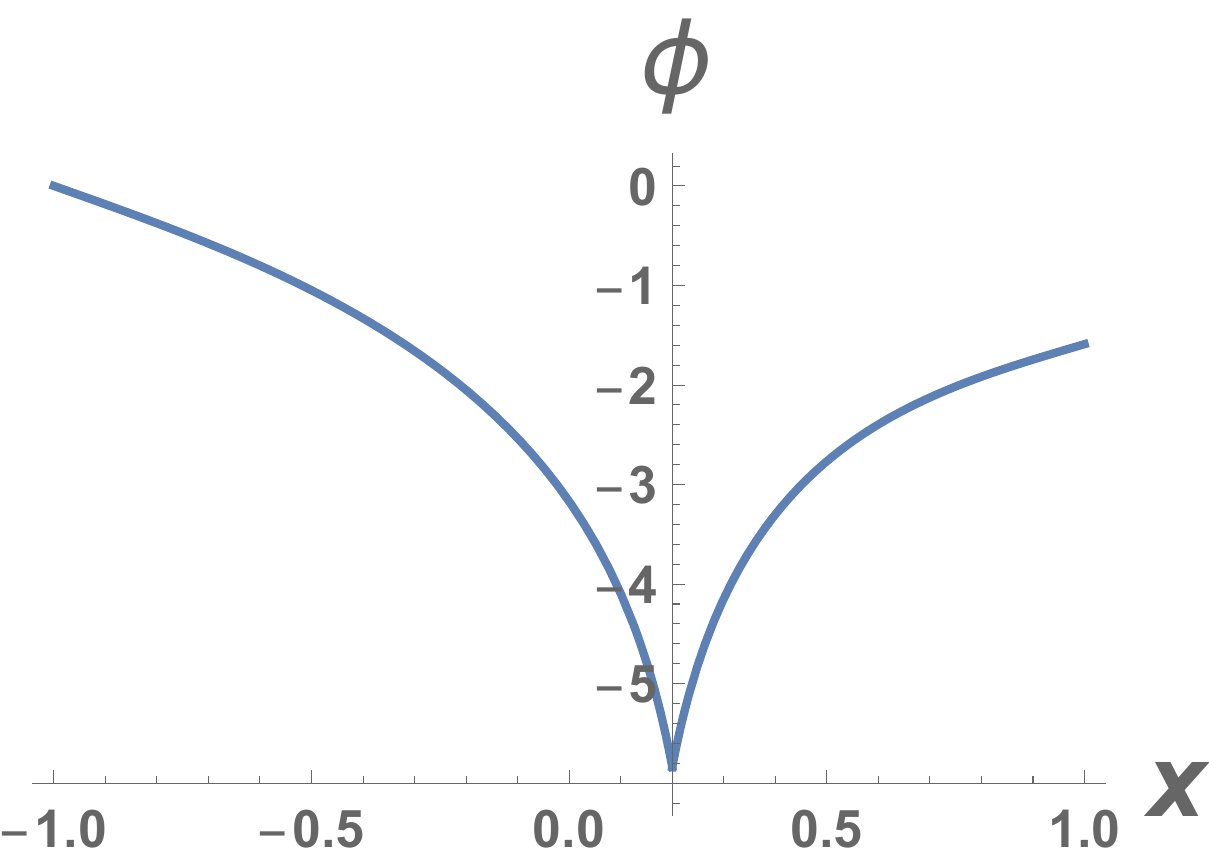} \includegraphics[width=5cm]{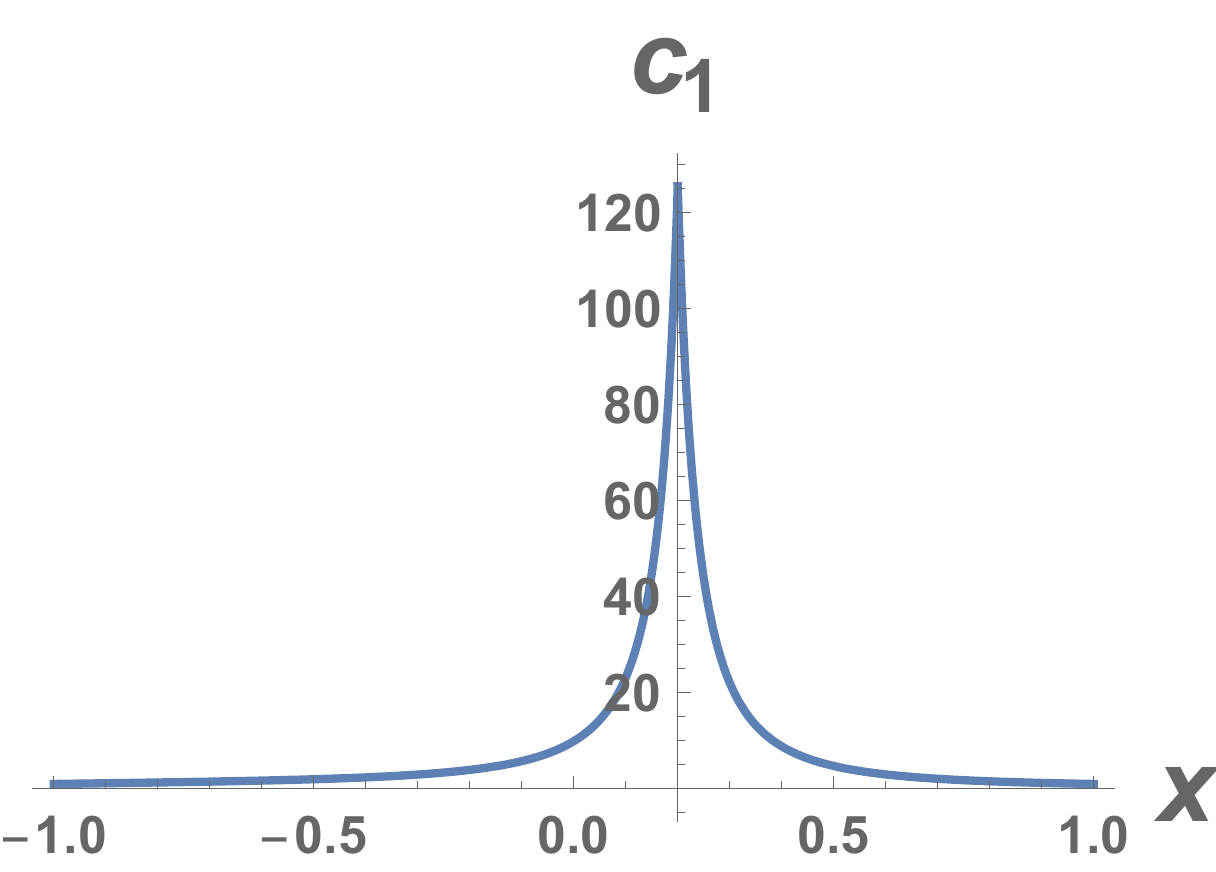}\\ \includegraphics[width=5cm]{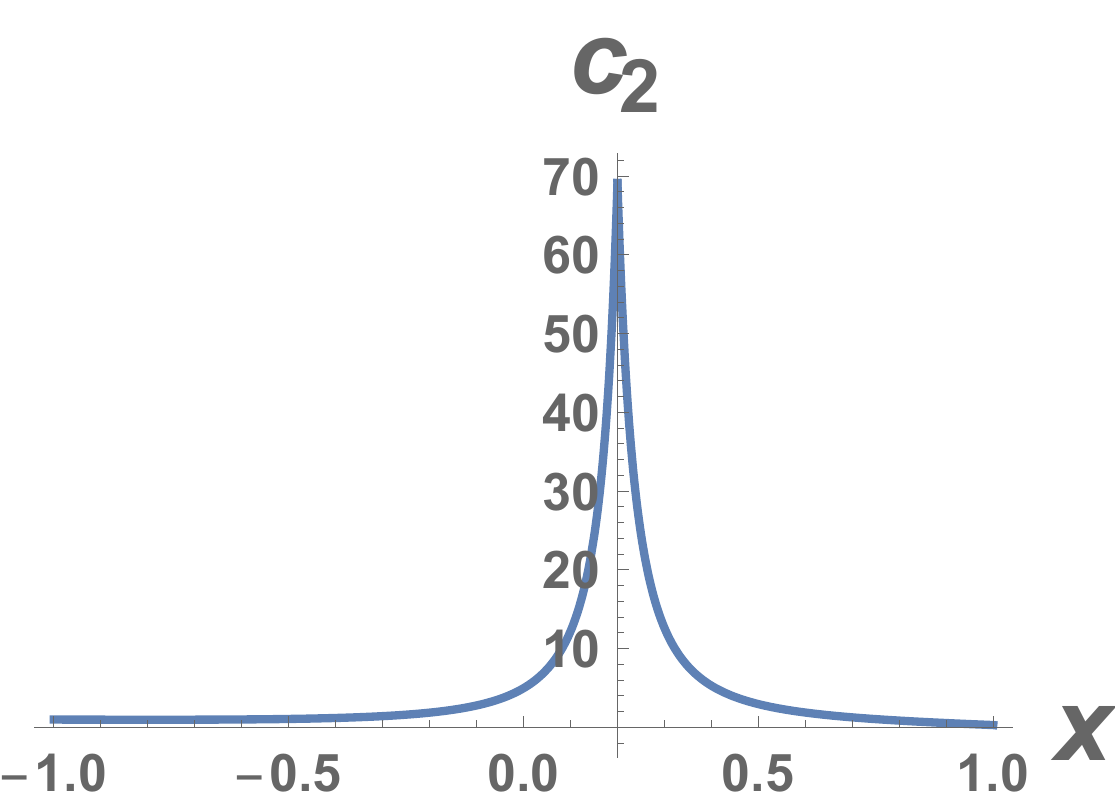}\includegraphics[width=5cm]{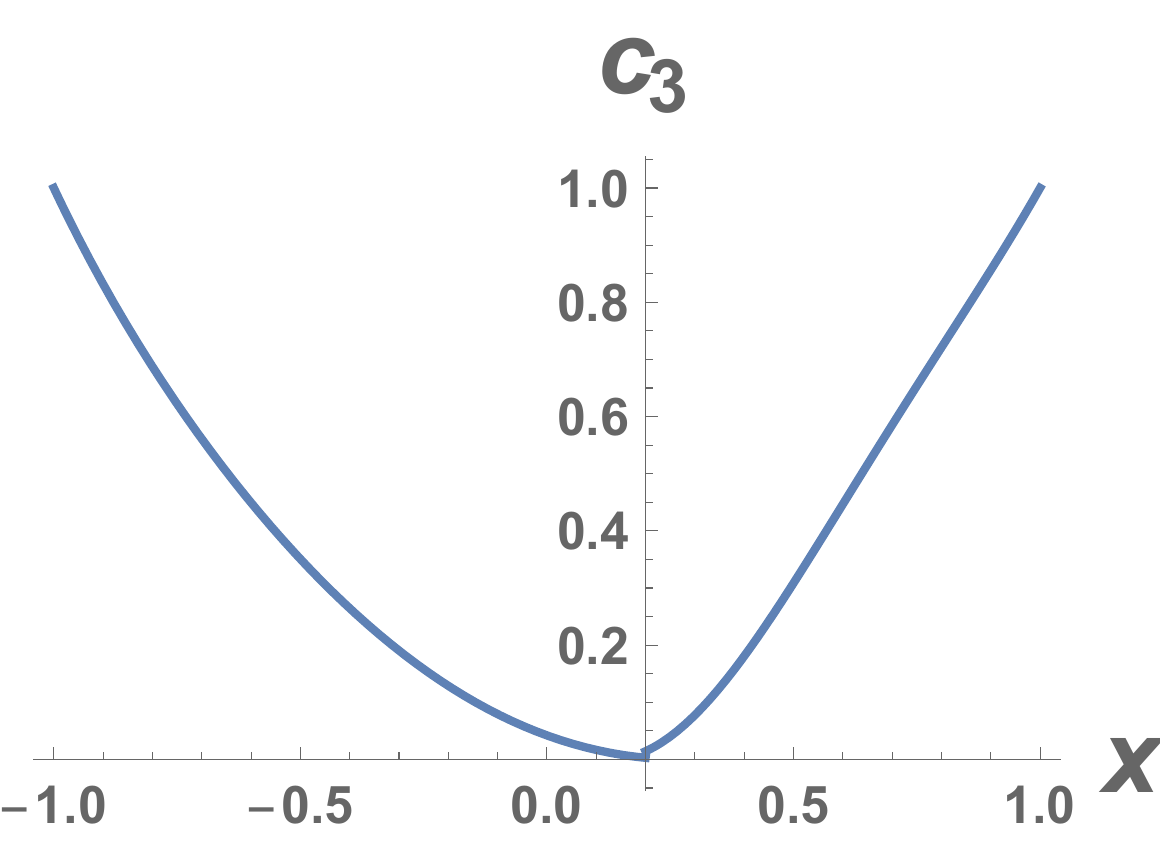}
\caption{Steady state $\phi$ and $c_i$ ($i=1,2,3$) for $V_1 = 1.59$. }
\label{fig11}
\end{center}
\end{figure}

\begin{figure}[h]
\begin{center}
\includegraphics[width=5cm]{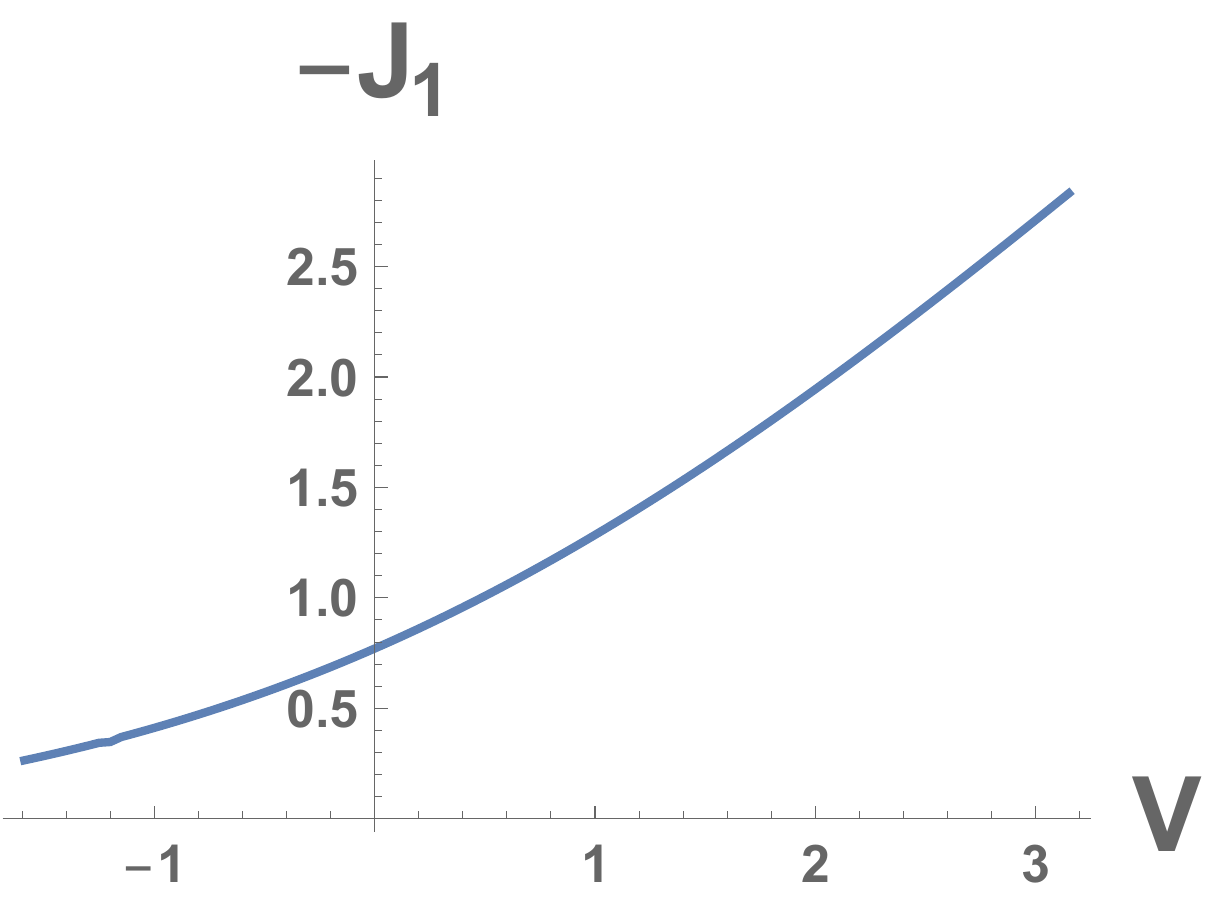}\quad \includegraphics[width=5cm]{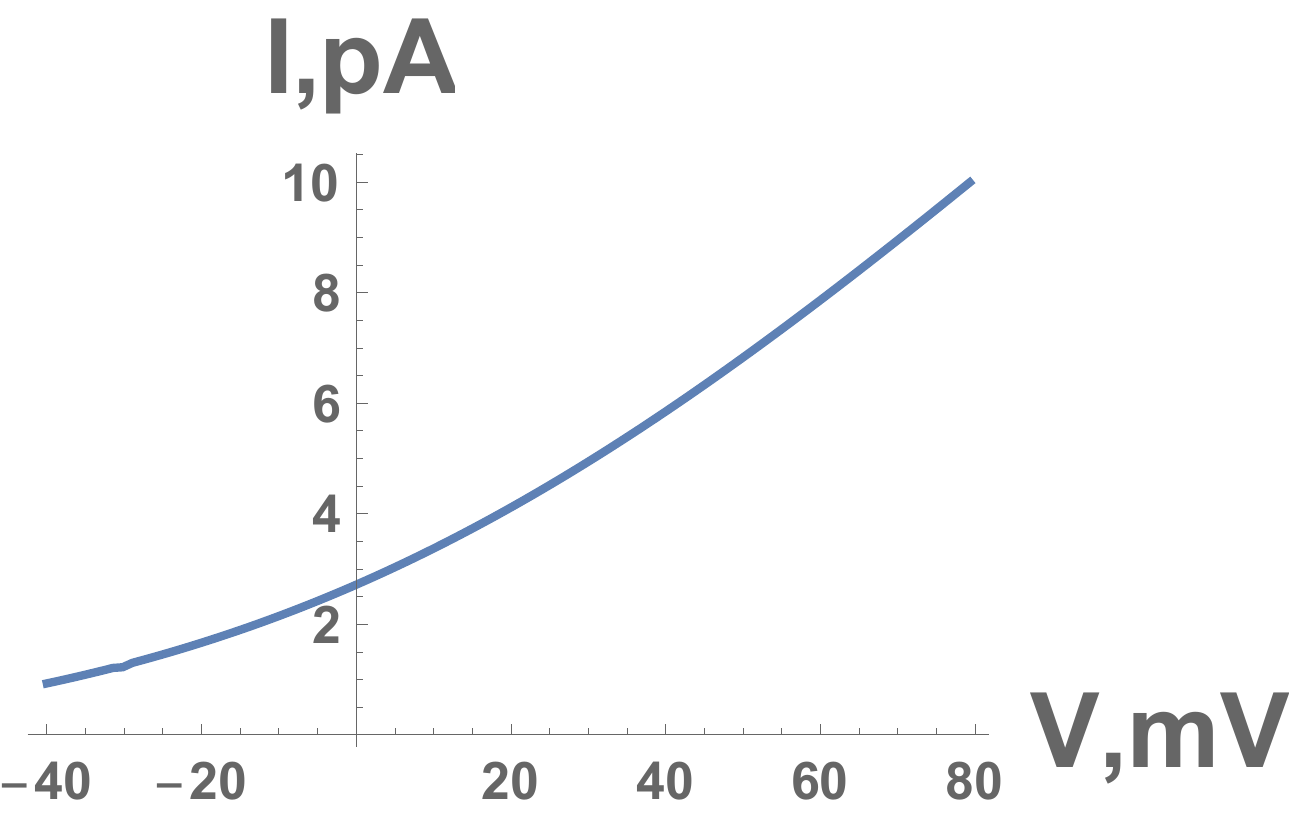}
\caption{The dependence of the dimensionless flux $J_1$ and the dimensional current $I$ on the voltage $V=V_0 + V_1$ at steady state. }
\label{fig12}
\end{center}
\end{figure}


Figure \ref{fig10} shows the semi-analytical approximation $\phi$ and $c_i$ ($i=1,2,3$) at steady state for $V_1 = 6.36$, which agrees with those in Figure \ref{fig7} except for $c_3$. For $V_1 = 6.36$, ionic flux $J_1$ is found to be $J_1\approx -2.855$ (also $\approx$ 10 pA in units), which is close to $-2.834$ in (\ref{eq34}) obtained by the finite difference method. The advantage of the method in this subsection is that the computation is extremely fast compared with the full finite difference method. It is much more efficient to use the semi-analytical approximation to compute the steady states (particularly the currents) with various different voltage jump $V_1$. 

Figure \ref{fig11} shows the results with $V_1 =1.59$ (i.e., 40 mV in physical units), and the flux is $J_1 \approx -0.264$ (i.e., 0.933 pA in units). In above computations, $J_2$ is very small since a small $D_2$ is used, therefore the dimensionless total current is almost the same as $J_1$.  Figure \ref{fig12} shows the dependence of  dimensionless flux $-J_1$ and the dimensional current on the voltage $V=V_0 + V_1$ at steady state. \\

{\noindent \bf Remark 4.} In the present simple model, the permanent charge is evenly distributed in the bubble only and the size effect of different ions are not considered. We do not expect our model to capture the current-voltage relation for large $V_1$, including the saturation phenomenon observed experimentally in the literature. To make our model more realistic, we need to know the distribution of permanent charge (i.e., acid base side chains) along the system. When that information is available, it can be incorporated into our model by adding permanent charge to the channel wall in the region $-s$ to $+s$ (see Fig. 1)  as in a practical implementation \cite{miedema2007}. The studies \cite{eisenberg2007,zhang2020} by  Weishi Liu and his group have illustrated the effects of permanent charge on current-voltage relation. With ionic size effect and the permanent charge, saturation phenomenon of current-voltage curves can be modelled as shown in \cite{song2019}. \\

\begin{figure}[h]
\begin{center}
\includegraphics[width=5cm]{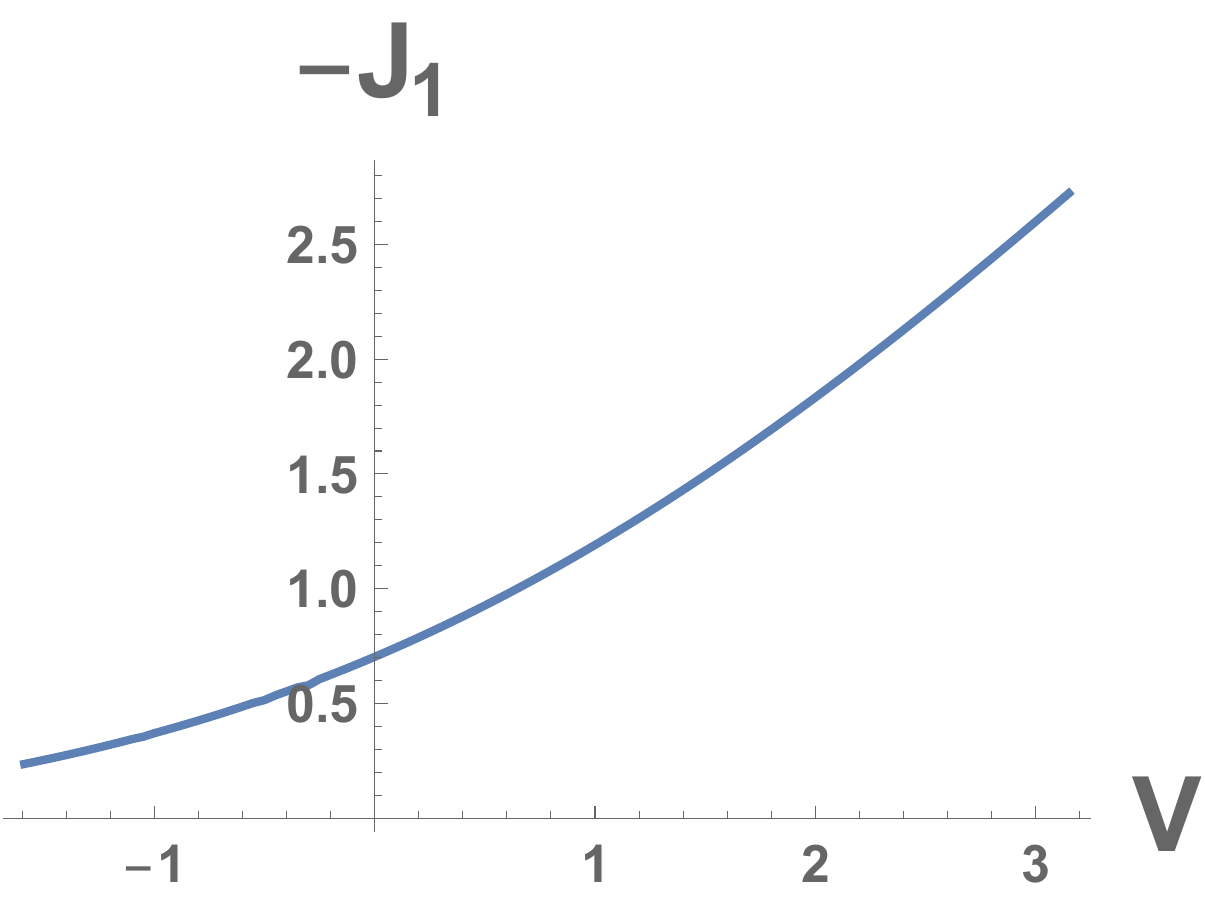}\quad \includegraphics[width=5cm]{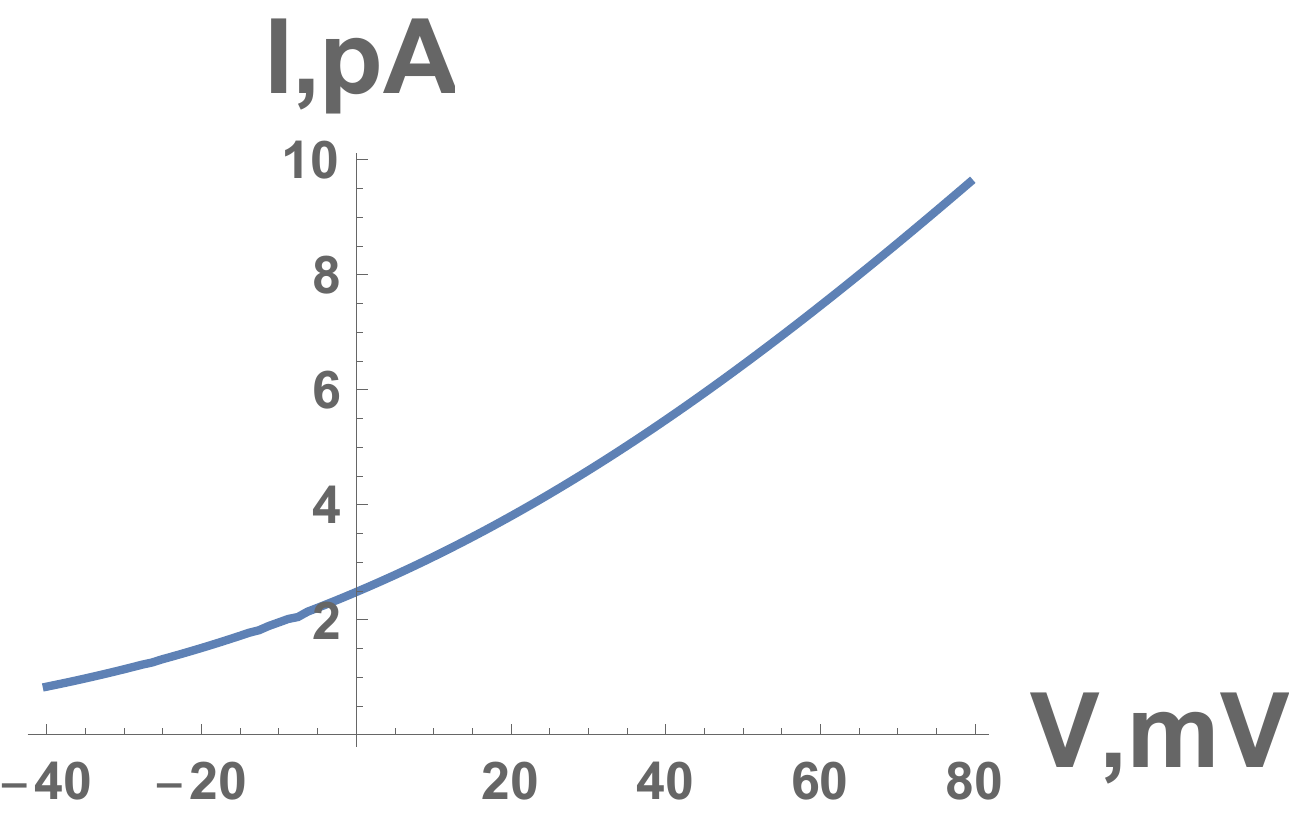}
\caption{The dependence of the dimensional flux $J_1$ and the dimensional current $I$ on the voltage $V=V_0 + V_1$ at steady state for the case that the dipole does not disappear after the bubble collapses.}
\label{fig13}
\end{center}
\end{figure}

{\noindent \bf Remark 5.} For the case that the dipole does not disappear after bubble collapses, Figure  \ref{fig13} shows the results for steady state flux, which are quite similar to those in Figure \ref{fig12} for the above case when dipole disappears after bubble collapses.

\section{Ensemble properties}

In this section, we extend our model by including stochastic effect in two aspects. We assume that the initial position of the bubble and the cross sectional area of the channel are both random and compute the ensemble averages of the macroscopic currents through the channel and estimate the Cole-Moore delay based on certain statistical distributions.

\begin{figure}[h]
\begin{center}
\includegraphics[width=6cm]{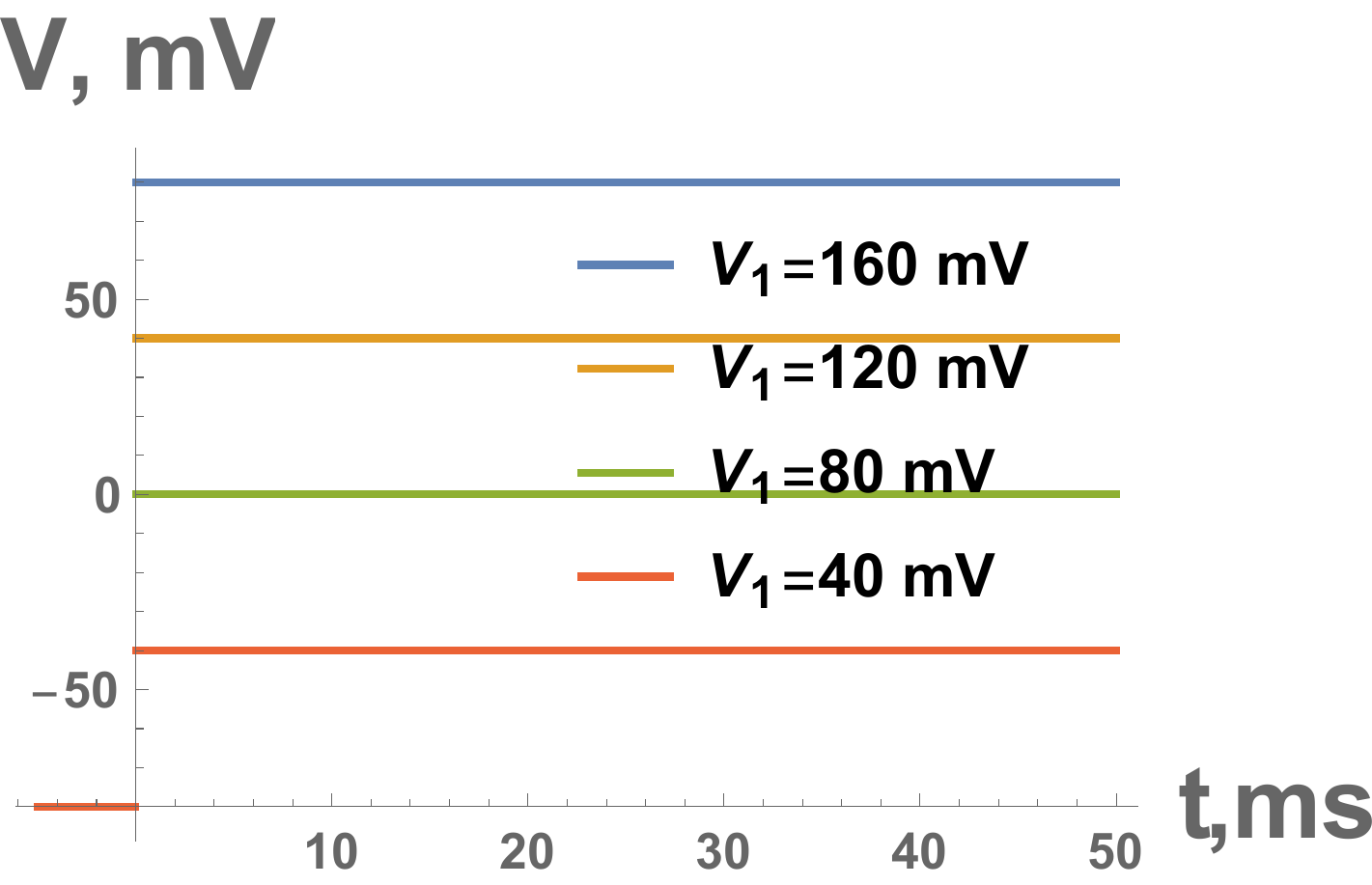} \\
\includegraphics[width=6cm]{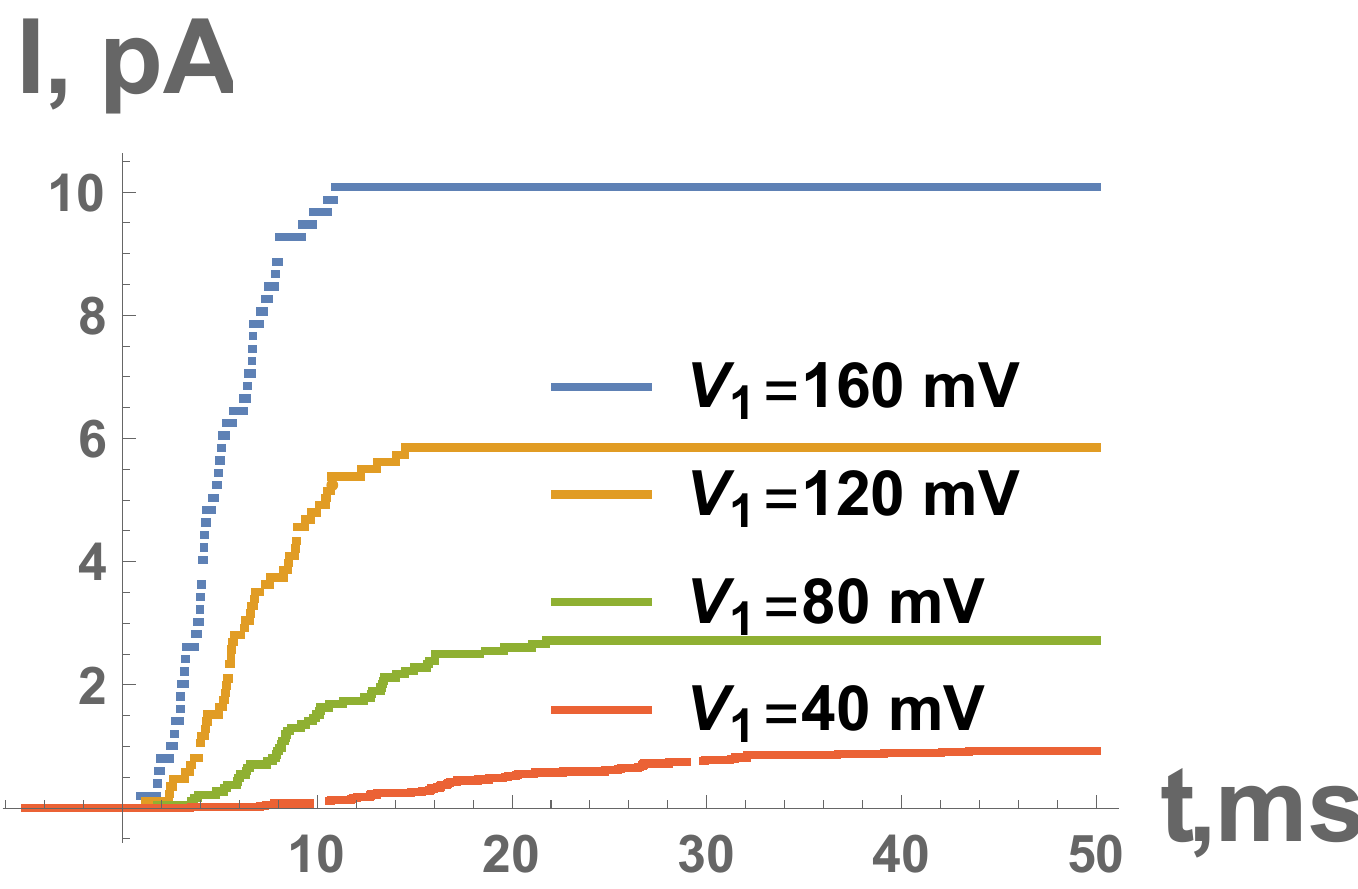} \includegraphics[width=6cm]{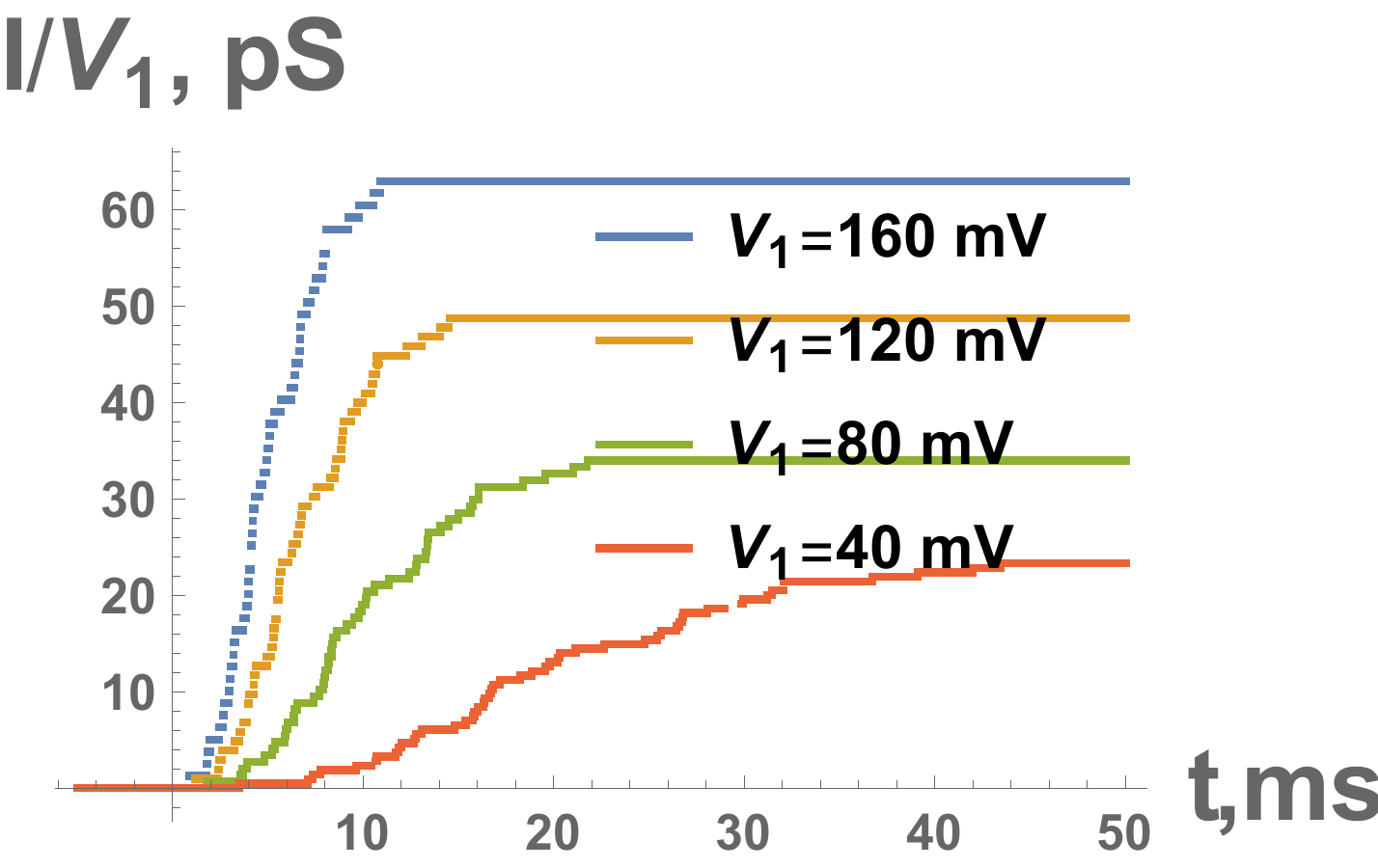}
\caption{The ensemble curves for the current $I$ and the ratio $I/V_1$ averaged by 50 random $s_b$ and with 4 different $V_1$.}
\label{fig14}
\end{center}
\end{figure}

First, we assume that the initial interface position $s_b$ is random, which could be due to the tiny fluctuations of strength of dipoles in different channels or due to the mechanism of bubble formation (which is not considered in the present work). For illustration, we consider that $s_b$ follows a normal distribution  
\begin{equation}
\label{eq43}
\begin{aligned}
s_b \sim N(\mu,\sigma^2),\quad \mu=0,\quad \sigma= 0.05,
\end{aligned}
\end{equation}
where the choice of $\sigma$ ensures that $s_b \in [-s,s]$ with $s=0.2$ for almost all the generated data. We can use the previous function $f(s_b)$ to compute the ensemble properties of the channel, since there is negligible effect on the curves of $f(s_b)$ with different starting value of $s_b$. With each different initial position $s_b$, the dynamics of the fluxes (particularly the time delay $t^\ast$ for opening of the channel) will be different. By taking the average of these fluxes, we get the ensemble curve for the dynamics of the current through the channel (i.e., fluxes of K$^+$). Figure \ref{fig14} (b,c) show the ensemble curves for the current $I$ and the ratio $I/V_1$  with 50 channels and with 4 different voltage jumps, which are given in Figure \ref{fig14}(a). Figure \ref{fig14} (b) shows similar trend and scale with experiments in Figure 2(a,c) in \cite{llano1988}.

\begin{figure}[h]
\begin{center}
\includegraphics[width=6 cm]{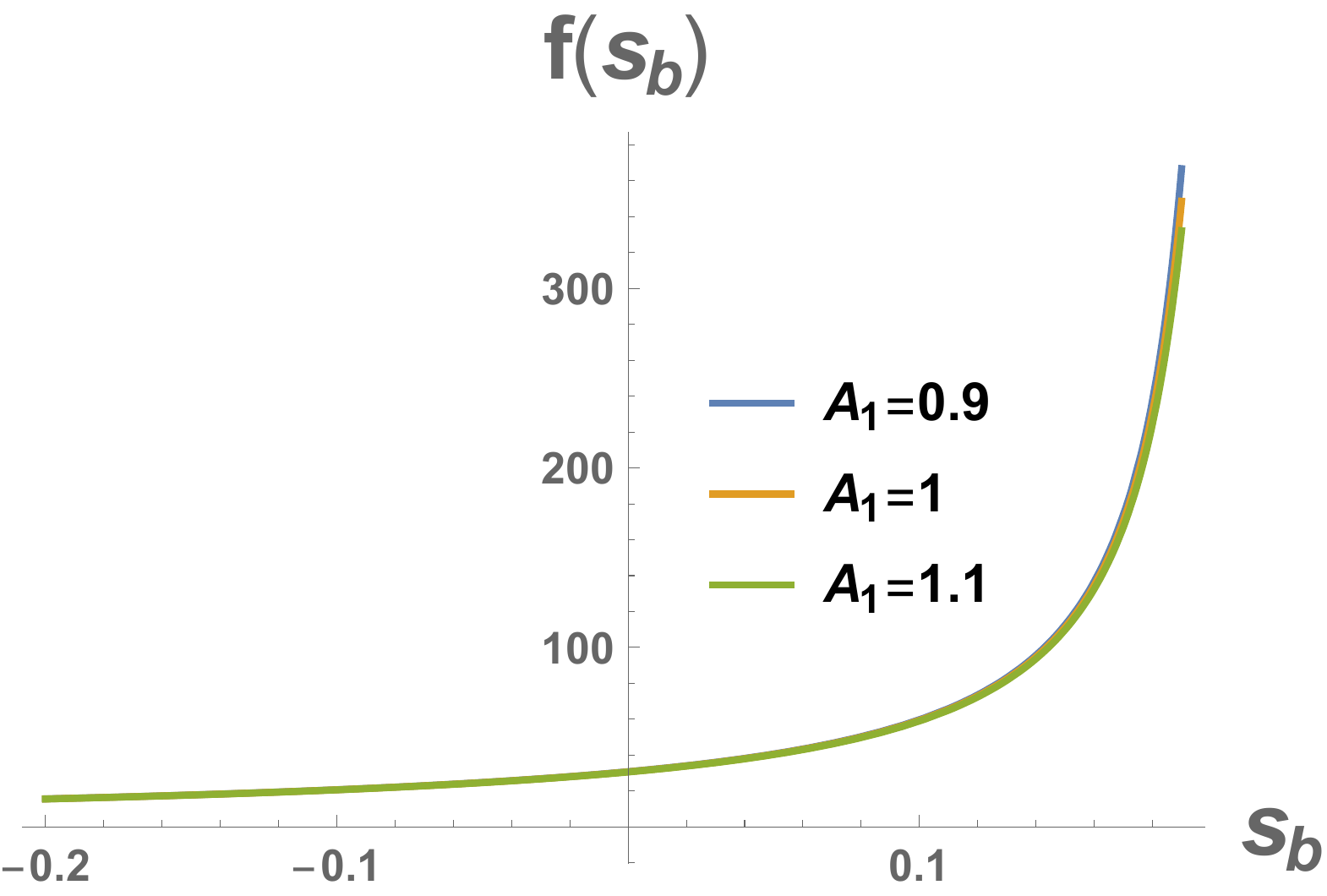}
\caption{The function $f(s_b)$  with 3 values of the parameter $A_1$.}
\label{fig15}
\end{center}
\end{figure}

\begin{figure}[h]
\begin{center}
\includegraphics[width=6 cm]{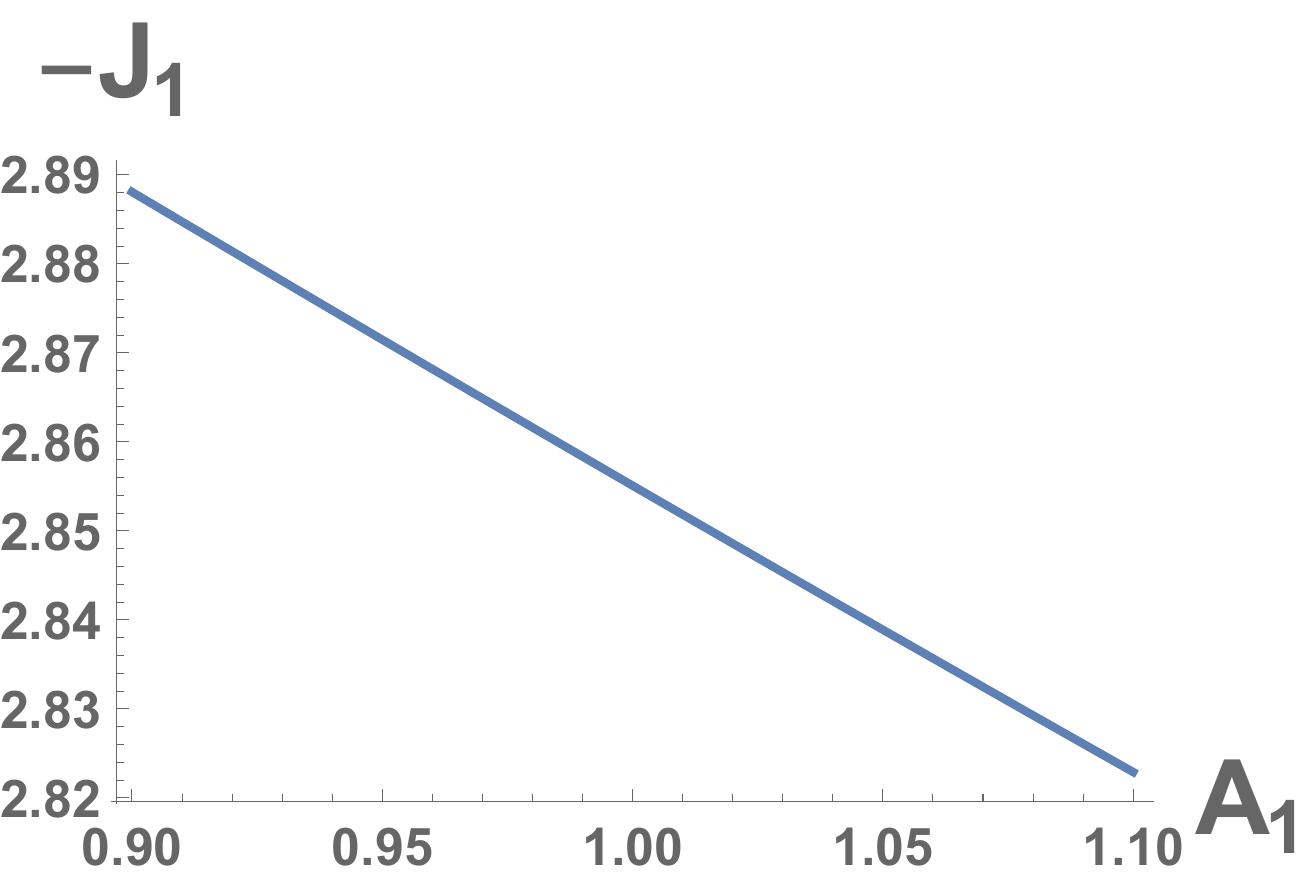} \includegraphics[width=6cm]{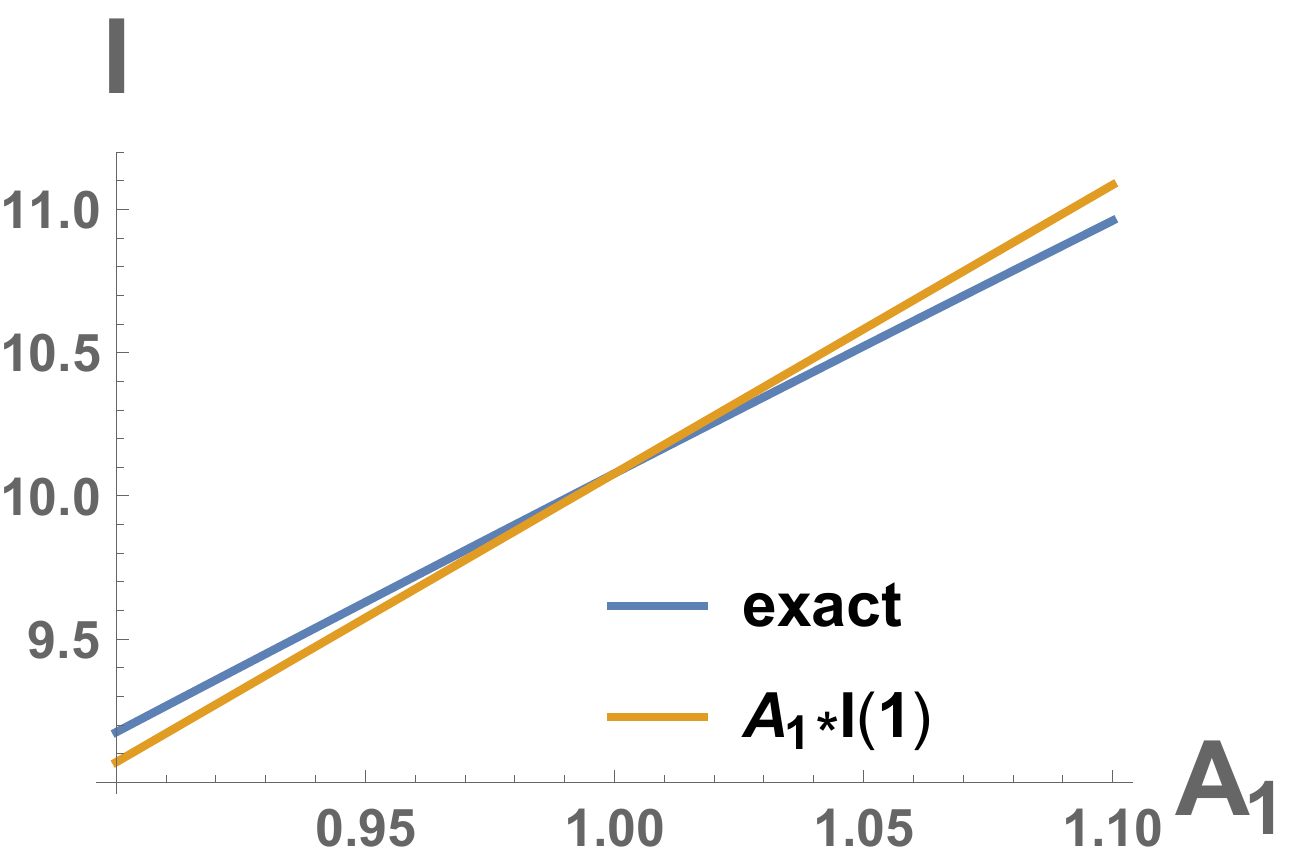}
\caption{The dependence of the flux $-J_1$ and the current $I$ on the parameter $A_1$.}
\label{fig16}
\end{center}
\end{figure}

\begin{figure}[h]
\begin{center}
\includegraphics[width=6cm]{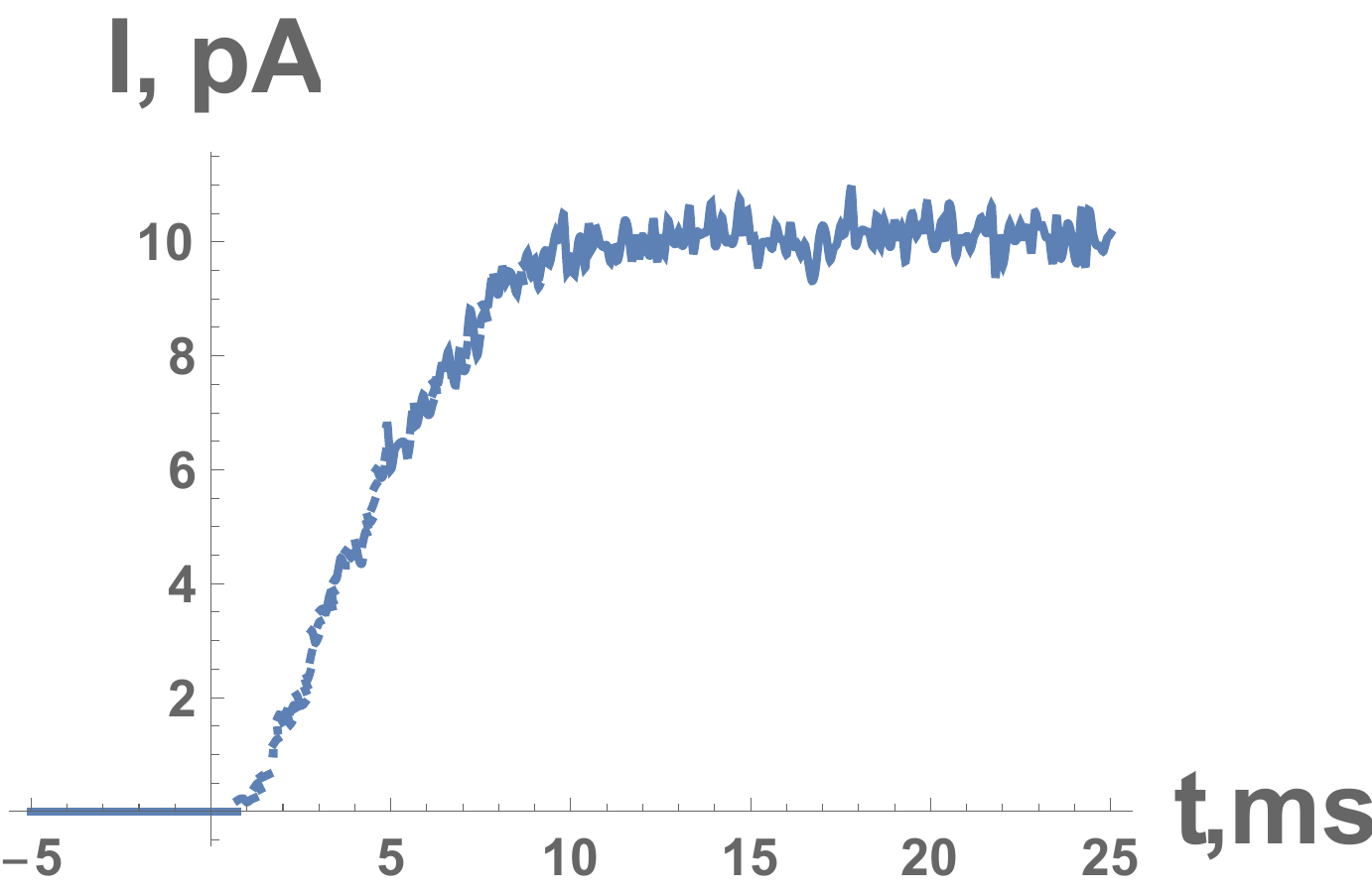}
\caption{The ensemble curves for the current with 50 channels, where $s_b\sim N(0, \sigma^2)$ and $A_1(t_i)\sim N(1,\sigma_A^2)$, with $\sigma=0.05, \sigma_A = 0.03$. }
\label{fig17}
\end{center}
\end{figure}

The case is more complicated when cross sectional area $A$ is random. We set $A=A_0 A_1$, where $A_1 \sim N(1, \sigma_A^2)$ with $\sigma_A=0.03$ and $A_0 =(0.7\mathrm{nm})^2 $ (the same as the value in Appendix A). The area $A$ will affect the dimensionless parameter $\beta$, and hence influences the effective permanent charge $q_b/\beta$. 

We start by examining the effect of $A_1$ on $t^\ast$. Figure \ref{fig15} shows $f(s_b)$ with 3 different values of $A_1$, indicating that the effect of $A_1$ on $f(s_b)$ and hence on $t^\ast$ is very small. Therefore, the previous curve $f(s_b)$ can be used to compute $t^\ast$ as an approximation. We study the effect of $A_1$ on the  flux $J_1$ or the current $I$ at steady state. Figure \ref{fig16}(a) shows the dependence of $-J_1$ on $A_1$, indicating that the magnitude of $J_1$ will slightly  decrease with increase of $A_1$.  Since the final dimensional current also depends on the scaling factor which contains $A_1$, Figure \ref{fig16}(b) shows the dependence of the current $I$ on the parameter $A_1$, indicating that the current increases with $A_1$. Figure \ref{fig16}(b) also shows the approximate current $A_1 I(1)$ where $I(1)$ is taken from previous computation with $A_1 = 1$, which is close to the exact curve. Therefore, the main effect of $A_1$ on the current is due to the scaling factor. We conclude that $A_1 I(1)$ can be used as an approximation for the current in the following figures. 

We fix $A_1$ for each channel during the evolution of the bubble, while allowing it (together with $s_b$) to vary randomly among 50 channels. The ensemble curves for the current  with  50 channels are very similar to those in Figure 14. We also consider the case that $A_1$ fluctuates randomly when the bubble evolves. We take $A_1(t_i)\sim N(1,\sigma_A^2)$ with $\sigma_A = 0.03$ for each discrete time $t=t_i$ and for each channel. Figure  \ref{fig17} shows the ensemble curve for the current with  50 channels, where  $s_b\sim N(0, \sigma^2)$ and $A_1(t_i)\sim N(1,\sigma_A^2)$, with $\sigma=0.05, \sigma_A = 0.03$, and 400 discrete $t_i$ are used for the time interval of 30 ms. Figure  \ref{fig17} shows similar trend and fluctuations with those in experiments, see Figure 2(a,c) in \cite{llano1988} and Figure 3.17 in \cite{Hille2001}.

\begin{figure}[h]
\begin{center}
\includegraphics[width=6cm]{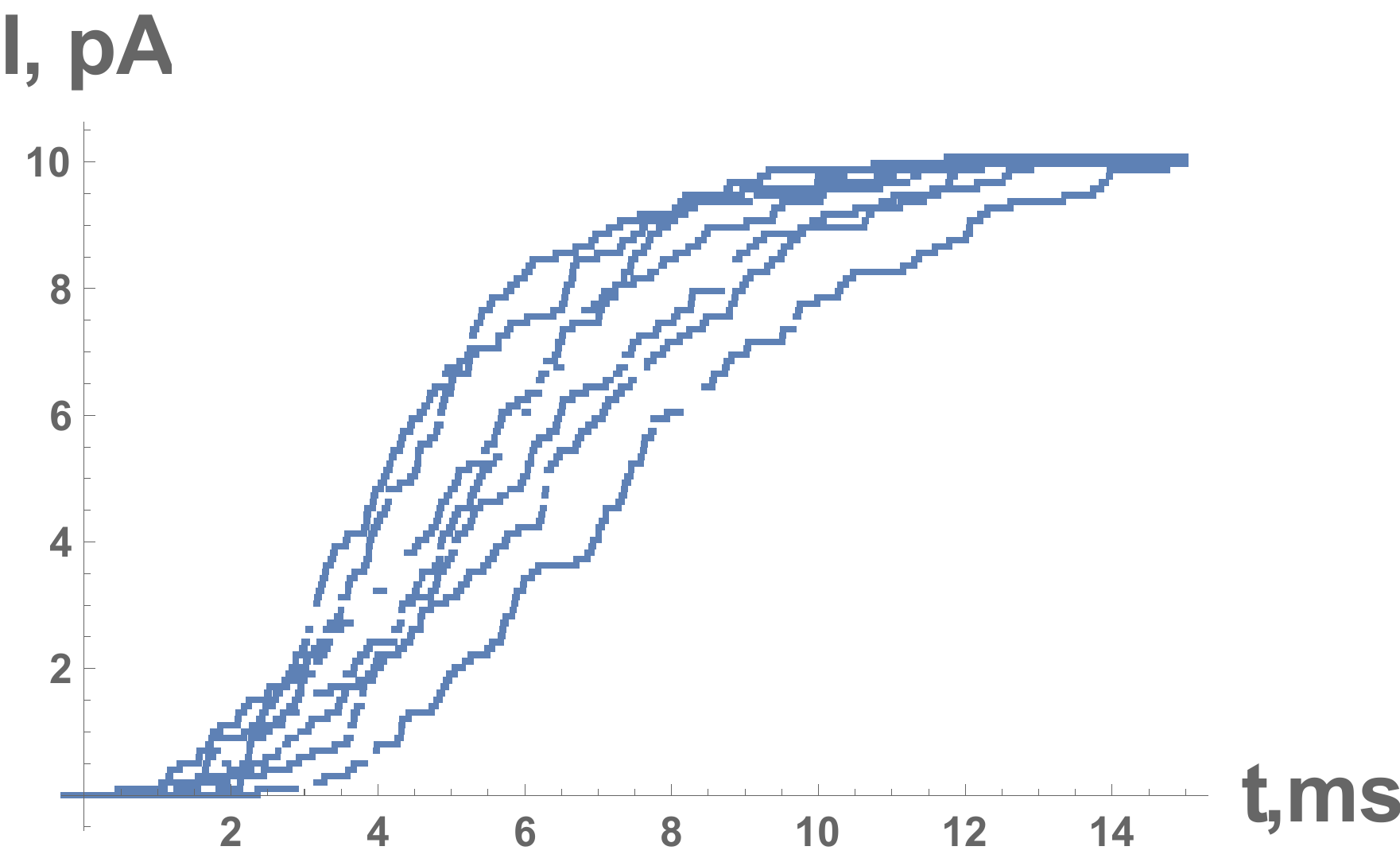} \includegraphics[width=6cm]{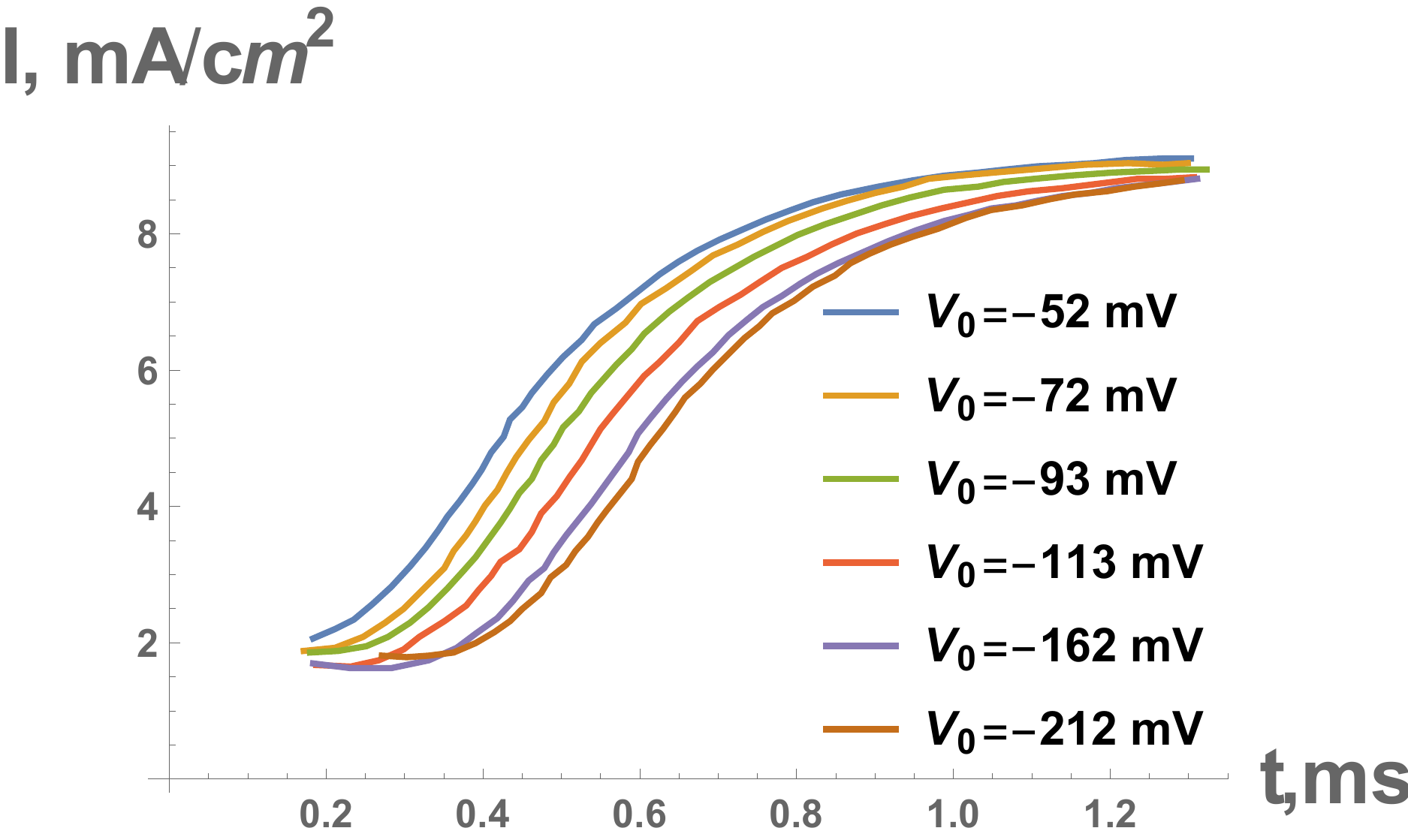} 
\caption{(a) The ensemble curves for the current with 100 channels and $V_1 = 160$ mV, for 7 different holding potentials $V_0$, (b) the ensemble curve reproduced based on the experiments in Figure 5(a) of \cite{cole1960}.}
\label{fig18}
\end{center}
\end{figure}

\begin{figure}[h]
\begin{center}
\includegraphics[width=6cm]{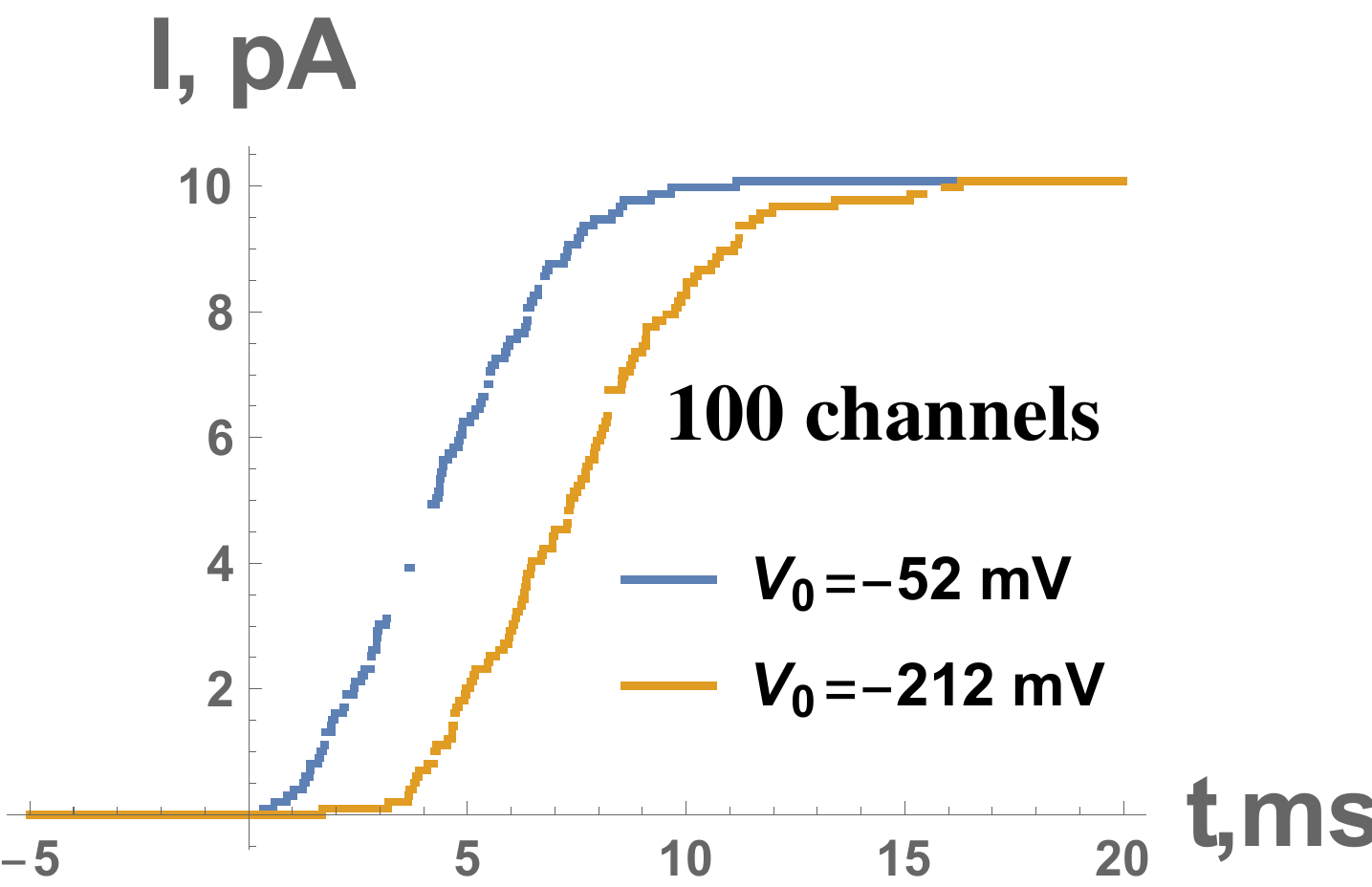} \includegraphics[width=6cm]{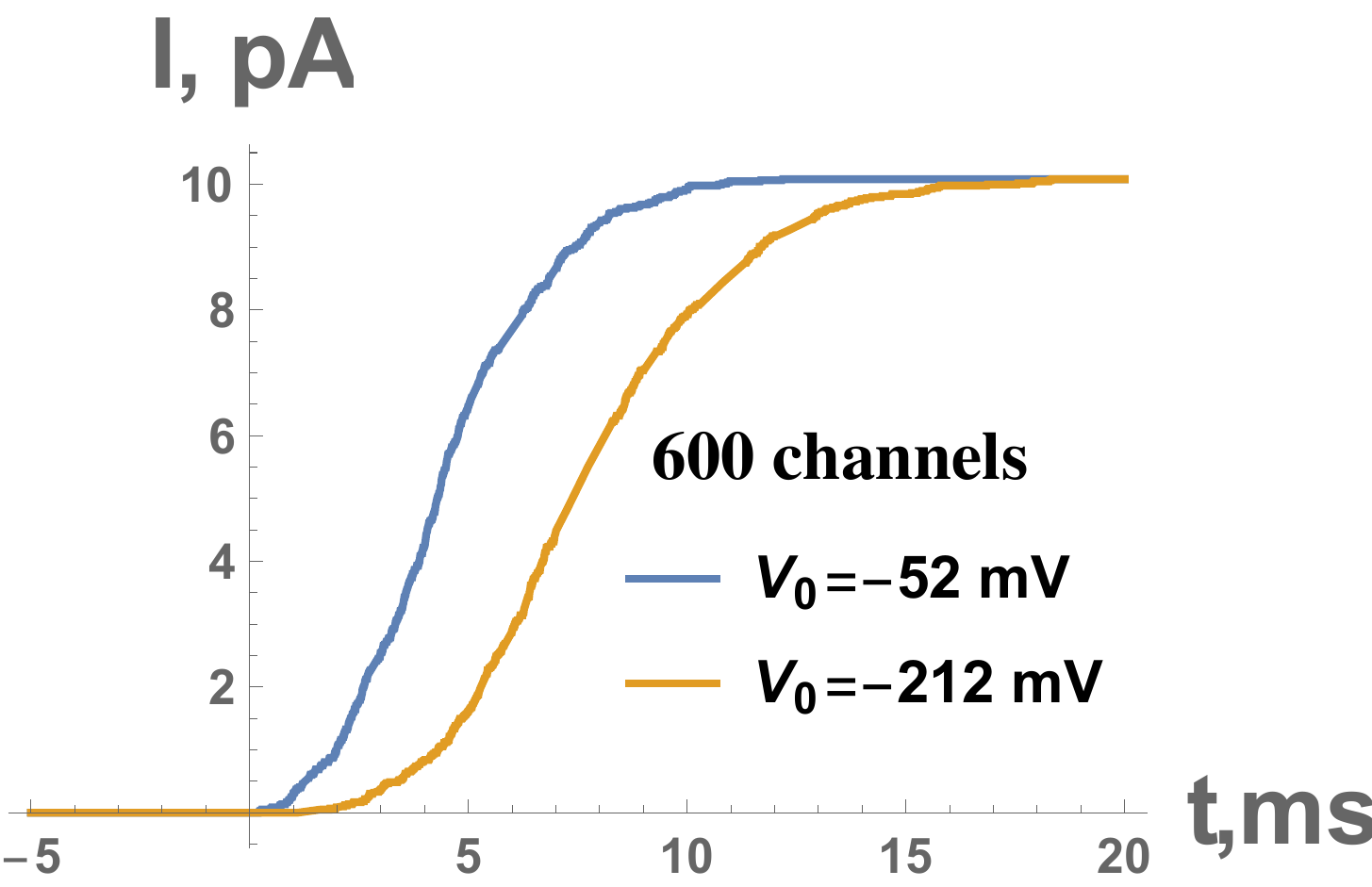} 
\caption{The ensemble curves for the current for  $V_1 = 160$ mV and 2 different holding potential $V_0$, (a) with 100 channels, and (b) with 600 channels.}
\label{fig19}
\end{center}
\end{figure}

To model the Cole-Moore delay \cite{cole1960}, we can treat the mean value $\mu$ and standard variation $\sigma$ in (\ref{eq43}) as a function of the holding potential $V_0$. For illustration, we take
\begin{equation}
\label{eq44}
\begin{aligned}
& s_b \sim N(\mu, \sigma^2),\\
& \mu(V_0)  = s \tanh  (k (V_0 -V_{0}^\ast)), \quad V_0^\ast = -80 \mathrm{mV}, \quad k=0.002 /\mathrm{mV},\\
& \sigma(V_0) = \sigma_0=0.05
\end{aligned}
\end{equation}
where $V_0$ is the initial holding potential, and $V_0^\ast$ is a reference value. Figure \ref{fig18}(a) shows the ensemble curves for the current with 100 channels and $V_1 = 160$ mV, for 7 different holding potential $V_0$ which are [-52, -72, -93, -113, -133, -162, -212]mV, corresponding to curves from left to right. The ensemble curves show similar features as experimental curves in Figure \ref{fig18}(b), which is reproduced from Figure 5(a) in \cite{cole1960}. Figure \ref{fig19} shows the ensemble curves for the current with 100 channels and 600 channels and with $V_1 = 160$ mV, for 2 different holding potential $V_0=-52,-212$ mV. It can be observed from the figures that the delay is longer when holding potential $V_0$ is smaller.

{\noindent \bf Remark 6.} We note that the Cole-Moore effect may also arise in the hydrophobic gasket of the voltage sensor region of the channel and show itself as a delay in gating current \cite{RN45984}. The bubble in the voltage sensor itself would not collapse, and the gating current would be given by our equations (\ref{eq50}-\ref{eq52}). We speculate that some of the gating current  could flow in the adjacent conduction pore, and open it, perhaps by collapsing a bubble in the conduction pore.

\section{Conclusion}

In this paper, we present a macroscopic bubble model for the gating of  K$_\mathrm{v}$ Channels. The time delay in the opening of a single channel is determined by the motion of the bubble before it collapses. The bubble motion is coupled with a Poisson-Nernst-Planck system, which is solved by a full numerical computation as well as a quasi-static approximation method. We also present a stochastic model for the bubble and channel sizes and the ensemble properties of the K$_\mathrm{v}$ channel are consistent with experimental observations. Furthermore, the Cole-Moore delay is explored by assuming the dependence of bubble properties on the holding potential.

Although the present simple model captures some key features in the ensemble properties, some parts are oversimplified and there is room for improvement. The permanent charges in the channel are lumped together in the model, and the distinction and effects of charges on bubble and on the channel wall could be examined in the future. The selectivity of channel is not considered in detail here, which depends on the ion sizes (which makes the PNP system very complicated). This is circumvented by assuming small diffusion constants of other ions except K$^+$ in the present work. The generalization to high-dimensional case is also interesting and nontrivial, since the bubble interface will have a curved shape and specific forces (e.g., some force due to the maxwell stress) can act on the interface.


\bibliographystyle{elsarticle-num.bst}
\bibliography{reference}

\begin{thebibliography}{10}
\expandafter\ifx\csname url\endcsname\relax
  \def\url#1{\texttt{#1}}\fi
\expandafter\ifx\csname urlprefix\endcsname\relax\def\urlprefix{URL }\fi
\expandafter\ifx\csname href\endcsname\relax
  \def\href#1#2{#2} \def\path#1{#1}\fi

\bibitem{biel2009}
M.~Biel, C.~Wahl-Schott, S.~Michalakis, X.~Zong, Hyperpolarization-activated
  cation channels: from genes to function, Physiological reviews 89~(3) (2009)
  847--885.

\bibitem{jensen2012}
M.~{\O}. Jensen, V.~Jogini, D.~W. Borhani, A.~E. Leffler, R.~O. Dror, D.~E.
  Shaw, Mechanism of voltage gating in potassium channels, Science 336~(6078)
  (2012) 229--233.

\bibitem{jacobson2010}
D.~A. Jacobson, F.~Mendez, M.~Thompson, J.~Torres, O.~Cochet, L.~H. Philipson,
  Calcium-activated and voltage-gated potassium channels of the pancreatic
  islet impart distinct and complementary roles during secretagogue induced
  electrical responses, The Journal of physiology 588~(18) (2010) 3525--3537.

\bibitem{Hille2001}
B.~Hille, Ion channels of excitable membranes, Sinauer Associates, Inc., 2001.

\bibitem{RN29636}
A.~Huxley, The quantitative analysis of excitation and conduction in nerve.
  from nobel lectures, physiology or medicine 1963–1970 (1972).

\bibitem{RN26041}
A.~F. Huxley, Kenneth Stewart Cole 1900-1984. A biographical memoir by Sir
  Andrew Huxley, National Academies Press, Washington DC, 1996.

\bibitem{RN12551}
A.~Huxley, From overshoot to voltage clamp, Trends in Neurosciences 25~(11)
  (2002) 553--558.

\bibitem{RN267}
A.~Hodgkin, Chance and Design, Cambridge University Press, New York, 1992.

\bibitem{RN229}
S.~J. Gould, The Flamingo's Smile: Reflections in Natural History, Norton, New
  York, 1987.

\bibitem{hodgkin1952}
A.~L. Hodgkin, A.~F. Huxley, A quantitative description of membrane current and
  its application to conduction and excitation in nerve, The Journal of
  physiology 117~(4) (1952) 500.

\bibitem{RN5856}
R.~MacKinnon, Nobel lecture. potassium channels and the atomic basis of
  selective ion conduction, Biosci Rep 24~(2)  75--100.

\bibitem{RN22885}
E.~Neher, Ion channels for communication between and within cells Nobel
  Lecture, December 9, 1991, World Scientific Publishing Co, Singapore, 1997,
  pp. 10--25.

\bibitem{RN26287}
D.~Boda, W.~Nonner, M.~Valisko, D.~Henderson, B.~Eisenberg, D.~Gillespie,
  \href{http://www.ncbi.nlm.nih.gov/pubmed/17526571}{Steric selectivity in na
  channels arising from protein polarization and mobile side chains}, Biophys J
  93~(6) (2007) 1960--80.
\newblock \href {https://doi.org/10.1529/biophysj.107.105478}
  {\path{doi:10.1529/biophysj.107.105478}}.
\newline\urlprefix\url{http://www.ncbi.nlm.nih.gov/pubmed/17526571}

\bibitem{RN45982}
T.~Dudev, C.~Lim, Factors governing the na+ vs k+ selectivity in sodium ion
  channels, Journal of the American Chemical Society 132~(7) (2010) 2321--2332.

\bibitem{RN45981}
T.~Dudev, C.~Lim, Ion selectivity strategies of sodium channel selectivity
  filters, Accounts of chemical research 47~(12) (2014) 3580--3587.

\bibitem{RN45983}
C.~Lim, T.~Dudev, Potassium versus sodium selectivity in monovalent ion channel
  selectivity filters, Springer, 2016, pp. 325--347.

\bibitem{RN11549}
F.~Bezanilla, How membrane proteins sense voltage, Nat Rev Mol Cell Biol 9~(4)
  323--32.
\newblock \href {https://doi.org/nrm2376 [pii] 10.1038/nrm2376}
  {\path{doi:nrm2376 [pii] 10.1038/nrm2376}}.

\bibitem{RN11546}
F.~Bezanilla, Ion channels: from conductance to structure, Neuron 60~(3)
  456--68.
\newblock \href {https://doi.org/S0896-6273(08)00900-8 [pii]
  10.1016/j.neuron.2008.10.035} {\path{doi:S0896-6273(08)00900-8 [pii]
  10.1016/j.neuron.2008.10.035}}.

\bibitem{RN29127}
F.~Bezanilla, \href{https://doi.org/10.1085/jgp.201812090}{Gating currents},
  The Journal of General Physiology 150~(7) (2018) 911--932.
\newblock \href {https://doi.org/10.1085/jgp.201812090}
  {\path{doi:10.1085/jgp.201812090}}.
\newline\urlprefix\url{https://doi.org/10.1085/jgp.201812090}

\bibitem{lacroix2014}
J.~J. Lacroix, H.~C. Hyde, F.~V. Campos, F.~Bezanilla, Moving gating charges
  through the gating pore in a kv channel voltage sensor, Proceedings of the
  National Academy of Sciences 111~(19) (2014) E1950--E1959.

\bibitem{catacuzzeno2019}
L.~Catacuzzeno, F.~Franciolini, Simulation of gating currents of the shaker k
  channel using a brownian model of the voltage sensor, Biophysical journal
  117~(10) (2019) 2005--2019.

\bibitem{catacuzzeno2020}
L.~Catacuzzeno, L.~Sforna, F.~Franciolini, Voltage-dependent gating in k
  channels: experimental results and quantitative models, Pfl{\"u}gers
  Archiv-European Journal of Physiology 472~(1) (2020) 27--47.

\bibitem{catacuzzeno2020b}
L.~Catacuzzeno, L.~Sforna, F.~Franciolini, R.~Eisenberg, Why are voltage gated
  na channels faster than k channels? one multi-scale hierarchical model,
  bioRxiv. Cold Spring Harbor Laboratory 11.

\bibitem{bassetto2021}
C.~A. Bassetto, J.~L. Carvalho-de Souza, F.~Bezanilla, Molecular basis for
  functional connectivity between the voltage sensor and the selectivity filter
  gate in shaker k+ channels, Elife 10 (2021) e63077.

\bibitem{RN30709}
L.~Catacuzzeno, F.~Franciolini, F.~Bezanilla, R.~S. Eisenberg,
  \href{https://dx.doi.org/10.1016/j.bpj.2021.08.015}{Gating current noise
  produced by brownian models of a voltage sensor}, Biophysical Journal
  120~(September 21, 2021) (2021) 1–19.
\newblock \href {https://doi.org/10.1016/j.bpj.2021.08.015}
  {\path{doi:10.1016/j.bpj.2021.08.015}}.
\newline\urlprefix\url{https://dx.doi.org/10.1016/j.bpj.2021.08.015}

\bibitem{catacuzzeno2021}
L.~Catacuzzeno, L.~Sforna, F.~Franciolini, R.~S. Eisenberg, Multiscale modeling
  shows that dielectric differences make nav channels faster than kv channels,
  Journal of General Physiology 153~(2).

\bibitem{horng2016}
T.-L. Horng, R.~S. Eisenberg, C.~Liu, F.~Bezanilla, Gating current models
  computed with consistent interactions, Biophysical Journal 110~(3) (2016)
  102a--103a.

\bibitem{horng2019}
T.-L. Horng, R.~S. Eisenberg, C.~Liu, F.~Bezanilla, Continuum gating current
  models computed with consistent interactions, Biophysical journal 116~(2)
  (2019) 270--282.

\bibitem{llano1986}
I.~Llano, R.~J. Bookman, Ionic conductances of squid giant fiber lobe neurons.,
  The Journal of general physiology 88~(4) (1986) 543--569.

\bibitem{llano1988}
I.~Llano, C.~K. Webb, F.~Bezanilla, Potassium conductance of the squid giant
  axon. single-channel studies., The Journal of general physiology 92~(2)
  (1988) 179--196.

\bibitem{kim2014}
I.~Kim, A.~Warshel, Coarse-grained simulations of the gating current in the
  voltage-activated kv1. 2 channel, Proceedings of the National Academy of
  Sciences 111~(6) (2014) 2128--2133.

\bibitem{bezanilla2002}
F.~Bezanilla, Voltage sensor movements, The Journal of general physiology
  120~(4) (2002) 465--473.

\bibitem{RN291}
A.~Hodgkin, A.~Huxley, B.~Katz, Ionic currents underlying activity in the giant
  axon of the squid, Arch. Sci. physiol. 3 (1949) 129--150.

\bibitem{RN135}
B.~Sakmann, E.~Neher, Single Channel Recording., 2nd Edition, Plenum, New York,
  1995.

\bibitem{RN25640}
J.~Zheng, M.~C. Trudeau, Handbook of ion channels, CRC Press, 2015.

\bibitem{hamill1981}
O.~P. Hamill, A.~Marty, E.~Neher, B.~Sakmann, F.~J. Sigworth, Improved
  patch-clamp techniques for high-resolution current recording from cells and
  cell-free membrane patches, Pfl{\"u}gers Archiv 391~(2) (1981) 85--100.

\bibitem{werry2013}
D.~Werry, J.~Eldstrom, Z.~Wang, D.~Fedida, Single-channel basis for the slow
  activation of the repolarizing cardiac potassium current, iks, Proceedings of
  the National Academy of Sciences 110~(11) (2013) E996--E1005.

\bibitem{cole1960}
K.~S. Cole, J.~W. Moore, Potassium ion current in the squid giant axon: dynamic
  characteristic, Biophysical Journal 1~(1) (1960) 1--14.

\bibitem{moore1960}
J.~W. Moore, K.~S. Cole, Resting and action potentials of the squid giant axon
  in vivo, The journal of general physiology 43~(5) (1960) 961--970.

\bibitem{taylor1960}
R.~E. Taylor, J.~W. Moore, K.~S. Cole, Analysis of certain errors in squid axon
  voltage clamp measurements, Biophysical journal 1~(2) (1960) 161--202.

\bibitem{RN45984}
M.~F. Priest, E.~E. Lee, F.~Bezanilla,
  \href{https://www.ncbi.nlm.nih.gov/pubmed/34779404}{Tracking the movement of
  discrete gating charges in a voltage-gated potassium channel}, Elife 10.
\newblock \href {https://doi.org/10.7554/eLife.58148}
  {\path{doi:10.7554/eLife.58148}}.
\newline\urlprefix\url{https://www.ncbi.nlm.nih.gov/pubmed/34779404}

\bibitem{langan2020}
P.~S. Langan, V.~G. Vandavasi, W.~Kopec, B.~Sullivan, P.~V. Afonne, K.~L.
  Weiss, B.~L. de~Groot, L.~Coates, The structure of a potassium-selective ion
  channel reveals a hydrophobic gate regulating ion permeation, IUCrJ 7~(5)
  (2020) 835--843.

\bibitem{schoppa1998}
N.~Schoppa, F.~Sigworth, Activation of shaker potassium channels: I.
  characterization of voltage-dependent transitions, The Journal of general
  physiology 111~(2) (1998) 271--294.

\bibitem{bezanilla1994}
F.~Bezanilla, E.~Perozo, E.~Stefani, Gating of shaker k+ channels: Ii. the
  components of gating currents and a model of channel activation, Biophysical
  journal 66~(4) (1994) 1011--1021.

\bibitem{tytgat1992}
J.~Tytgat, P.~Hess, Evidence for cooperative interactions in potassium channel
  gating, Nature 359~(6394) (1992) 420--423.

\bibitem{delemotte2011}
L.~Delemotte, M.~Tarek, M.~L. Klein, C.~Amaral, W.~Treptow, Intermediate states
  of the kv1. 2 voltage sensor from atomistic molecular dynamics simulations,
  Proceedings of the National Academy of Sciences 108~(15) (2011) 6109--6114.

\bibitem{delemotte2017}
L.~Delemotte, M.~A. Kasimova, D.~Sigg, M.~L. Klein, V.~Carnevale, M.~Tarek,
  Exploring the complex dynamics of an ion channel voltage sensor domain via
  computation, BioRxiv (2017) 108217.

\bibitem{peyser2012}
A.~Peyser, W.~Nonner, Voltage sensing in ion channels: Mesoscale simulations of
  biological devices, Physical Review E 86~(1) (2012) 011910.

\bibitem{dryga2012}
A.~Dryga, S.~Chakrabarty, S.~Vicatos, A.~Warshel, Coarse grained model for
  exploring voltage dependent ion channels, Biochimica et Biophysica Acta
  (BBA)-Biomembranes 1818~(2) (2012) 303--317.

\bibitem{song2020}
Z.~Song, X.~Cao, T.-L. Horng, H.~Huang, Electric discharge of electrocytes:
  Modelling, analysis and simulation, Journal of Theoretical Biology 498 (2020)
  110294.

\bibitem{song2019}
Z.~Song, X.~Cao, T.-L. Horng, H.~Huang, Selectivity of the kcsa potassium
  channel: Analysis and computation, Physical Review E 100~(2) (2019) 022406.

\bibitem{cao2020}
X.~Cao, Z.~Song, T.-L. Horng, H.~Huang, Electric potential generation of
  electrocytes: Modelling, analysis, and computation, Journal of Theoretical
  Biology 487 (2020) 110107.

\bibitem{song2018}
Z.~Song, X.~Cao, H.~Huang, Electroneutral models for a multidimensional dynamic
  poisson-nernst-planck system, Physical Review E 98~(3) (2018) 032404.

\bibitem{eisenberg2007}
B.~Eisenberg, W.~Liu, Poisson--nernst--planck systems for ion channels with
  permanent charges, SIAM Journal on Mathematical Analysis 38~(6) (2007)
  1932--1966.

\bibitem{zhang2020}
L.~Zhang, W.~Liu, Effects of large permanent charges on ionic flows via
  poisson--nernst--planck models, SIAM Journal on Applied Dynamical Systems
  19~(3) (2020) 1993--2029.

\bibitem{colquhoun1981}
D.~Colquhoun, A.~Hawkes, On the stochastic properties of single ion channels,
  Proceedings of the Royal Society of London. Series B. Biological Sciences
  211~(1183) (1981) 205--235.

\bibitem{colquhoun1995}
D.~Colquhoun, A.~G. Hawkes, The principles of the stochastic interpretation of
  ion-channel mechanisms, in: Single-channel recording, Springer, 1995, pp.
  397--482.

\bibitem{eisenberg2017}
B.~Eisenberg, X.~Oriols, D.~Ferry, Dynamics of current, charge and mass,
  Computational and Mathematical Biophysics 5~(1) (2017) 78--115.

\bibitem{eisenberg2018}
B.~Eisenberg, N.~Gold, Z.~Song, H.~Huang, What current flows through a
  resistor?, arXiv preprint arXiv:1805.04814.

\bibitem{miedema2007}
H.~Miedema, M.~Vrouenraets, J.~Wierenga, W.~Meijberg, G.~Robillard,
  B.~Eisenberg, A biological porin engineered into a molecular, nanofluidic
  diode, Nano letters 7~(9) (2007) 2886--2891.

\end{thebibliography}


\appendix 

\section{Parameter values}

We adopt the following values for the physical parameters \cite{llano1986,llano1988}
\begin{equation}
\label{eq45}
\begin{aligned}
& k_B = 1.38 \times 10^{-23} \mathrm{J}/\mathrm{K}, \quad e_0 = 1.602 \times 10^{-19} \mathrm{C}, \quad \epsilon_0 = 8.854 \times 10^{-12} \mathrm{C}/(\mathrm{V} \cdot \mathrm{m}),\\
& T = 292.15 \mathrm{K}, \quad k_B T/e_0 \approx 25.17 \textrm{mV}, \quad \epsilon_{r0} = 2, \quad \epsilon_{r1} =40, \\
&L = 0.75 \textrm{nm},\quad s= 0.15 \textrm{nm}, \quad A = (0.7 \textrm{nm})^2, \quad c_0 = 560 ~ \textrm{mM} \approx 3.37\times 10^{26} /\textrm{m}^3,\\
& c_1^L= 10 \textrm{mM}, \quad c_2^L = 550 \textrm{mM}, \quad c_3^L = 560 \textrm{mM},\\
& c_1^R= 400 \textrm{mM}, \quad c_2^R = 160 \textrm{mM}, \quad c_3^R = 560 \textrm{mM},\\
&D_0 = D_1 = 10^{-10} \textrm{m}^2/\textrm{s}, \quad D_2=D_3 = 10^{-12} \textrm{m}^2/\textrm{s}, \quad , D_b = 10^{-19} \textrm{m}^2/\textrm{s},\\
& A J_0 = A D_0 c_0  /L \approx 2.2 \times10^{7} /\textrm{s}, \quad e_0 A J_0  \approx 3.53 ~ \textrm{pA},\\
&t_0 = \frac{L^2}{D_0} = 5.625\times 10^{-9} \textrm{s}, \quad  V_0 =-80 \textrm{mV}, \quad V_1 = 160  \textrm{mV}.
\end{aligned}
\end{equation}

The dimensionless quantities are 
\begin{equation}
\label{eq46}
\begin{aligned}
& \epsilon = \frac{\epsilon_0 k_B T}{e_0^2  c_0 L^2} \approx 7.3 \times 10^{-3}, \quad \beta = LA c_0 \approx 0.12,\\
& c_1^L \approx 0.018, \quad c_2^L \approx 0.982 , \quad c_3^L = 1,\\
& c_1^R \approx 0.71,  \quad c_2^R \approx 0.29, \quad c_3^R =1,\\
& D_1 = 1, \quad D_2 =D_3= 0.01, \quad D_b = 10^{-9},\\
& V_0 = -3.18,\quad V_1 = 6.36.
\end{aligned}
\end{equation}

\section{Continuity of the total current}

\subsection{The continuous system}

The total current consists of three different types of current in different regions
\begin{itemize}
\item[(i)] the current from the change of electric field (for the whole interval/channel)
\item[(ii)] the current from the ionic fluxes (outside of the bubble)
\item[(iii)] the current from the motion of the bubble charge (in the bubble)
\end{itemize}
We will illustrate the continuity of the total current by the dimensional system in Section 2.1. For the region outside of the bubble ($-L<x<s_b, s<x<L$), we define the total current (per unit cross-sectional area) as $I_{total}^{pnp}$
\begin{equation}
\label{eq47}
\begin{aligned}
I_{total}^{pnp} (x,t) = \epsilon_0 \epsilon_r \partial_t E + \sum_{i=1}^3 e_0 z_i J_i = -\epsilon_0 \epsilon_r \partial_{tx} \phi + e_0 (J_1 + J_2-J_3).
\end{aligned}
\end{equation}
Taking the time derivative of $(\ref{eq1})_1$ and using $(\ref{eq1})_2$, we get
\begin{equation}
\label{eq48}
\begin{aligned}
\partial_x I_{total}^{pnp}=0,
\end{aligned}
\end{equation}
which implies the continuity of current outside of the bubble.

In the bubble, the define the total current  (per unit cross-sectional area) as
\begin{equation}
\label{eq49}
\begin{aligned}
I_{total}^{bubble} (x,t) = \epsilon_0 \epsilon_r \partial_t E + \partial_t {Q}_b = -\epsilon_0 \epsilon_r \partial_{tx} \phi + \partial_t {Q}_b,
\end{aligned}
\end{equation}
where ${Q}_b$ is the total bubble charge (per unit area) stored in the interval $[s_b,x]$ (it is the magnitude of total negative charge)
\begin{equation}
\label{eq50}
\begin{aligned}
{Q}_b = \int_{s_b}^x \frac{q_b}{V_b} dx = \frac{q_b (x-s_b)}{A (s-s_b)}.
\end{aligned}
\end{equation}
If ${Q}_b$ increases, that means some positive current of the bubble charge goes across the interface at $x$. Another interpretation is based on the velocity of the cross sectional surface at $x$
\begin{equation}
\label{eq51}
\begin{aligned}
v(x) = \frac{(s-x)}{(s-s_b)} \frac{d s_b}{dt},
\end{aligned}
\end{equation}
and one can easily verify that 
\begin{equation}
\label{eq52}
\begin{aligned}
\partial_t {Q}_b = J_b = -\frac{q_b}{V_b} v(x).
\end{aligned}
\end{equation}
Taking the time derivative of (\ref{eq2}), we get the continuity of the total current in the bubble
\begin{equation}
\label{eq53}
\begin{aligned}
\partial_x I_{total}^{bubble} =0.
\end{aligned}
\end{equation}

\subsection{The discrete numerical scheme}

The quantities $q_k^{n+1}$ and $\epsilon_{r,k+1/2}$ are defined as
\begin{equation}
\label{eq54}
\begin{aligned}
& q_k^{n+1} = \int_{x_{k-1/2}}^{x_{k+1/2}} \frac{1}{\beta} \frac{{q}_b}{(s-s_b)} dx,
\end{aligned}
\end{equation}
and 
\begin{equation}
\label{eq55}
\begin{aligned}
&   \epsilon_{r,k+1/2} = \frac{h}{\frac{h_1}{40} + \frac{h_2}{2}}, \quad h_1 = s_b -x_k, \quad h_2 = x_{k+1} - s_b,\quad \mathrm{if} \, s_b \in [x_k, x_{k+1}].
\end{aligned}
\end{equation}
Next we show the continuity of the total current in the discrete scheme. The  equation of $\phi$ can be written as
\begin{equation}
\label{eq56}
\begin{aligned}
& J_{\phi,k+1/2}^{n+1}- J_{\phi,k-1/2}^{n+1} = h c_{1,k}^{n+1}+  h c_{2,k}^{n+1}-h c_{3,k}^{n+1} - h q_{k}^{n+1},\\
& J_{\phi,k+1/2}^{n+1}= -\epsilon \epsilon_{r,k+1/2} \frac{\phi_{k+1}^{n+1} -\phi_k^{n+1}}{h}.
\end{aligned}
\end{equation}
Summing over $k=1,..,N$ gives
\begin{equation}
\label{eq57}
\begin{aligned}
& J_{\phi,N+1/2}^{n+1}- J_{\phi,1/2}^{n+1} = h \sum_{k=1}^{N} c_{1,k}^{n+1}+   \sum_{k=1}^{N} h c_{2,k}^{n+1} -\sum_{k=1}^{N} h c_{3,k}^{n+1}  - \sum_{k=1}^{N} h q_{k}^{n+1}.
\end{aligned}
\end{equation}
Repeat it for $J^n$ with the time step $t_n$ , take the difference, divide it by $\Delta t$, and then we get
\begin{equation}
\label{eq58}
\begin{aligned}
& \frac{J_{\phi,N+1/2}^{n+1} - J_{\phi,N+1/2}^{n}}{\Delta t}  -\frac{J_{\phi,1/2}^{n+1} - J_{\phi,1/2}^n}{\Delta t}\\
=&  h \sum_{k=1}^{N} \frac{c_{1,k}^{n+1}  -c_{1,k}^{n} } {\Delta t} +  h \sum_{k=1}^{N} \frac{c_{2,k}^{n+1}  -c_{2,k}^{n} } {\Delta t}  -  h \sum_{k=1}^{N} \frac{c_{3,k}^{n+1}  -c_{3,k}^{n} } {\Delta t} \\
& -  \frac{h}{\Delta t} \left( \sum_{k=1}^{N} q_{k}^{n+1}- \sum_{k=1}^{N} q_{k}^{n}\right)\\
=& -\left( J_{1,N+1/2}^{n+1} - J_{1,1/2}^{n+1} \right) -  \left( J_{2,N+1/2}^{n+1} - J_{2,1/2}^{n+1}  \right)+  \left( J_{3,N+1/2}^{n+1} - J_{3,1/2}^{n+1}  \right) \\
& -  \frac{h}{\Delta t}  \left(Q_b^{n+1} - Q_b^{n}\right).
\end{aligned}
\end{equation}
Rearranging the terms leads to
\begin{equation}
\label{eq59}
\begin{aligned}
& \frac{J_{\phi,N+1/2}^{n+1} - J_{\phi,N+1/2}^{n}}{\Delta t} + \left( J_{1,N+1/2}^{n+1}+ J_{2,N+1/2}^{n+1} - J_{3,N+1/2}^{n+1}\right)  +  \frac{h}{\Delta t}  \left(Q_b^{n+1} - Q_b^{n}\right)\\
=& \frac{J_{\phi,1/2}^{n+1} - J_{\phi,1/2}^n}{\Delta t} + J_{1,1/2}^{n+1}  + J_{2,1/2}^{n+1}- J_{3,1/2}^{n+1},
\end{aligned}
\end{equation}
where the three terms on the left hand side are the discrete version of the three types of currents defined in (i), (ii), (iii) in the previous subsection. 
If the sum is over the entire interval (i.e., $N+1/2$ is the right end), the term $Q_b^{n+1} - Q_b^{n}$ disappears since the total bubble charge $Q_b^n$ is conserved by definition, and the total current is conserved at the two ends.

\section{Derivation for quasi-static state and steady state}

\subsection{The quasi-static state}

For the quasi-static state, we ignore the dipole and $V_0$ (equivalently the $\phi$ is shifted up by a constant $V_0$ and continuity condition of $\phi$ will be used at interface).  We first consider the right part $s<x<1$. We get
\begin{equation}
\label{eq60}
\begin{aligned}
& c_1 = c_1^R e^{-(\phi-V_1)},\quad c_2 = c_2^R e^{-(\phi-V_1)},\quad c_3 =  e^{\phi-V_1},\\
& \epsilon \epsilon_{r1} \phi'' = e^{\phi-V_1} - e^{-(\phi-V_1)},
\end{aligned}
\end{equation}
where $c_3^R =c_1^R + c_2^R=1$ have been used. Integrating once gives
\begin{equation}
\label{eq61}
\begin{aligned}
& \frac{1}{2}\epsilon \epsilon_{r1} [(\phi'(x))^2 - (\phi'(s+))^2 ] =e^{\phi-V_1}+ e^{-(\phi-V_1)}  - B_2,
\end{aligned}
\end{equation}
where 
\begin{equation}
\label{eq62}
\begin{aligned}
& B_2 = e^{\phi_s-V_1}+ e^{-(\phi_s-V_1)},\quad \phi'(s+)= \frac{\epsilon_{r0}}{\epsilon_{r1}} \phi'(s-) = \frac{\epsilon_{r0}}{\epsilon_{r1}} \tilde{\phi}_s.
\end{aligned}
\end{equation}
Then, we obtain
\begin{equation}
\label{eq63}
\begin{aligned}
 (\phi'(x))^2 = G_1(\phi; \phi_s,\tilde{\phi}_s)& =\left(\frac{\epsilon_{r0}}{\epsilon_{r1}} \tilde{\phi}_s\right)^2 + \frac{2}{\epsilon \epsilon_{r1}}[e^{\phi-V_1}+ e^{-(\phi-V_1)}  - B_2]
\end{aligned}
\end{equation}
and
\begin{equation}
\label{eq64}
\begin{aligned}
& x=  \int_{\phi_s}^\phi \frac{1}{\sqrt{G_1(\phi; \phi_s,\tilde{\phi}_s)}} d\phi + s.
\end{aligned}
\end{equation}

For the left part $-1<x<s_b$, we get
\begin{equation}
\label{eq65}
\begin{aligned}
& \epsilon \epsilon_{r1} \phi'' = e^{\phi} - e^{-\phi},\\
& \frac{1}{2}\epsilon \epsilon_{r1} [(\phi'(x))^2 - (\phi'(s_b-))^2 ] =e^{\phi}+ e^{-\phi}  - B_3,
\end{aligned}
\end{equation}
with
\begin{equation}
\label{eq66}
\begin{aligned}
& B_3 = e^{\phi_{s_b}}+ e^{-\phi_{s_b}},\quad \phi_{s_b} = B_1 (s_b-s)^2 + \tilde{\phi}_s (s_b-s)+\phi_s\\
&\phi'(s_b-)= \frac{\epsilon_{r0}}{\epsilon_{r1}} \phi'(s_b+) = \frac{\epsilon_{r0}}{\epsilon_{r1}} [\tilde{\phi}_s+ 2 B_1 (s_b - s)],
\end{aligned}
\end{equation}
where $B_1$ is given in (\ref{eq35}). Then, we get
\begin{equation}
\label{eq67}
\begin{aligned}
 (\phi'(x))^2 = G_2(\phi; \phi_s,\tilde{\phi}_s)& =  (\phi'(s_b-))^2 + \frac{2}{\epsilon \epsilon_{r1}}[e^{\phi}+ e^{-\phi} - B_3]
\end{aligned}
\end{equation}
and
\begin{equation}
\label{eq68}
\begin{aligned}
& x=  -\int_{\phi_{s_b}}^\phi \frac{1}{\sqrt{G_2(\phi; \phi_s,\tilde{\phi}_s)}} d\phi + s_b.
\end{aligned}
\end{equation}

\subsection{The steady state}

Now we consider the steady state.  Define 
\begin{equation}
\label{eq69}
\begin{aligned}
& p= c_1 + c_2, \quad J_p = J_1+ \frac{J_2}{D_2},
\end{aligned}
\end{equation}
then the two equations for $c_1$ and $c_2$ lead to 
\begin{equation}
\label{eq70}
\begin{aligned}
& -J_p = p'(x) + p \phi'(x).
\end{aligned}
\end{equation}
Let $V=V_0 + V_1$. For the right part $x>s$, we get
\begin{equation}
\label{eq71}
\begin{aligned}
& c_3 = c_{3}^R e^{\phi-V} =e^{\phi-V}.
\end{aligned}
\end{equation}
Multiplying $\phi'$ in the equation of $\phi$ (i.e., equation $(\ref{eq40})_1$, and the delta function is put into the jump conditions) gives
\begin{equation}
\label{eq72}
\begin{aligned}
& -\epsilon \epsilon_{r1} \phi''(x) \phi'(x) = p \phi' -c_3 \phi' = -J_p -p'  -c_3',
\end{aligned}
\end{equation}
and integrating gives
\begin{equation}
\label{eq73}
\begin{aligned}
& p = \frac{1}{2} \epsilon \epsilon_{r1} \left[ (\phi'(x))^2  -  (\phi'(1))^2 \right] - J_p (x-1) - c_3 + 2,
\end{aligned}
\end{equation}
where the boundary conditions at $x=1$ have been used. Substituting into $(\ref{eq40})_1$, we get
\begin{equation}
\label{eq74}
\begin{aligned}
  -\epsilon \epsilon_{r1} \phi''(x) =\frac{1}{2} \epsilon \epsilon_{r1} \left[ (\phi'(x))^2  -  (\phi'(1))^2 \right] - J_p (x-1) - 2 (e^{\phi-V} -1),
\end{aligned}
\end{equation}
for $x>s$. Similarly for the left part $x<s$, we have
\begin{equation}
\label{eq75}
\begin{aligned}
& c_3 = c_3^L e^{\phi},\quad c_1 = \frac{1}{2} \epsilon \epsilon_{r1} \left[ (\phi'(x))^2  -  (\phi'(-1))^2 \right] - J_p(x+1) - c_3 +2,\\
 & -\epsilon \epsilon_{r1} \phi''(x) = \frac{1}{2} \epsilon \epsilon_{r1} \left[ (\phi'(x))^2  -  (\phi'(-1))^2 \right] - J_p (x+1) - 2 (e^{\phi} -1).
\end{aligned}
\end{equation}

\end{document}